\documentclass[preprint,twocolumn]{aastex63}
\pdfoutput=1 
\usepackage{amsmath,amstext}
\usepackage{nicefrac}
\usepackage{wasysym} 						
\usepackage{verbatim}
\usepackage{soul}
\usepackage[T1]{fontenc}
\usepackage{apjfonts} 
\usepackage{hyperref}
\usepackage{multirow}
\usepackage{natbib}

\newcommand{\etal}{~et~al. } 

\shorttitle{}
 \shortauthors{Meinke \etal}

\begin{document}


\title{Evidence of Extended Dust and Feedback around $z\approx1$ Quiescent Galaxies via Millimeter Observations}

\author{Jeremy Meinke}
\affiliation{Department of Physics, Arizona State University, P.O. Box 871504, Tempe, AZ 85287, USA}

\author{Seth Cohen}
\affiliation{School of Earth and Space Exploration, Arizona State University, P.O. Box 876004, Tempe, AZ 85287, USA}

\author{Jenna Moore}
\affiliation{School of Earth and Space Exploration, Arizona State University, P.O. Box 876004, Tempe, AZ 85287, USA}

\author{Kathrin B\"ockmann}
\affiliation{Universitat Hamburg, Hamburger Sternwarte, Gojenbergsweg 112, 21029, Hamburg, Germany}

\author{Philip Mauskopf}
\affiliation{Department of Physics, Arizona State University, P.O. Box 871504, Tempe, AZ 85287, USA}
\affiliation{School of Earth and Space Exploration, Arizona State University, P.O. Box 876004, Tempe, AZ 85287, USA}

\author{Evan Scannapieco}
\affiliation{School of Earth and Space Exploration, Arizona State University, P.O. Box 876004, Tempe, AZ 85287, USA}


\begin{abstract}

We use public data from the South Pole Telescope (SPT) and Atacama Cosmology Telescope (ACT) to measure radial profiles of the thermal Sunyaev-Zel'dovich (tSZ) effect and dust emission around massive quiescent galaxies at $z\approx1.$  Using survey data from the Dark Energy Survey (DES) and Wide-Field infrared Survey Explorer (WISE), we selected $387,627$ quiescent galaxies within the ACT field, with a mean stellar $\log_{10}(M_{\star}/\rm{M_{\odot}})$ of $11.40$. A subset of $94,452$ galaxies, with a mean stellar $\log_{10}(M_{\star}/\rm{M_{\odot}})$ of $11.36,$ are also covered by SPT.  In $0.5\arcmin$ bins around these galaxies, we detect the tSZ profile at levels up to $11\sigma$, and dust profile up to $20\sigma.$  Both profiles are extended, and the dust profile slope at large radii is consistent with galaxy clustering.  We analyze the thermal energy and dust mass versus stellar mass via integration within $R=2.0\arcmin$ circular apertures and fit them with a forward-modeled power-law to correct for our photometric stellar mass uncertainty.  At the mean log stellar mass of our overlap and wide-area samples, respectively, we extract thermal energies from the tSZ of $E_{\rm{pk}}=6.45_{-1.52}^{+1.67}\times10^{60}~{\rm{erg}}$ and $8.20_{-0.52}^{+0.52}\times10^{60}~{\rm{erg}},$ most consistent with moderate to high levels of active galactic nuclei feedback acting upon the circumgalactic medium.  Dust masses at the mean log stellar mass are $M_{\rm{d,pk}}=6.23_{-0.67}^{+0.67}\times10^{8}\rm{~M_{\odot}}$ and $6.76_{-0.56}^{+0.56}\times10^{8}\rm{~M_{\odot}},$ and we find a greater than linear dust-to-stellar mass relation, which indicates that the more massive galaxies in our study retain more dust.  Our work highlights current capabilities of stacking millimeter data around individual galaxies and potential for future use.

\end{abstract}



\pagenumbering{arabic}
\pagestyle{plain}


\section{Introduction}

Much is still unknown about the evolution of our universe's most massive galaxies and the processes that shaped them.   These elliptical galaxies are comprised of a central massive black hole, surrounded by a bulge of old, red stars. An additional mechanism is needed to explain the lack of young stars in these galaxies,  \citep{Silk98,Somerville2015}, and the prevailing consensus is that star formation is quenched by feedback on the surrounding environment by active galactic nuclei (AGN) \citep{Granato2004,Scannapieco2004,Croton2006,Bower2006}.  
Observations of galaxy stellar mass are well explained by AGN feedback, showing a `downsizing' or drop in star formation rate for progressively lower masses with decreasing redshift \citep{Cowie1996,Treu2005,Drory2008}, which is contrary to hierarchical models of galaxy formation with no feedback present \citep{Rees1977,White1991}.

Yet, many aspects of AGN feedback remain uncertain, with two commonly proposed feedback models.  In `quasar mode' feedback, the circumgalactic medium (CGM) surrounding the galaxy is impacted by a powerful outburst when the supermassive black hole is accreting most rapidly.  In this case, the CGM is heated such that the gas cooling time is much longer than the Hubble time, suppressing further star formation until today.  These models are supported by observations of high-velocity flows of ionized gas associated with the black holes accreting near the Eddington rate \citep{Harrison2014, Greene2014,Lansbury2018,Miller2020}. Unfortunately, uncertainty arises in the mass and energy flux from such quasars due to uncertain estimates of the outflowing material's distance from the central source \citep{Wampler1995,deKool2001,Chartas2007,Feruglio2010,Dunn2010,Veilleux2013,Chamberlain2015}.  

Second, in `radio mode' feedback, cooling material is more gradually prevented from forming stars by jets of relativistic particles that arise during periods of lower accretion rates.  Here, the CGM is maintained at a roughly constant temperature and entropy, as low levels of gas cooling are continually balanced by energy input from the relativistic jets.
Such models are supported by AGN observations of lower power jets of relativistic plasma \citep{Fabian2012}.  These couple efficiently to the volume-filling hot atmospheres of galaxies clusters \citep{McNamara2000,Churazov2001,McNamara2016}, but 
may or may not be significant for balancing cooling in less massive gravitational potentials \citep{Werner2019}.
 
One of the most promising methods for distinguishing between these models is by looking at anisotropies in the cosmic microwave background (CMB) photons passing through hot, ionized gas. Sufficiently heated gas will impose observable redshift-independent fluctuations in the CMB known as the thermal Sunyaev-Zel'dovich (tSZ) effect \citep{Sunyaev1972}.  The resulting CMB anisotropy has a distinctive frequency dependence, which causes a deficit of photons below and an excess above $\nu_{\text{null}}=217.6\,\text{GHz}$. The change in CMB temperature $\Delta T$ as a function of frequency due to the (non-relativistic) tSZ effect is given by
\begin{equation} 
\frac{\Delta{T}}{T_{\text{CMB}}}=y\left(x\frac{e^x+1}{e^x-1}-4\right),
\label{eq:DeltaT}
\end{equation}
where the dimensionless Compton-$y$ parameter is defined as
\begin{equation}
 y\equiv\int{dl}\,\sigma_T\frac{n_{e}k\left(T_{e}-T_{\rm CMB}\right)}{m_{e}c^{2}}, 
 \label{eq:y}
\end{equation}
 where $\sigma_T$ is the Thomson cross-section, $k$ is the Boltzmann constant, $m_e$ is the electron mass,
 $c$ is the speed of light, $n_e$ is the electron number density, $T_e$ is the electron temperature, $T_{\text{CMB}}=2.725$ K is the CMB temperature used throughout this paper, $l$ is the line-of-sight distance over which the integral is performed, and $x$ is the dimensionless frequency given by $x\equiv{h}\nu/kT_{\text{CMB}}=\nu/56.81\,\text{GHz}$, with Planck constant $h$. 
 
Proportional to both $n_e$ and $T,$ the Compton-$y$ parameter provides a measure of the total pressure along the line-of-sight.   Therefore by integrating the tSZ signal over a patch of sky, $y(\boldsymbol{\theta})$, we can obtain the volume integral of the pressure, and calculate the total thermal energy $E_{\rm{th}}$ in the CGM associated with a source \citep[e.g.][]{Scannapieco2008,Mroczkowski2019}.  Detailed in \citet{Spacek2016}, this gives 
\begin{equation}
E_{\rm{th}}=2.9\times10^{60} {\rm erg} \, \left(\frac{D_{\rm a}}{\text{Gpc}}\right)^{2}
\frac{\int y(\boldsymbol{\theta})~d\boldsymbol{\theta}}{10^{-6}\text{~arcmin}^{2}}.
\label{eq:EthrmT}
\end{equation}
where $D_{\rm a}$ is the angular diameter distance in Gpc and the integrated compton-$y$ is in units of $10^{-6}\text{~arcmin}^{2}$.  Throughout this work, we adopt a $\Lambda$CDM cosmological model with parameters \citep[within limits from][]{Planck2020-VI}, $h=0.68$, $\Omega_0=0.31$, $\Omega_\Lambda=0.69$, and $\Omega_{b}=0.049$, where $h$ is the Hubble constant in units of $100$ km s$^{-1}$ Mpc$^{-1}$, and $\Omega_0$, $\Omega_\Lambda$, and $\Omega_b$ are the total matter, vacuum, and baryonic densities, respectively, in units of the critical density. 

The relationship of eq.~(\ref{eq:EthrmT}) means that improvements in the sensitivity and angular resolution of tSZ measurements translate directly to better constraints on thermal energy.   Thus, cosmic structures with higher gas thermal energies, galaxy clusters, are most easily detected and indeed, have been the focus of tSZ measurements over the last decade \citep[e.g.][]{Planck2011-VIII,Reichardt2013,Planck2016-XXIV,Hilton2018,Lokken2022}.   

Further challenges arise when going to lower mass halos.  Bright targets such as quasars with abundant amounts of outflowing gas are detectable in tSZ  on an individual basis using ALMA \citep{Lacy2019,Brownson2019}.  However, averaging over many objects is currently required for appreciable detection of most samples.  \citet{Chatterjee2010} stacked quasars and galaxies with data from the Wilkinson Microwave Anisotropy Probe (WMAP) and Sloan Digital Sky Survey (SDSS) to find a tentative $\approx{2}\sigma$ tSZ signal suggesting AGN feedback;  \citet{Hand2011} used data from SDSS and the Atacama Cosmology Telescope (ACT) to see a $\approx{1}\sigma-3\sigma$ tSZ signal around galaxies; \citet{Gralla2014} found a $\approx 5\sigma$  detection for AGNs with ACT;  \citet{Ruan2015} used SDSS and Planck to find $\approx~3.5\sigma-5.0\sigma$ tSZ signals around both quasars and galaxies;  \citet{Crichton2016} used SDSS and ACT to find a $3\sigma-4\sigma$ SZ signal around quasars;  \citet{Hojjati2016} found a $\approx{7}\sigma$ tSZ detection suggestive of AGN feedback with data from Planck and the Red Cluster Sequence Lensing Survey; and \citep{Hall2019} used ACT, Herschel, and the Very Large Array data to measure the tSZ effect around $\approx100,000$ optically selected quasars, finding a $3.8\sigma$ signal that provided a joint constraint on AGN feedback and mass of the $z\gtrsim2$ quasar host halos.

Recent measurements have also been made around massive galaxies.   \citet{Greco2015} used SDSS and Planck data to compute the average tSZ signal from a range of over 100,000 `locally brightest galaxies' (LBGs) at $z\lesssim0.5$. This sample was large enough to derive constraints on $E_{\rm{th}}$  as a function of galaxy stellar mass $M_\star$ for objects with $M_\star\gtrsim2\times10^{11}~\text{M}_{\odot}$. At redshifts $0.5\lesssim{z}\lesssim{1.5}$ \citet{Spacek2016,Spacek2017} studied the tSZ signal from massive quiescent galaxies. These are prime candidates for which AGN feedback is thought to quench star formation and where a significant excess tSZ signal is expected to be produced in the CGM \citep[e.g.][]{Scannapieco2008}. \citet{Spacek2016} performed a stacking analysis with the 150 and 220 GHz South Pole Telescope's (SPT) 2011 data release, using a 43 deg$^2$ overlap with VISTA Hemisphere Survey and Blanco Cosmology Survey data to select samples of up to 3394, finding a $\approx2-3 \sigma$ signal hinting at non-gravitational heating. While \citet{Spacek2017} used SDSS and the Wide-Field Infrared Survey Explorer (WISE) data overlapping with $312~\deg^2$ of 2008/2009 ACT data at 148 and 220 GHz, finding a marginal detection that was consistent with gravitational-only heating models.  With the latest SPT release covering $2500~\deg^2$, \citet{Meinke2021} stacked nearly $140,000$ quiescent galaxies selected in a similar process from the Dark Energy Survey (DES) and WISE, to obtain a combined $10.1\sigma$ detection of tSZ at $z\approx1$.  They found the signal was most consistent with moderate forms of AGN feedback models.

Other measurements with the latest \textit{Planck} $y$-maps have been successfully conducted on nearby targets. Support for AGN feedback in local galaxy groups was found by \citet{Pratt2021}.  While \citet{Bregman2022} observed a $4.0\sigma$ detection of the tSZ effect in $11$ local $L^*$ spiral galaxies.

The recent ACT DR5 data release \citep{MallabyKay2021} has unlocked additional parts of the sky for detailed analysis. \citet{Schaan2021} and \citet{Amodeo2021} combined microwave maps from ACT and Planck with galaxy catalogs from the Baryon Oscillation Spectroscopic Survey (BOSS), to study the gas associated with these galaxy groups. They constrained the gas density profile through measurements of the tSZ signal at $\approx{10}\sigma$ and a weaker detection of the kinetic Sunyaev-Zel'dovich effect \citep[kSZ,][]{Sunyaev1980},
which is caused by peculiar motions. They were able to compare these results to cosmological simulations \citep{Battaglia2010,Springel2018} to find that the feedback employed in these models was insufficient to account for the gas heating observed at $\approx$ Mpc scales.  Meanwhile \citet{Calafut2021} and \citet{Vavagiakis2021} used SDSS and ACT to detect kSZ measurements consistent with one another.  \citet{Vavagiakis2021} also found up to a $12\sigma$ detection of the tSZ in their galaxy groups and clusters. A novel oriented stacking method was also used in \citet{Lokken2022} on DES clusters to identify tSZ associated with the cosmic web.  These are just a first step in a new wave of tSZ and kSZ analyses as more data becomes available.

A significant difficulty in accurate tSZ detection is the presence and removal of dust.  This becomes all the more important for higher redshift samples in far-infrared and millimeter bands.  Many tSZ studies have sought to simply remove this contaminant source, although there have also been an increasing number of mid- and far-infrared (MIR; FIR) studies with a primary emphasis on the dust associated with galaxies \citep{Berta2016,Gobat2018}.  Dust is an excellent tracer of galaxy characteristics such as star formation and gas, and is a key component in understanding galaxy dynamics \citep{Satini2014,Calura2016,Donevski2020}.  Despite having a lower star formation rate, dust in quiescent galaxies is still significant.  A recent study by \citet{Magdis2021} highlights a noticeable increase in dust-to-stellar mass ratio for quiescent galaxies between $z=0$ and $z=1$.

Here we expand upon the work of \citet{Meinke2021} by including the recent millimeter-wave data from ACT DR5 and conducting a more detailed analysis of dust.  Using the same quiescent galaxy selection method with DES and WISE, we now analyze data from where the SPT and ACT telescopes overlap within $\approx2,100~\deg^2$ in the Southern Hemisphere.  An ACT-only analysis is also conducted over the wider ACT field, which shares $\approx4,600~\deg^2$ with DES and WISE.  We apply a two-component fit to separate the tSZ and dust components, both in bins by radial profile and stellar mass.  We compare these profiles to expectations and other relevant studies, detecting signals up to $11\sigma$ tSZ and $20\sigma$ dust in the centermost radial bins.  Divided into stellar mass bins, we calculate the thermal energy and dust mass versus stellar mass.  We then compare our thermal energies to current simple feedback models to provide needed constraints for future simulations.

In Section~\ref{sec:data} we describe all data sets used for our analysis.  In Section~\ref{sec:sample_selection} we outline our galaxy selection procedure, and the overall properties of the massive, moderate-redshift, quiescent galaxies we use for stacking. In Section~\ref{sec:Analysis}, we detail all considerations and stacking processes used (Section~\ref{subsec:NeighborSources}-\ref{subsec:profileFits}), followed by our various results extracted from both the dust and tSZ associated with our samples (Section~\ref{subsec:dust}-\ref{subsec:Feedback}).  Discussions are given in Section~\ref{sec:Discussion}.

\section{Data}
\label{sec:data}
 
\begin{figure*}[ht]
	\centering
	\includegraphics[width=0.8\linewidth]{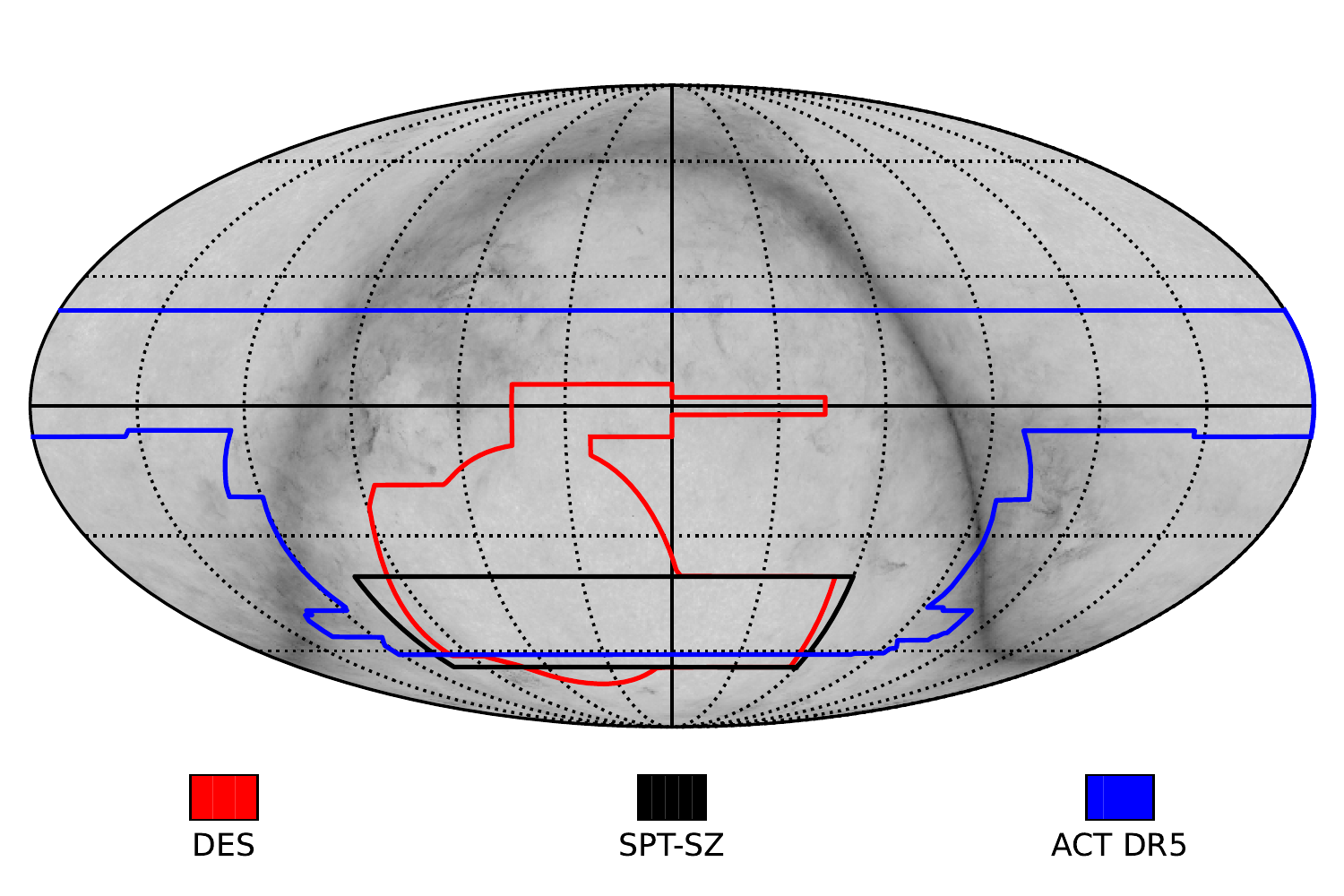}
	\caption{Mollweide (equatorial) projected sky footprints showing the coverage of DES (\textit{red}), SPT-SZ (\textit{black}), and ACT (\textit{blue}) surveys used in this analysis.  The \textit{Planck} HFI $353$ GHz is shown in the background.  This was made with the help of publicly available resources at \url{https://lambda.gsfc.nasa.gov/toolbox/footprint/}.}
	\label{fig:footprints}
\end{figure*}

Our analysis uses five public datasets: two for galaxy selection, and three to conduct our stacking analysis upon. For selection, we make use of optical and near-infrared data from DES data release 1 \citep{Abbott2018}, which are already matched to AllWISE data spanning  $3-25~\mu{m}$ \citep{Schlafly2019}.  We select and carry out photometric fitting of passive galaxies at $0.5\lesssim{z}\lesssim{1.5}$ that requires this large span of wavelengths.  Finally, the maps we stack include millimeter-wave observations from both the SPT-SZ \citep{Bocquet2019} and ACT surveys \citep{Naess2020}, along with a Planck component-separated CMB map \citep{Planck2020-I}.  
The datasets are described in more detail below.  Footprints of DES, SPT-SZ and ACT DR5 are shown in Fig.~\ref{fig:footprints}.

\subsection{DES}
DES DR1 consists of optical and near-infrared imaging from 345 nights between August 2013 to February 2016 by the Dark Energy Camera mounted on the 4-m Blanco telescope at Cerro Tololo Inter-American Observatory in Chile. The data covers $\approx{5000}~\deg^2$ of the South Galactic Cap in five photometric bands: grizY. These five bands have point-spread functions of $\rm{g}=1.12\arcsec$, $\rm{r}=0.96\arcsec$, $\rm{i}=0.88\arcsec$, $\rm{z}=0.84\arcsec$, and $\rm{Y}=0.90\arcsec$ FWHM \citep{Abbott2018}.   The survey has exposure times of 90s for griz and 45s for Y band, yielding a typical single-epoch PSF depth at S/N = 10 for $\rm{g}\lesssim23.57$, $\rm{r} \lesssim23.34$, $\rm{i}\lesssim22.78$, $\rm{z}\lesssim22.10$ and $\rm{Y}\lesssim20.69$ mag \citep{Abbott2018}. Here and below, all magnitudes are quoted in the AB system \citep[i.e.][]{Oke1983}.

\vspace{1cm}

\subsection{WISE}
The AllWISE catalog is derived from data from the 40 cm diameter Wide-field Infrared Survey Explorer (WISE) NASA Earth orbit mission \citep{wise10,neowise11}. WISE carried out an all-sky survey in 2010 of the sky in bands W1, W2, W3 and W4, centered at 3.4, 4.6, 12 and 22 $\mu$m, respectively \citep{Schlafly2019}. AllWISE uses the post-cryogenic data of the WISE mission to produce a deeper coverage in W1 and W2, which are the two bands used here. 

The added sensitivity of AllWISE extends the detection limit of luminous distant galaxies because their apparent brightness at 4.6 $\mu$m (W2) no longer declines significantly with increasing redshift. The increased sensitivity yields better detection of those galaxies for redshift $z > 1$, which are the primary focus of this analysis. 

\subsection{SPT-SZ}

The SPT-SZ survey \citep{Chown2018} covered $2,500~\deg^2$ of the southern sky between 2007 to 2011 in three different frequencies: 95 GHz and 150 GHz, which lie on either side of the maximum tSZ intensity decrement ($\approx128$ GHz), and 220 GHz, which is very near the tSZ null frequency, $\nu_{\rm{null}}=217.6$ GHz.  The South Pole Telescope (SPT) is a 10 m telescope located within 1 km of the geographical South Pole and consists of a 960-element bolometer array of superconducting transition edge sensors. 

The SPT maps used in this analysis are publicly available\footnote{\url{https://lambda.gsfc.nasa.gov/product/spt/index.cfm}} combined maps of SPT and all-sky Planck satellite (with similar bands at 100, 143, and 217 GHz).  Each combined map has a provided beam resolution of $1.85\arcmin$ FWHM, and is given in a HEALPix (Hierarchical Equal Area isoLatitude Pixelation) format with $N_{\text{side}}=8192$ \citep{Chown2018}.

\subsection{ACT}

The DR5 data release from the Atacama Cosmology Telescope (ACT) contains combined maps from observations during 2008-2018 \citep[ACT-MBAC and ACTpol,][]{Naess2020,MallabyKay2021}.  These are publicly available\footnote{\url{https://lambda.gsfc.nasa.gov/product/act/actpol_prod_table.cfm}} and cover $\approx18,000\deg^2$, predominantly in the Southern Hemisphere.  ACT uses a 6 m telescope with transition edge bolometer detectors.  The provided maps include three frequency bands centered near 90, 150, and 220 GHz.  For our purpose, we use the combined ACT+\textit{Planck}, day+night, source-free frequency maps. These have provided FWHM resolutions of 2.1\arcmin, 1.3\arcmin, and 1.0\arcmin, respectively. ACT maps differ from SPT and \textit{Planck} by projection; instead given in CAR (Plate-Carr\'ee), cylindrical coordinates of right ascension and declination.

\begin{deluxetable*}{ccccccc}[t] 
	\tablecaption{Galaxy catalogs used in this analysis with redshifts and stellar mass statistics.  \label{tab:catalogs}}
	\tablehead{
		\colhead{Sample Name} & \colhead{Map Fields} & \colhead{$N$} & \colhead{$\widetilde{z}$} & \colhead{$\overline{z}$} & \colhead{$\log_{10}(\widetilde{M_{\star}}/\text{M}_{\odot})$} & \colhead{$\log_{10}(\overline{M_{\star}}/\text{M}_{\odot})$}
	}
	\startdata
	Overlap Sample & SPT, ACT & $94,452$ & $1.031$ & $1.063$ & $11.36$ & $11.41$ \\
	Wide-Area Sample & ACT & $387,627$ & $1.037$ & $1.066$ & $11.40$ & $11.44$ \\
	\enddata
	\tablecomments{Both catalogs were selected from DES and WISE as described in Section~\ref{sec:sample_selection}. \vspace{-0.2cm}}
\end{deluxetable*}

\subsection{Planck}

The \textit{Planck} Satellite was launched in 2009 by the European Space Agency and operated from 30 to 857 GHz in 9 total frequency bands.  Taking measurements until 2013, \textit{Planck} proved invaluable to the study of CMB anisotropies and the early Universe.  Its third and ultimate data release in 2018 included full-sky frequency and component-separated maps \citep{Planck2020-I}.  Of importance to us are the \textit{Planck} CMB maps generated from various component separation techniques \citep{Planck2020-IV}.  Here we have elected to use the \textit{Planck} SMICA (Spectral Matching Independent Component Analysis) SZ-free CMB map with SZ sources projected out, to safely remove large-scale CMB anisotropies around our sample area.  This map has a resolution of $5.0\arcmin$ FWHM, provided in HEALPix format with $N_{\text{side}}=2048$.  All of the \textit{Planck} products mentioned are publicly available\footnote{\url{https://irsa.ipac.caltech.edu/data/Planck/release_3/docs/}}.

\section{Defining the Galaxy Sample}\label{sec:sample_selection}

\subsection{Selection}

We carried out our initial galaxy selection  using the DES database server at NOAO, called NOAO-Lab.  In order to start with a manageable sample, we applied a cut in color-color space designed to select old galaxies with low star-formation rates at approximately $1.0\leq{z}\leq{1.5}$ in the initial database query, as previously shown in \citet{Meinke2021}. We used mag\_auto from the DES in {\it grizy} bands, along with {\it W1} and {\it W2} PSF-magnitudes (converted to AB-system) from AllWISE \citep{wise10,neowise11} joined to the main DES table. The bands and color-selection used here are slightly different than \citet{Spacek2017} used in SDSS Stripe 82.

The NOAO Data lab allows direct queries in SQL via Jupyter notebook on their server. The lines we used to make the color selection were {\tt ((mag\_auto\_z\_dered-(w1mpro+2.699))} {\tt <= \\ (1.37*mag\_auto\_g\_dered-1.37* \\ mag\_auto\_z\_dered-0.02))} and \\ {\tt ((mag\_auto\_z\_dered-(w1mpro+2.699))>=2.0)}.

\begin{figure*}[ht]
	\gridline{\fig{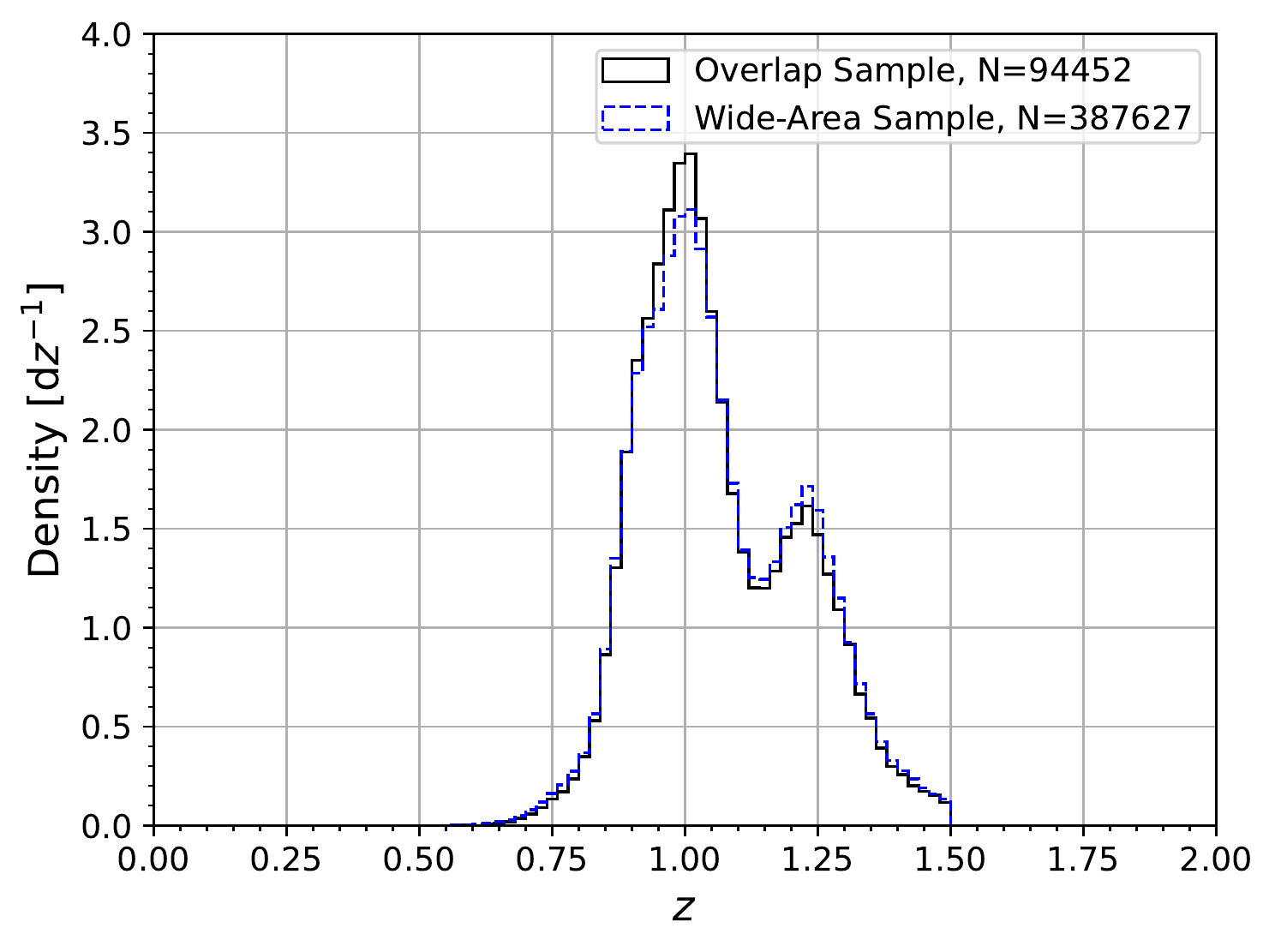}{0.5\textwidth}{	(a)} \fig{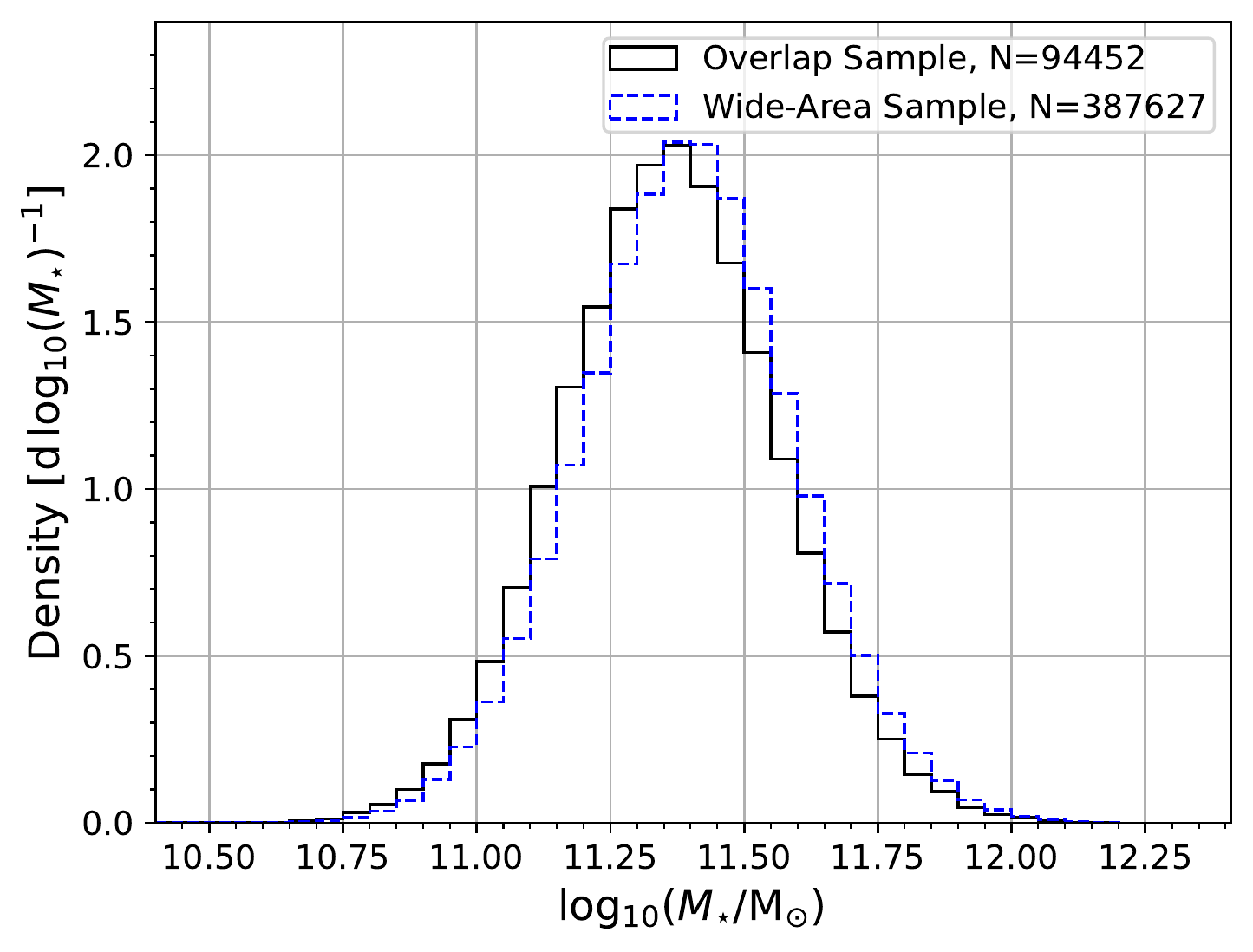}{0.5\textwidth}{	(b)}}
	\caption{ (a) Redshift and (b) $\log_{10}$ stellar mass distributions of our Overlap Sample (\textit{black}) that overlaps with both SPT and ACT fields, and a Wide-Area Sample (\textit{blue, dashed}) that utilizes the larger ACT field. Distributions shown are after SED selection, normalized by count $N$ and bin-width.}
	\label{fig:z_mass_hists}
\end{figure*}

\subsection{Photometric Fitting}

After the galaxies were selected, photometric redshifts were computed using EAZY \citep{eazy08} and the seven broad bands {\it grizyW1W2}. In calling EAZY, we used the CWW$+$KIN \citep{cww80,kinney96} templates, and did not allow for linear combinations. Since we are looking for red galaxies and have a gap in wavelength coverage between {\it y}-band and {\it W1}, we were worried that allowing combinations of templates would yield unreliable redshifts, where e.g., a red template was fit to the IR-data and a blue one was fit to the optical data and they met in the wavelength gap.

Once the redshifts were measured, we fit the spectral energy distributions (SEDs) using our own code, following the method in \citet{Spacek2017}, to which the reader is referred for more details. Briefly, a grid of BC03 \citep{bc03} models with exponentially declining star formation rates (SFRs) was fit over a range of stellar ages, SFHs (i.e., $\tau$), and dust-extinction values ($0<A_V<4$).  Our code uses BC03 models assuming a Salpeter initial mass function (IMF), but to facilitate comparisons with the literature, we convert all stellar masses to the value assuming a Chabrier IMF ($0.24$ dex offset; \citet{Santini2015}).  As in \citet{Spacek2017}, we choose as our final sample all galaxies with age$>1$ Gyr, $SSFR<0.01{\rm Gyr}^{-1}$, $0.5<z_{\rm{phot}}<1.5$, and reduced $\chi^{2}<5$.   Final redshift and stellar mass distributions are shown in Fig.~\ref{fig:z_mass_hists}.

Table~\ref{tab:catalogs} outlines the two different final catalogs used in this study.  Shared between both SPT and ACT fields is an `Overlap Sample' consisting of $94,452$ quiescent galaxies.  Meanwhile, selection of galaxies in the entire ACT field produces a larger `Wide-Area Sample' of $387,627$ galaxies.  Unlike \citet{Meinke2021}, we do not directly remove any galaxies near source contaminants in order to limit potential radial profile biases.  However both SPT and ACT maps are provided with bright sources already masked, as discussed further below. 

\section{Analysis}\label{sec:Analysis}

\subsection{Neighboring Sources}\label{subsec:NeighborSources}
The SPT-SZ maps contain an applied mask of all bright 150 GHz sources greater than $50$ mJy. This was done in \citet{Chown2018}, through the removal of all signal within $5\arcmin$ and apodization with a $5\arcmin$ Gaussian beam.  For our purposes these locations result in a large hole that potentially skews measurements.  We avoid them by using the SPT-SZ provided mask to remove any targets within $20\arcmin$ of a masked pixel.  The statistics for our Overlap Sample as listed in Table~\ref{tab:catalogs} are determined after the removal process has occurred.  The random catalog in the overlap field, described in Section~\ref{subsec:RandomCatalog}, also applies this removal process.

Similarly, we have chosen to use the source-free ACT maps.  They however differ from SPT-SZ, as all sources removed were done so using a finer matched filter and fitting procedure \citep{Naess2020}.  We have found this source removal process has a minimal effect on our stacking results.

\subsection{Map Processing}\label{subsec:MapProcessing}
The SPT and ACT maps span similar frequency bands and regions of the sky, making them ideal products for tSZ and dust comparisons.  However, we employ multiple steps to further process the maps into similar formats and ensure all likely systemic differences are minimized.  Notably:

\begin{figure*}[ht]
	\centering
	\includegraphics[width=1.0\linewidth]{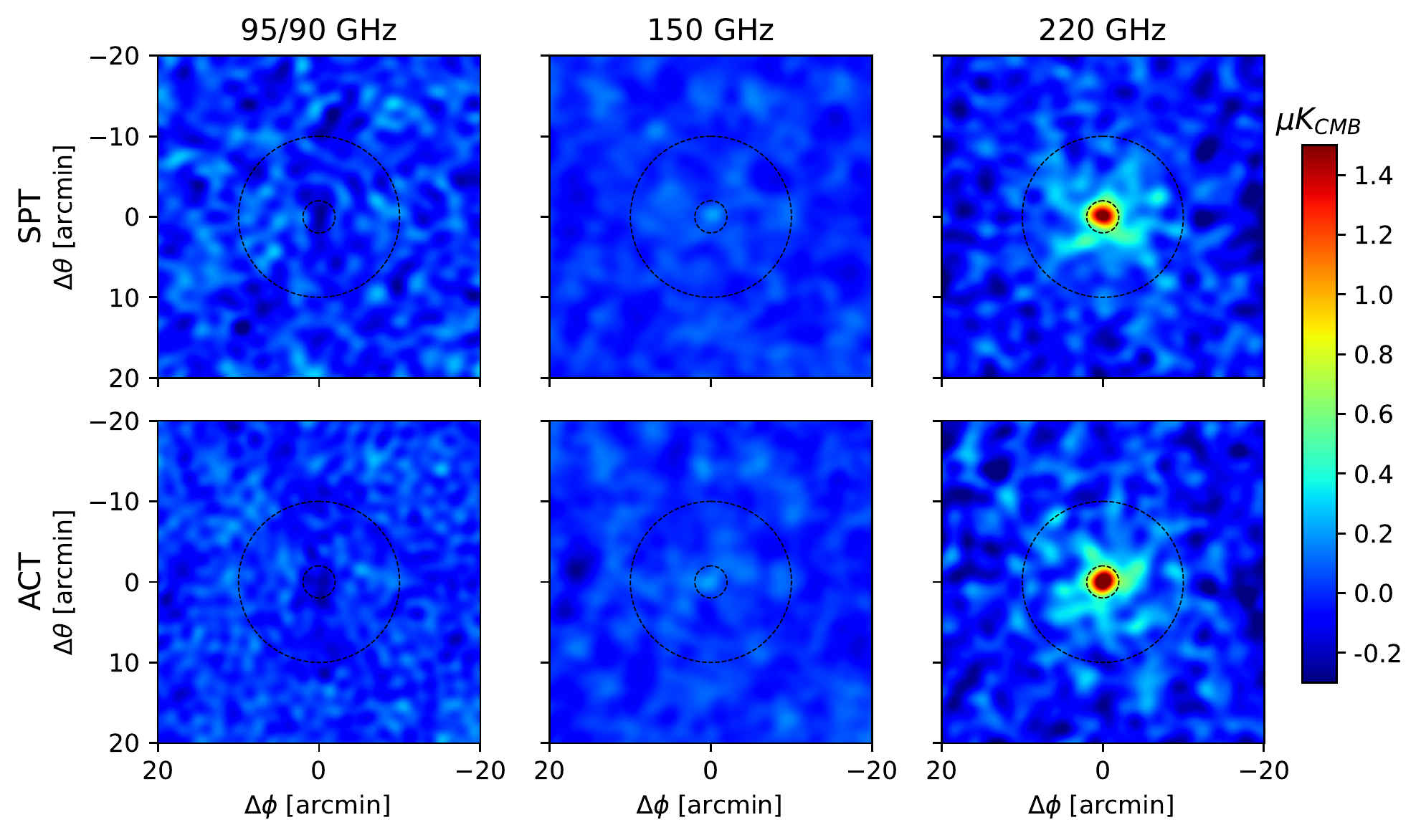}
	\caption{Overlap Sample galaxy stacks ($N=94,452$) for their respective SPT and ACT frequency maps, processed according to Section~\ref{subsec:MapProcessing}. A gradient was also removed from each image.  Dashed circles correspond to radii of $2.0\arcmin$ and $10\arcmin$.}
	\label{fig:FreqStack_SPTACT}
\end{figure*}

\begin{itemize}
	\item  The maximum spherical harmonic or Legendre polynomial degree $\ell_{\rm{max}}$, differs between the provided maps of SPT ($\ell_{\rm{max}}=10,000$) and ACT ($\ell_{\rm{max}}=30,000$).  For consistency, we elect to use the smaller limit of $\ell_{\rm{max}}=10,000$ on each, cutting all higher-order terms within ACT.  This removes ACT fluctuations at near pixel-size scales and introduces greater correlation between neighboring pixels, but otherwise does not significantly influence our results.
	\item  Respective beam functions of all frequencies were replaced with a Gaussian beam of $2.10\arcmin$ FWHM. This corresponds to the lowest resolution map (ACT $90$GHz).  The operation was done on the spherical harmonics ($a_{\ell{m}}$), with the aforementioned cutoff at $\ell_{\rm{max}}=10,000$.
	\item To remove any potential discrepancies due to projection differences, all ACT maps in their original Plate-Car\'ee projection were converted into the SPT's HEALPix format with $N_{side}=8192$. They were first transformed to spherical harmonics, beam and pixel window function corrections applied, and transformed into the final HEALPix map.
	\item  For each frequency map, the SMICA CMB map was masked with the corresponding instrument's boundary mask and converted into spherical $a_{\ell{m}}$ coefficients.  The pixel window function was replaced with the $N_{side}=8192$ HEALPix pixel window function of the final map format.  The CMB map was then subtracted from the desired frequency map(s).  This approach is akin to a high-pass filter, removing all large-scale CMB anisotropies to help reduce overall noise at small angular scales and correlation at larger scales.
	\item The HEALPix projection does not lend itself to uniform stacking of individual pixels and we also seek to place our target galaxies in the direct centers of our measurements. Thus, we make cutouts centered on each target galaxy using a gnomonic-projected grid with a pixel resolution of $0.05\arcmin$.  A HEALPix map with $N_{side}=8192$ has pixel side lengths of roughly $0.18\arcmin$, so we are purposely oversampling for finer alignment.  Bilinear interpolation was used to prevent any artificial beam effects from the pixel window function and allow additional precision in positioning.  Final image cutouts of our Overlap Sample are shown in Fig.~\ref{fig:FreqStack_SPTACT} in both SPT and ACT processed maps.  As outlined in the following subsection, we conducted final measurements on each individual galaxy cutout and then averaged together.
\end{itemize}

\subsection{Radial Profile}\label{subsec:RadialProfile}
With the smoothed and CMB-subtracted frequency maps, we measure the radial profile around all galaxies in our catalog.  We choose to create radial bins with uniform widths of $0.50\arcmin$, out to a radius of $20.0\arcmin$.  For our mean redshift of roughly $\approx1.1$ this translates to a furthest comoving distance of $21\text{  Mpc}\approx14h^{-1}\text{ Mpc}$.  Gnomonic projection cutouts were made around each galaxy with a pixel size of $0.05\arcmin$.  Cutouts were mean subtracted, and radial bin averages as described above were measured on each catalog location individually.  All samples of interest were then averaged with equal weight to create a final radial profile per map.  

With three frequencies, we are able to fit both the tSZ and the dust that obscures it.  However, any attempts to fit potential mean offsets from CMB or foreground signals would result in overfitting.  For this reason we assume all profiles go to zero at large radii. We calculate the average signal in the three largest bins ($18.5-20.0\arcmin$) and subtract it as an offset from the entire radial profile for each frequency map.  This method also subtracts any large-scale extragalactic background light (EBL) that might have further biased results. We recognize this subtraction likely truncates a non-zero signal, but at $20\arcmin$ consider it negligible in amplitude and detection.  For completeness, we test the effect by comparing different numbers of furthest bin subtractions from one ($19.5-20.0\arcmin$) to ten ($15.0-20.0\arcmin$), which results in a shift of $<0.5\sigma$ for $95$ and $150$~GHz radial bin measurements, and $<1.0\sigma$ for $220$~GHz.  The $220$~GHz causes the most noticeable shift due to it containing the highest S/N at large radii as a result of extended dust emission.

\begin{figure*}[ht]
	\gridline{\fig{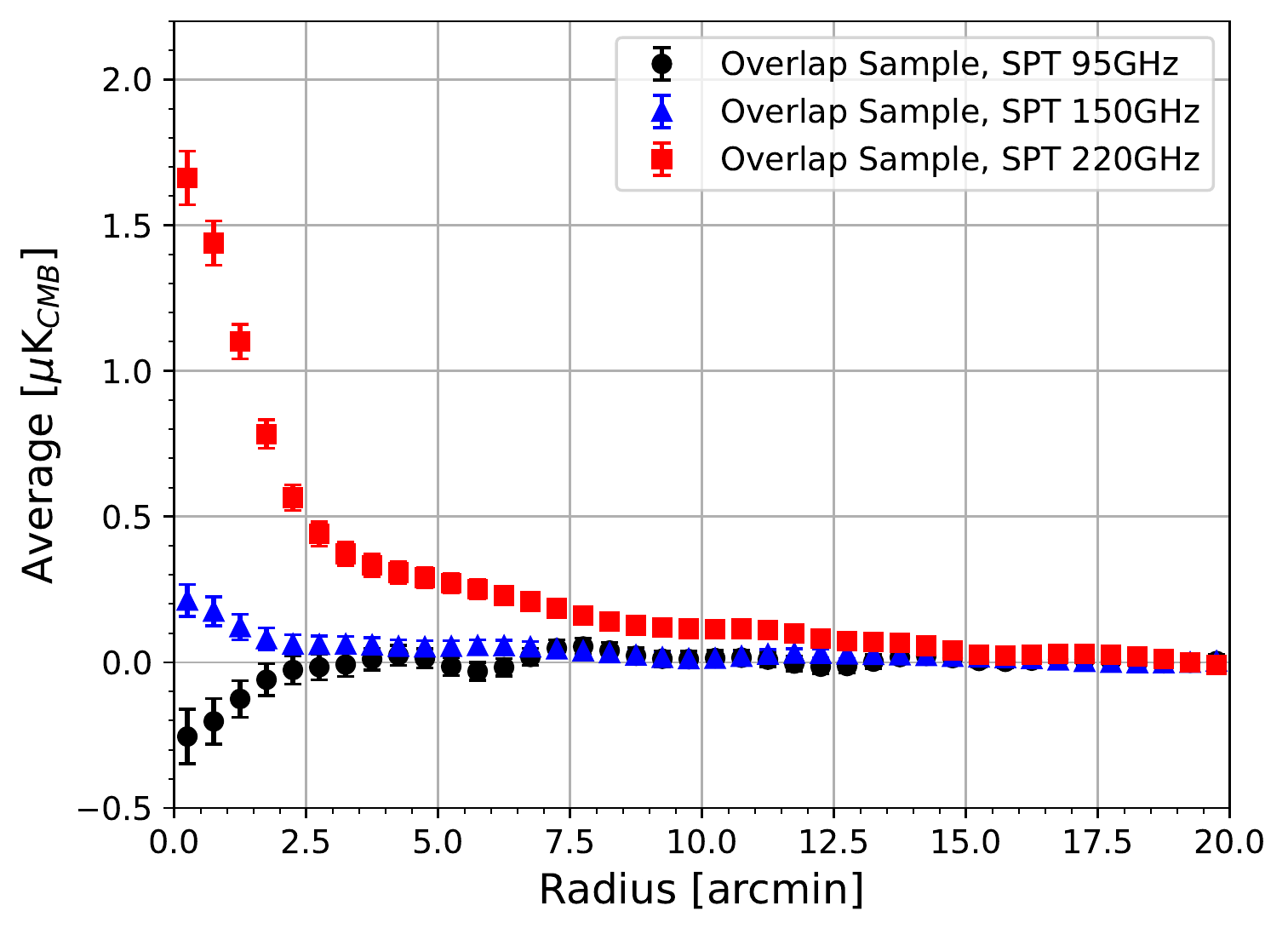}{0.5\textwidth}{  (a)}
		\fig{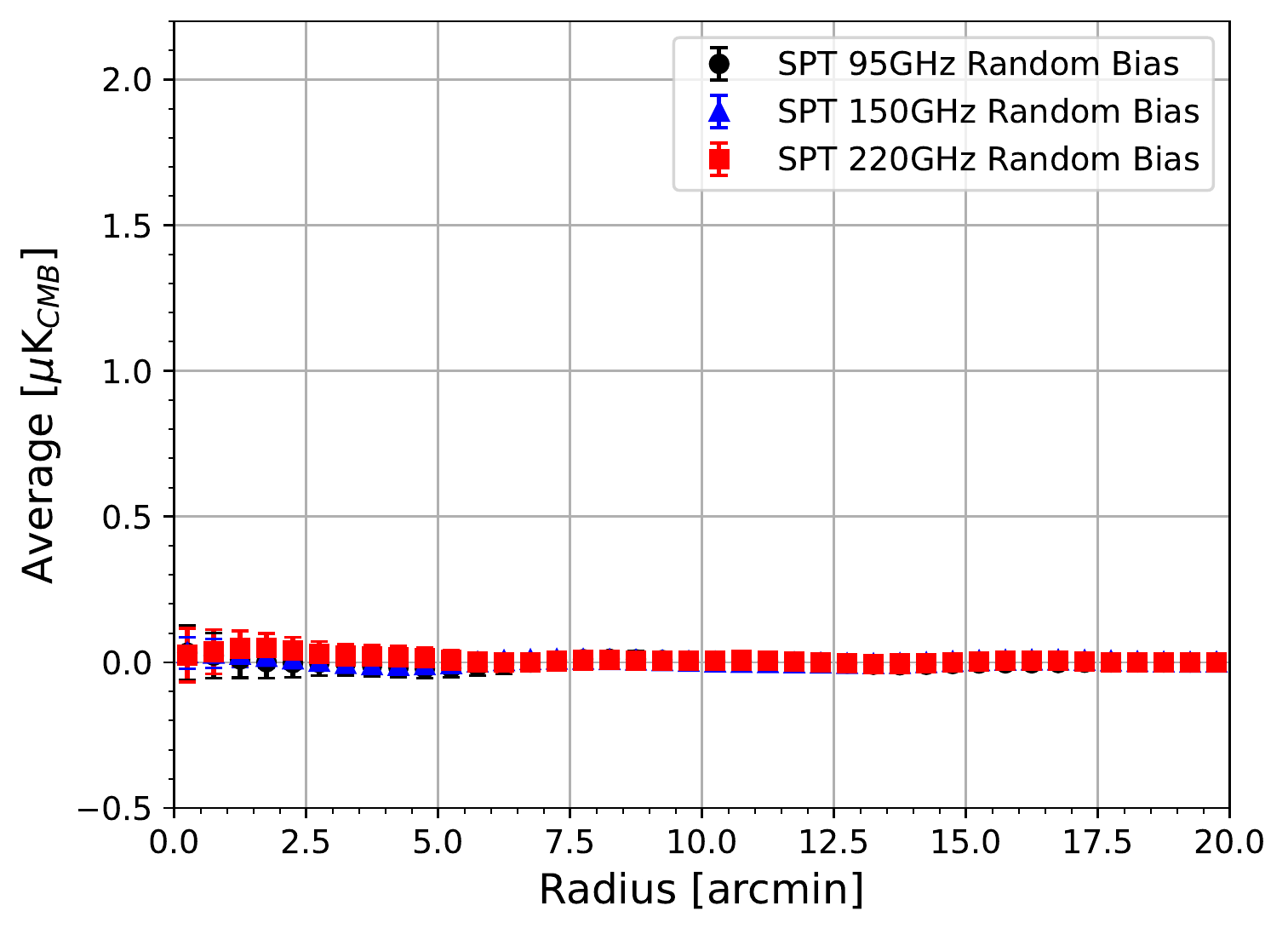}{0.5\textwidth}{  (b)}}
	\caption{Measured radial profiles as detailed in Section~\ref{subsec:RadialProfile} for: (a) all $N=94,452$ galaxies within the overlap field as measured on the SPT frequency maps of $95$, $150$, and $220$ GHz.  (b) Profile of estimated bias in the same SPT maps and overlap field, calculated from bootstrap resampling a catalog of randomly generated positions (Section~\ref{subsec:RandomCatalog}). \vspace{0.1cm}}
	\label{fig:SPT_profiles}
\end{figure*}

Fig.~\ref{fig:SPT_profiles} shows these described radial profiles for the $N=94,452$ Overlap Sample galaxies as measured on the SPT maps, alongside a bootstrap resampled random catalog profile to highlight the lack of any unexpected bias.  Our method for calculating uncertainty and random catalog are outlined in the Sections below.

\subsection{Uncertainties}\label{subsec:Uncertainties}

Correct evaluation of our results requires an accurate calculation of uncertainties.  This not only pertains to the error within a radial bin, but also correlation between its neighbors.  We employ a bootstrap resampling procedure to construct a covariance matrix across all maps and radial average bins.  This is done by resampling our galaxy catalog with replacement and with the same number of objects as the original.  We repeat this process for a large number of resamples ($4,000$) and measure the radial profile in identical fashion to Section~\ref{subsec:RadialProfile}.  The offset correction done by subtraction of three largest radial bins' average ($18.5\arcmin-20.0\arcmin$) likely skews these calculations and results in underestimated noise near large radii.  For this reason and low overall $S/N$ at large radii, we elect to not use any radial bins above $15\arcmin$.

The covariance matrix per frequency map is determined from the corresponding distribution of bootstrapped profiles.  The tSZ and dust covariance matrices are also calculated via fitting each bootstrap resample to the two-component fit outlined below in Section~\ref{subsec:TwoCompFitting} and shown in Appendix~\ref{appendix:tSZ_dust_corr}.

This bootstrapped covariance estimation assumes the noise is independent between each galaxy.  However for our sample, radial measurements out to a radius of $20\arcmin$ will on average have a few dozen catalog neighbors.  A spatial overlap will thus cause correlation between these neighboring galaxies.  This concern has been noted by others, such as \citet{Schaan2021}, that found bootstrap resampling produced $\approx10\%$ underestimation of error at $\geq6\arcmin$ in their circular apertures. 

This effect will also impact our analysis, and its importance will depend on our choice of aperture and the fact that we subtract the large-scale CMB.  In our case, our radial profile $S/N$ drops by roughly a factor of three between the center and $6\arcmin$, with the tSZ falling below $2\sigma$ by $8\arcmin$.  As a result, any profile fits should be largely controlled by the inner radial bins where the effects of underestimated error are minor.  To quantify this, we generated $400$ mock skies with basic Gaussian noise and measured at identical locations to our samples that showed an underestimation of roughly $10\%$ in variance (or $4.9\%$ error).  Thus, we elected to scale all our bootstrapped frequency covariances by $10\%$, while recognizing larger radial bins may still be slightly underestimated.

If instead we were to apply a $10\%$ error at $6\arcmin$ with a linear scaling relation versus radius, the noise of reported radial profile slopes is increased by up to $50\%$.  However, all other values reported below would remain within quoted margins of uncertainty.

\subsection{Random Catalog Comparison}\label{subsec:RandomCatalog}
To validate our procedure outlined above, we also generate random samples of $1,000,000$ points uniformly distributed within the SPT and ACT catalog footprints.  From these, we measure the radial profile (following Section~\ref{subsec:RadialProfile}) and bootstrap resample subsets with the same size as our desired galaxy catalog(s).  The resultant bootstrap mean corresponds to the expected bias of our sample's background.  Fig.~\ref{fig:SPT_profiles}b) shows our bias result of the SPT maps within the SPT-ACT overlap field.  Throughout all radial bins the random bootstrap mean stays within $1\sigma$ of zero, indicating no additional bias is present.

\subsection{Fitting Procedure}

\begin{figure*}[ht]
	\gridline{\fig{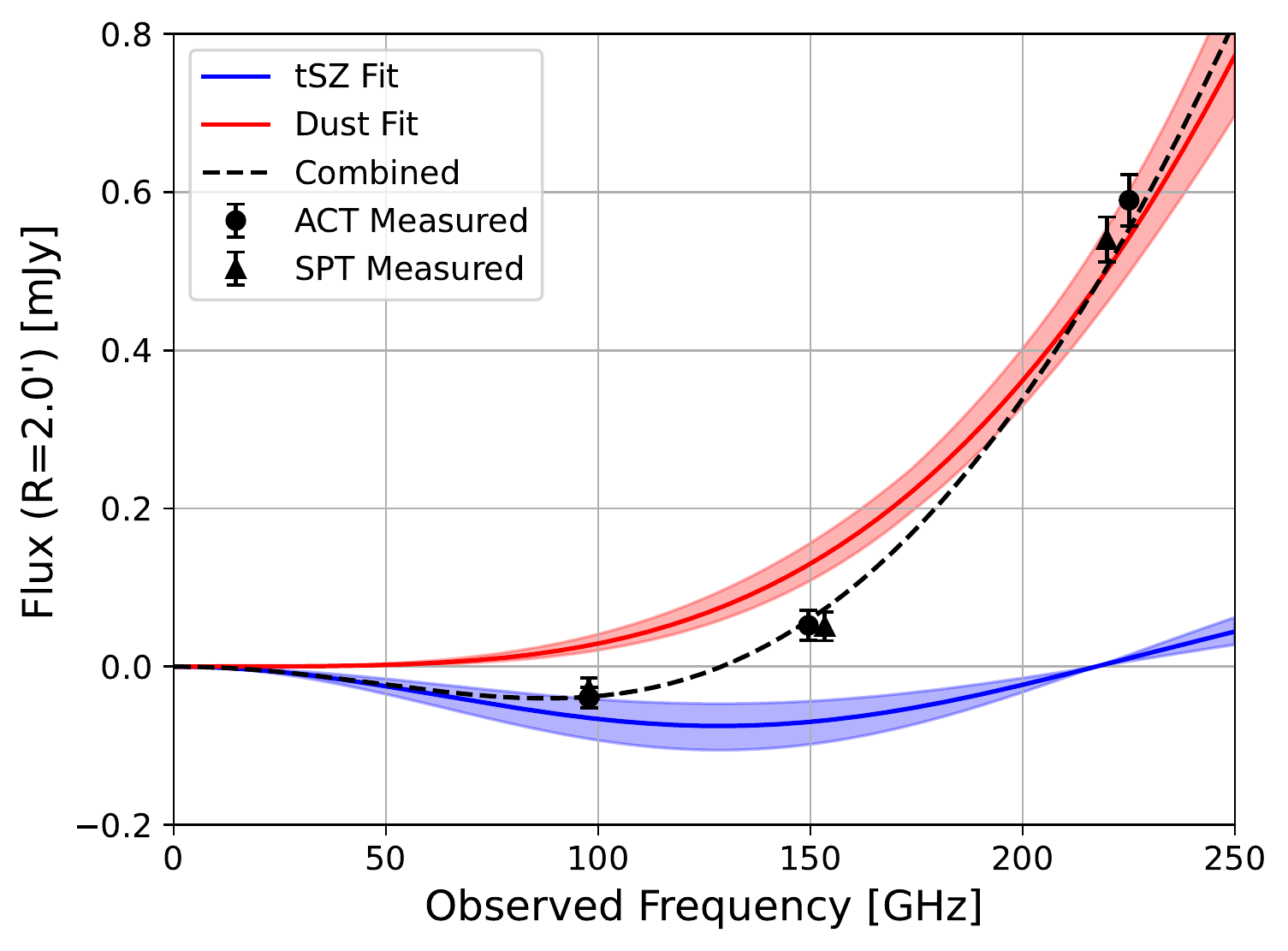}{0.5\textwidth}{  (a)}
		\fig{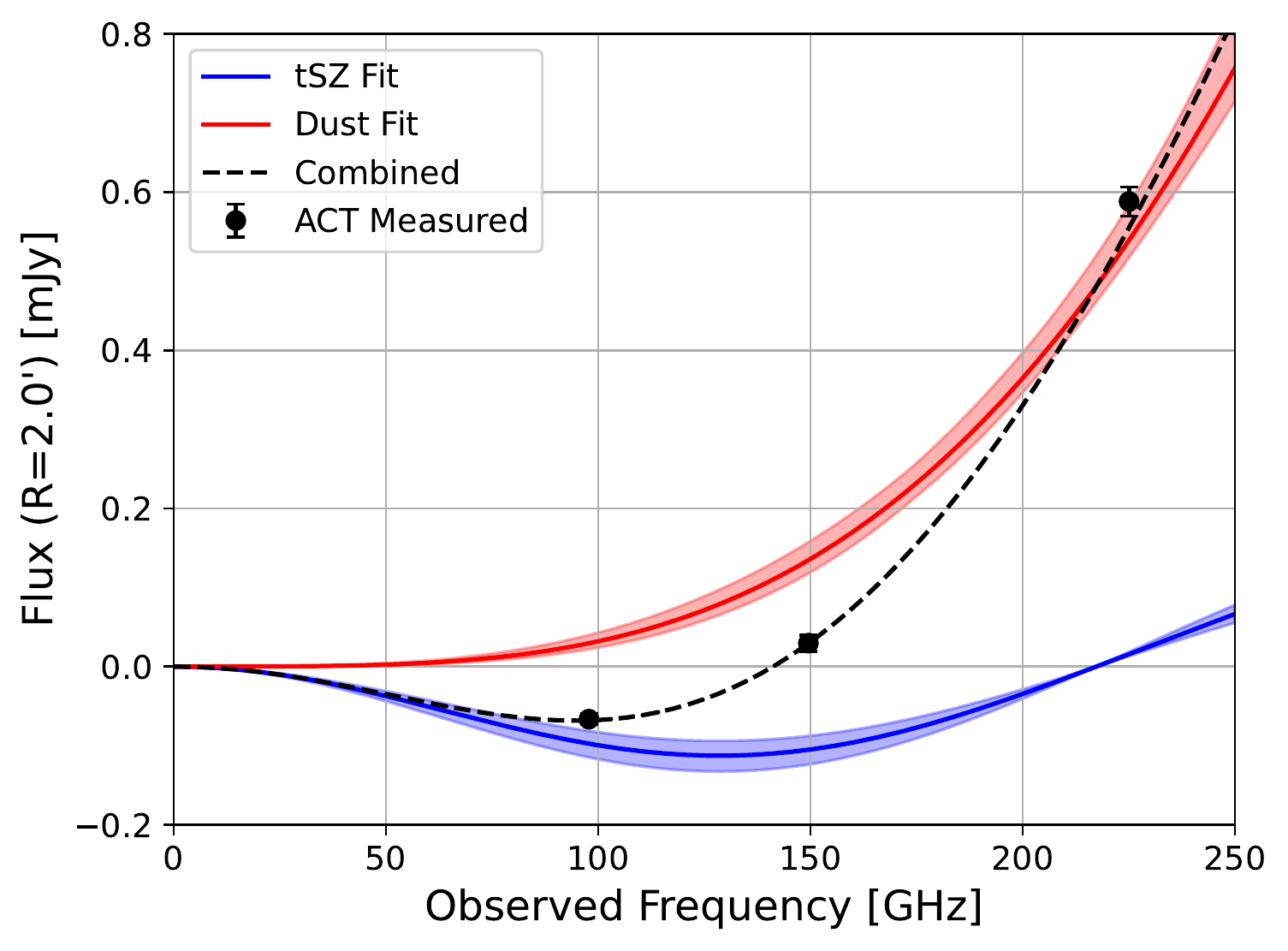}{0.5\textwidth}{  (b)}}
	\caption{Observed intensity spectrum within a $2.0\arcmin$ circular aperture for: (a) Overlap Sample ($N=94,452$) and (b) Wide-Area Sample ($N=387,627$).  Measured ACT (\textit{black, circles}) and SPT (\textit{black, triangles}) values are placed at each frequency map bandcenter, determined by integrating over their respective response.  Dust (\textit{red}) and tSZ (\textit{blue}) are shown with shaded $2\sigma$ bounds determined from the two-component fit of eq.~(\ref{eq:2compFit}).}
	\label{fig:Spectrum_Intensity_mJy}
\end{figure*}

All fits reported are conducted via Bayesian estimation with the assumption that our measurements are normally distributed but not necessarily independent.  The likelihood function is related to our fit residuals ($\mathbf{X}_i-\mathbf{\hat{X}}_i$) and covariance matrix ($\mathbf{C}$) as

\begin{equation}
\mathbf{\mathcal{L}}(\psi|X)=p(X|\psi)\propto\exp\left[-\frac{1}{2} (\mathbf{X}_i-\mathbf{\hat{X}}_i)^T\times\mathbf{C}^{-1}\times(\mathbf{X}_i-\mathbf{\hat{X}}_i)\right],
\end{equation}
incorporating parameters with discrete predefined ranges and priors $p(\psi)$. The posterior distributions are obtained as

\begin{equation}
p(\psi|X)=\frac{p(X|\psi)p(\psi)}{\int{p}(X|\psi^\prime)p(\psi^\prime) d\psi^\prime},
\end{equation}
where we normalize across all combinations of fit parameters ($\psi^\prime$).  This is calculated for the $\psi$-dimensional array for all possible parameter combinations and implemented via our own custom Python code.  Each parameter's reported best fit is classified as the median ($50$th percentile) after the posterior is marginalized over all other parameter ranges.  Similarly, the $1\sigma$ bounds are calculated as the $16$th and $84$th percentiles.

\subsection{Two Component Fitting}\label{subsec:TwoCompFitting}

From our aperture measurements, we used a two-component fitting model consisting of tSZ ($y$) and the dust spectral intensity at $\nu_0=353$ GHz in the source's rest frame $I_r(\nu_0)$ with units [W Hz$^{-1}$m$^{-2}$sr$^{-1}$],

\begin{equation}\label{eq:2compFit}
\delta{T}(\nu)=y\,{g}(\nu)\,{T_{CMB}}+\frac{I_{r}(\nu_0)}{(1+z)^2}\,\frac{I_{o}(\nu)}{I_{r}(\nu_0)}\left.\frac{dT}{dB(\nu,T)}\right\vert_{T_{\rm CMB}},
\end{equation}
where $g(\nu)=[{x}~(e^x+1)/(e^x-1)-4]$ of the tSZ signal (eq.~\ref{eq:DeltaT}) and $B(\nu,T)$ is the Planck function. The $(1+z)^{-2}$ term arises from redshift corrections due to time dilation and energy.  $I_{o}(\nu)$ is the specific dust intensity in the observed frequency band $\nu$. It is converted to the rest frame band $\nu\,(1+z)$,

\begin{equation}
I_{o}(\nu)=(1+z)~I_{r}\left[\nu~(1+z)\right],
\end{equation}
where we assume a gray-body dust spectrum for $I_{r}$ with a dust temperature ($T_{d}$) and spectral emissivity index ($\beta$). Thus, the intensity term from eq.~(\ref{eq:2compFit}) can be written as

\begin{equation}
\frac{I_{o}(\nu)}{I_{r}(\nu_0)}=(1+z)\left[\frac{\nu~(1+z)}{\nu_0}\right]^{\beta}\frac{B[\nu~(1+z),T_{d}]}{B(\nu_0,T_{d})},
\end{equation}
normalized with respect to $I_r(\nu_0)$.  This normalization term helps define a reference frequency for all measurements while reducing the correlation between dust temperature and intensity amplitude when near the Rayleigh-Jeans limit. Equation~\ref{eq:2compFit} is integrated over each respective map's frequency band response. The SPT bands were extracted from \citet{Chown2018}, as the SPT+Planck maps are dominated by the SPT response for most of our angular scales.  Full ACT bandpasses were available as a function of position, detector array, and multipole $\ell$. We average each ACT response across our field of observation, all detectors, and with a cut of $2,000<\ell<=\ell_{\rm{max}}$.  The $\ell=2,000$ minimum was chosen to reflect our angular scales of interest and subtraction of the large-scale CMB (Section~\ref{subsec:MapProcessing}).  The observed flux in mJy integrated within a simple circular aperture of $R=2.0\arcmin$ radius is shown in Fig.~\ref{fig:Spectrum_Intensity_mJy} for our Overlap and Wide-Area samples, respectively.  This circular aperture is further used in our stellar mass binning shown in Section~\ref{subsec:stellar_mass_binning}.

The two component fit described above was also applied to each set of frequency measurements per radial bin for all listed catalogs in Table~\ref{tab:catalogs}.  We assume priors as outlined in Table~\ref{tab:twoCompFitPriors} for all fits.  Uniform priors are set for the Compton-$y$ ($0\leq{y}\leq$$4\times10^{-7}$) and dust intensity in the $220$ GHz rest frame ($0\leq{I_{r}}(\nu_{0})\leq4\times10^{-24}$ W Hz$^{-1}$m$^{-2}$sr$^{-1}$).  In the event of fits near zero indicating low signal to noise, we shift these uniform priors to include slight negative values. Thus, in the absence of a signal we will then correctly produce a result centered about zero.  Gaussian priors were assumed for the additional parameters of dust emissivity ($\beta=1.75\pm0.25$) and dust temperature ($T_{\rm{d}}=20\pm3$ K).  These Gaussian priors were chosen to align within standard ranges \citep{Dunne2001-SCUBA-II,Draine2011,Addison2013,Magdis2021}, but were not set as free uniform parameters due to our limited number of maps to fit.  The resultant dust parameter fits are found to be highly constrained to within $1.5\sigma$ of the prior mean.  This method allows us to include additional uncertainty associated with our lack of information about the dust in our sample(s), while still ensuring our two-component fit does not encounter problems with overfitting.

\begin{deluxetable}{ccc}
	\tablecaption{Two component fit parameters (from eq.~\ref{eq:2compFit}) and given priors used on each catalog and radial bin.  \label{tab:twoCompFitPriors}}
	\tablehead{
		\colhead{Parameter} & \colhead{Description} & \colhead{Prior}
	}
	\startdata
	$y$ & Compton-$y$ [unitless] & [$0^{\dagger},~4\times10^{-7}$] \\
	$I_r(\nu_{0})$ & Dust Intensity [W Hz$^{-1}$m$^{-2}$sr$^{-1}$] & [$0^{\dagger},~4\times 10^{-23}$] \\
	$\beta$ & Dust Emissivity [unitless] & ${G}(1.75,0.25^2)$ \\
	$T_{\rm{d}}$ & Dust Temperature [K] & ${G}(20,3^2)$ \\
	\enddata
	\tablecomments{Gaussian ${G}(\mu,\sigma^2)$ priors are assumed for dust emissivity $\beta=1.75\pm0.25$, and temperature $T_{\rm{d}}=20\pm3$ K. \textsuperscript{\textdagger}A realistic lower limit of zero is used on the uniform free parameters unless the fit is poor and near zero. In which case, the lower limit is shifted negative to allow for accurate fitting around zero and avoid artificially inflated values. \vspace{-0.6cm}}
\end{deluxetable}

Our samples were selected with low SFRs and thus should have minimal radio sources at these frequencies. However if non-negligible radio contamination was present in the lower frequency bands, our two-component fit would then underestimate the tSZ signal. Meanwhile the dust fit would be either over- or under-estimated, dependent upon the radio source's spectrum into the higher bands.

\subsection{Profile Fits}\label{subsec:profileFits}

As detailed above, we obtain profiles for both the tSZ and dust responses per radial average bin from our frequency maps.  The dominant source is expected to be a central point source associated with our target sample.  However, we also expect an extended secondary profile term due to spatial correlations with neighboring galaxies.  

A few different profile models could be considered, such as a generalized Navarro-Frenk-White (NFW) profile like that conducted by \citep{Amodeo2021}, or basic power-law models for two-point correlation clustering measurements \citep{Coil2017}.  However, our $2.1\arcmin$ beam and $z\approx1$ redshift would result in highly degenerate and correlated NFW parameter fits, while a power-law model cannot easily be forward-modeled with the beam since it diverges to infinity as $r\xrightarrow{}0$.  As we are primarily interested in the power-law slope at radii away from the center, we opt for a simple pseudo-power-law approximation that can be made using a type of King or isothermal model \citep{King1962}:
\begin{equation}
	f(r)=\frac{A_{\rm{k}}}{r_{0}}\left(1+\frac{r^2}{r_{0}^{2}}\right)^{-\frac{\gamma}{2}},
\end{equation}
with an amplitude $A_{\rm{k}}$, comoving core radius $r_0$, and that now instead converges to $A_{\rm k}/r_{0}$ as $r\xrightarrow{}0$.  Converted to a function of projected angle ($\theta$) through the line-of-sight, this gives
\begin{equation}\label{eq:projectedKing}
f(\theta)={A_{\rm{k}}}\frac{\Gamma(\frac{1}{2})\Gamma(\frac{\gamma-1}{2})}{\Gamma(\frac{\gamma}{2})}\left(1+\frac{\left(D_{\rm c}\theta\right)^2}{r_0^2}\right)^{\frac{1-\gamma}{2}},
\end{equation}
where $D_{\rm c}$ is the comoving distance.  This profile is best defined as a function of angle $\theta$, as it must be convolved with the beam for accurate comparison to our measured values. For a combined model of a point source plus King ($\delta+f$) convolved with the beam ($b$) can be described as,
\begin{equation}
	F(\theta):=\iint_{-\infty}^{\infty}\left[\delta(\boldsymbol{\theta}^\prime)+f(\boldsymbol{\theta}^\prime)\right]b(\boldsymbol{\theta}-\boldsymbol{\theta}^\prime)d\boldsymbol{\theta}^\prime.
\end{equation}

Our final beam as described in Section~\ref{subsec:MapProcessing} is a Gaussian with FWHM$=2.1\arcmin$, but with an $\ell_{\rm{max}}=10,000$ cutoff.  Compared to convolution with a perfect Gaussian beam this can produce a $10\%$ difference for a central point source, but has a negligible effect on our broader King profile of eq.~(\ref{eq:projectedKing}).
For this reason we elect to assume a perfect Gaussian beam to simplify the King convolution, but maintain the exact beam (with $\ell_{\rm{max}}$ cut) for the point source defined below as $b(\theta)$.  These yield a profile function with one integral that we compute numerically,
\begin{equation}\label{eq:convolvedDust}
	F(\theta)=A_{\rm{ps}}b(\theta)+\int_{0}^{\infty}\exp\left(-\frac{\theta^2+{\theta^\prime}^2}{2\sigma_{\rm beam}^2}\right){J_{0}}\left(i\frac{\theta\theta^\prime}{\sigma_{\rm beam}^2}\right)\frac{f(\theta^\prime)}{\sigma_{\rm beam}^2}\theta^\prime{d}\theta^\prime,
\end{equation}
where $J_0$ is the Bessel function of the first kind and the size of our Gaussian beam as $\sigma_{\rm beam}=0.8918\arcmin$.This eq.~(\ref{eq:convolvedDust}) allows us to set a lower bound for the profile's central point source component and examine the extended profile slope.

\subsection{Dust}\label{subsec:dust}

\begin{deluxetable*}{ccccc}[ht]
	\tablecaption{Dust profile fit parameters for eq.~(\ref{eq:convolvedDust}), applied priors, and resultant fits for our Overlap and Wide-Area samples.  \label{tab:dust_profile_fit}}
	\tablehead{
		\colhead{Parameter} & \colhead{Description} & \colhead{Prior} & \colhead{Overlap Sample} & \colhead{Wide-Area Sample}
	}
	\startdata
	$A_{\rm ps}$ [$10^{-23}$ W Hz$^{-1}$m$^{-2}$sr$^{-1}$] & Point Source Amplitude & [$0,~4.0$] & $2.14_{-0.22}^{+0.24}$ & $2.32_{-0.18}^{+0.22}$ \\
	$A_{k}$ [$10^{-24}$ W Hz$^{-1}$m$^{-2}$sr$^{-1}$] & King Amplitude & [$0,~4.0$] & $1.38_{-0.20}^{+0.24}$ & $1.56_{-0.18}^{+0.22}$ \\
	$\gamma$ [unitless] & King Slope  & [$1.0,~4.0$] & $2.60_{-0.15}^{+0.16}$ & $2.95_{-0.14}^{+0.16}$ \\
	$r_{0}$ [Comoving Mpc] & Core Radius & $3.0$ & -- & -- \\
	\enddata
	\tablecomments{The King amplitude and slope will be positively correlated. We set the core radius to a constant larger than the beam due to its inherent degeneracy with the amplitudes. \vspace{-0.6cm}}
\end{deluxetable*}

\begin{figure}[ht]
	\centering
	\includegraphics[width=1\linewidth]{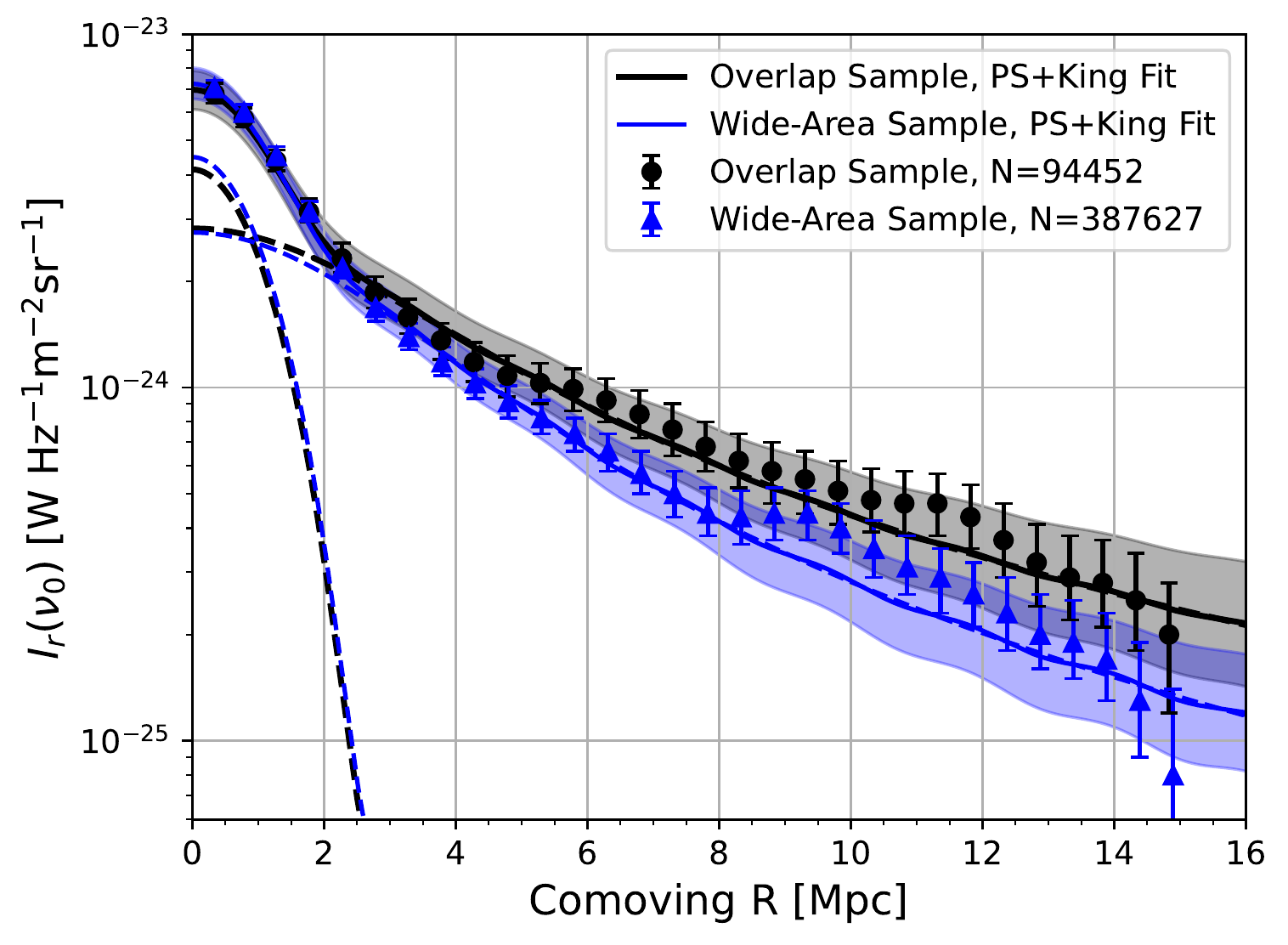}
	\caption{Dust radial profile and best fit point source + King model as defined in Section~\ref{subsec:dust}, shown here with core radius $r_0=3.0$ comoving Mpc for Overlap (\textit{black}) and Wide-Area (\textit{blue}) samples.  Shaded regions represent $2\sigma$ uncertainty of the combined fit. Dashed lines correspond to the separate best fit point source and King components.}
	\label{fig:dust_profile}
\end{figure}

Our resultant dust from the two-component fit per radial bin is shown in Fig.~\ref{fig:dust_profile}. We observe up to a $16\sigma$ and $20\sigma$ detection of dust in the center bins of our Overlap and Wide-Area samples respectively.  Beyond the beam's FWHM, detection in both cases monotonically decreases to roughly $5\sigma$ at $10\arcmin$ and down further to $2\sigma$ at $15.0\arcmin$ where noise begins to dominate.  Of particular interest is the shape of our dust profile, which has a definitive central source similar to the beam along with a sloped extended signal.

We expect the dust profile to consist of an unresolved central source associated with our target galaxies, and a secondary extended profile tied to the two-point correlation function of neighboring galaxies. We fit the convolved point source plus King model of eqs.~(\ref{eq:projectedKing}) and~(\ref{eq:convolvedDust}) to our dust profile up to $15\arcmin$ ($\approx15.2$ comoving Mpc).  This cutoff is meant to avoid incorporating low $S/N$ radial bins and reduce any residual impact from the offset correction discussed in Section~\ref{subsec:Uncertainties}.  

We assume fit parameters with priors as outlined in Table~\ref{tab:dust_profile_fit}.  As core radius ($r_{0}$) has inherent degeneracy with the amplitudes we instead hold $r_{0}$ as a constant larger than the beam, selecting $r_{0}=3.0$ comoving Mpc.  Due to this degeneracy and inability to resolve our central source, this fit is not an attempt to fully separate the one- and two- component contributions within the profile.  However, it provides us the opportunity to determine other characteristics such as the extended profile slope at larger radii.

The radial bin dust profile and resultant fits are shown in Fig.~\ref{fig:dust_profile} for a core radius of $r_{0}=3.0$ comoving Mpc.  We have separately checked the impact of different core radii.  For instance, if a core radius of $r_{0}=5.0$ comoving Mpc was chosen instead, it would result in a $\approx10\%$ increase of our dust's point source ($A_{\rm{ps}}$), with a $\approx5-10\%$ decrease in the King ($A_{\rm{k}}$) amplitude.  Increasing the core radius also has a noticeable effect on the King slope, due to heightened dependence on the noisier high radial bins and our limited range of $15\arcmin$.  A core radius of $r_{0}=5.0$ comoving Mpc produces steeper dust slopes ($\gamma$) by a factor $\approx25\%$.

Our dust profile fits are shown in Table~\ref{tab:dust_profile_fit} and Fig.~\ref{fig:dust_profile} for both catalogs.  The resultant posterior distributions are shown in the corner plot of Fig.~\ref{fig:dust_profile_fit_corner} within Appendix~\ref{appendix:profile_fit_posteriors}.  The point source ($A_{\rm{ps}}$) and King ($A_{\rm{k}}$) amplitudes are fit at a $9.3-11.5\sigma$ and $6.2-7.8\sigma$ level, respectively, and are consistent (within $2\sigma$) between galaxy samples.  The best fit King slopes are $2.60_{-0.15}^{+0.16}$ and $2.95_{-0.14}^{+0.16}$ for our Overlap and Wide-Area samples.  These are slightly steeper than reported power law slopes from galaxy clustering studies \citep[$\gamma=1.5-2.0$,][]{Eftekharzadeh2015,Coil2017,Amvrosiadis2018}, likely as a result of the difference between our King model and a power-law, which diverge near and below the core radius.  

Since our King model is designed to level off as it nears the core radius, it would have to fit a steeper slope to be comparable with that of a power-law.  Additionally, our necessary zeroing of the frequency profiles at large radii (Section~\ref{subsec:RadialProfile}) results in an underestimation of the dust by a small constant which would contribute to steeper slope fits.  To test this, we incorporated an additional constant offset term in our profile fit and found it to be insignificant.  The best fit offset was less than our measured signal at $15\arcmin$, within $1\sigma$ of zero, and simply increased the fit uncertainty of our other parameters while marginally decreasing the slope $\gamma$ by $<5\%$.  We account for some noise underestimation from our bootstrap resampling, as discussed in Section~\ref{subsec:Uncertainties}, but an even further increase in noise at large radii would also primarily result in a lower signal-to-noise fit of the King slope.

Thus, we can still conclude that our extended dust has a shape consistent with that expected from the two-point correlation function of neighboring galaxies and structure.  Overall, we have shown here that at our $z\approx1$ redshifts, dust in the millimeter bands contains useful insights into intergalactic structure and can be detected at a high significance.

\subsection{Dust Mass}\label{subsec:dust_mass}
Also of interest is the mean dust mass associated with our galaxy samples, which can be estimated from the rest-frame dust intensities $I_{r}(\nu_0)$ found from eq.~(\ref{eq:2compFit}).  The dust mass follows,
\begin{equation}\label{eq:dust_mass}
	M_{\rm{d}}=\frac{D_{\rm{c}}^{2}\int{I_{r}}(\nu_0){d}\Omega}{\kappa(\nu_{0})B(\nu_0,T_{\rm{d}})},
\end{equation}
where $\kappa(\nu_{0})$ is the dust mass opacity coefficient or absorption cross-section per unit mass [m$^2$ kg$^{-1}$] at our reference frequency of $353$GHz.  We take $T_{d}=20\pm3$ K as used previously in our two-component fit.  The final error is determined by standard error propagation of both $I_{r}(\nu_0)$ and $T_{\rm{d}}$. 

Unfortunately $\kappa(\nu_{0})$ is overall poorly constrained. Further potential uncertainty arises as $\kappa(\nu_{0})$ values in literature are often derived from dust observations or models designed for the Milky Way or other local galaxies, which may slightly differ compared to our $z\approx1$ quiescent samples.  At $\nu_{0}=353$~GHz, or $\lambda_{0}=850$ $\mu\text{m}$, commonly used $\kappa(\nu_{0})$ values range from $0.04-0.15$~m$^{2}$kg$^{-1}$ \citep{Draine2001,Dunne2001-SCUBA-II,Draine2003,Dunne2003,Casey2012}.  Thus, we take a conservative approach and assume a center value of $\kappa(\nu_{0})=0.08$~m$^{2}$kg$^{-1}$, while acknowledging this can fluctuate by a factor of two.

As evident by the previous subsection, we observe a dust profile containing both a central point source and extended neighboring structures.  However, our beam introduces difficulty in accurate separation of them.  As a lower bound for the  expected central dust, we take the fit point source component: $\int{I_{r}}(\nu_0){d}\Omega=A_{\rm{ps}}\int{b}(\theta){d}\Omega$, integrated over the beam solid angle.  In contrast, we also integrate within a $R=2.0\arcmin$ circular aperture instead, assuming that the central point source will dominate any extended dust structure within this radius.  Our results for each catalog are shown in Table~\ref{tab:DustMass_FromProfileFits}.

\begin{deluxetable}{ccccc}[th]
	\tablecaption{Dust mass associated with our central point source fit shown in Fig.~\ref{fig:dust_profile} and Table~\ref{tab:dust_profile_fit}, and for all dust within $R=2.0\arcmin$. Dust-to-stellar mass ratio is also shown. \label{tab:DustMass_FromProfileFits}}
	\tablehead{
		\colhead{Parameter} & \multicolumn{2}{c}{Overlap Sample} & \multicolumn{2}{c}{Wide-Area Sample} \\
		\colhead{} & \colhead{$A_{\rm{ps}}$} & \colhead{$R=2.0\arcmin$} & \colhead{$A_{\rm{ps}}$} & \colhead{$R=2.0\arcmin$}
	}
	\startdata
	$\log_{10}(M_{\rm{d}}/\text{M}_{\odot})$ & $8.43_{-0.12}^{+0.10}$ & $8.82_{-0.11}^{+0.09}$ & $8.46_{-0.12}^{+0.09}$ & $8.83_{-0.11}^{+0.09}$ \\
	$\log_{10}(M_{\rm{d}}/\overline{M_{\star}})$ & $-2.98_{-0.12}^{+0.10}$ & $-2.59_{-0.11}^{+0.09}$ & $-2.98_{-0.12}^{+0.09}$ & $-2.61_{-0.11}^{+0.09}$\\
	\enddata
	\tablecomments{For a $\kappa(\nu_{0})=0.08$~m$^{2}$kg$^{-1}$, which we recognize might fluctuate by a further factor of two or $0.30$ dex. \vspace{-0.6cm}}
\end{deluxetable}

The lower limit to our dust mass - extracted solely from the profile's point source component ($A_{\rm{ps}}$) in Section~\ref{subsec:dust} - indicates consistent dust masses of $8.43_{-0.12}^{+0.10}$ and $8.46_{-0.12}^{+0.09}$ $\log_{10}(\text{M}_{\odot})$ for the complete Overlap and Wide-Area samples, respectively.  In comparison, an upper limit to the dust mass - simply integrating within a radius of $R=2.0\arcmin$ - produces dust masses $0.39$ and $0.37$ dex larger.  The ratio of dust mass to stellar mass show even greater consistency between catalogs, ranging from $-2.98$ (lower limit using $A_{\rm{ps}}$) to $-2.59$ (upper limit using $R=2.0\arcmin$) orders of magnitude.  For smaller sample sizes when profiles cannot be well-constrained, such as when binning by stellar mass, the circular $R=2.0\arcmin$ aperture is still possible.  We employ this generalized method in Section~\ref{subsec:stellar_mass_binning} to analyze our dust-to-stellar mass relation.  

While these dust masses are on the high side expected for galaxies with low SFRs, other studies have found similar results for massive galaxies with increasing redshift \citep{Satini2014,Calura2016,Gobat2018}.  There are also indications that this increase in dust-to-stellar mass with redshift is more extreme for quiescent galaxies than dusty star-forming ones \citep{Donevski2020,Magdis2021}.  The additional uncertainty from $\kappa(\nu_{0})$ prevents us from drawing any strong conclusions.  However, as our dust masses appear to be within an acceptable range compared to these previous studies, we can treat them as another verification of our stacking and analysis process.  Determination of dust mass in this manner also highlights the potential for similar use in future sub-mm and FIR investigations. 

\subsection{Compton-$y$}\label{subsec:compton_y}

In comparison with the dust measured above, we expect our tSZ profile to be similar but not identical in shape. Unlike dust, we expect the tSZ from our target galaxies to have a broader one-halo distribution associated with hot ionized gas, which spans throughout the CGM out to $\approx0.5-1.0$ comoving Mpc.  With our $2.1\arcmin$ FWHM beam, most of this central component will still be unresolved.  We also expect a steeper profile slope, as the extended tSZ is a tracer for hot gas that is less prevalent in lower-mass neighbors.

\begin{deluxetable*}{ccccc}[t]
	\tablecaption{Compton-y profile fit parameters for eq.~(\ref{eq:convolvedDust}), given priors, and resultant fits on our Overlap and Wide-Area samples.  \label{tab:tSZ_profile_fit}}
	\tablehead{
		\colhead{Parameter} & \colhead{Description} & \colhead{Prior} & \colhead{Overlap Sample} & \colhead{Wide-Area Sample}
	}
	\startdata
	$A_{\rm{ps}}$ [$10^{-7}$] & Point Source Amplitude & [$0,~8$] & $2.0_{-1.1}^{+1.3}$ & $2.2_{-0.8}^{+0.8}$ \\
	$A_{\rm{k}}$ [$10^{-8}$] & King Amplitude & [$0,~20$] & $8.2_{-4.0}^{+5.1}$ & $7.4_{-1.7}^{+2.3}$ \\
	$\gamma$ [unitless] & King Slope  & [$1.0,~10.0$] & $6.6_{-2.1}^{+2.1}$ & $4.1_{-0.5}^{+0.7}$ \\
	$r_{0}$ [Comoving Mpc] & Core Radius & $3.0$ & -- & -- \\
	\enddata
	\tablecomments{We set the core radius to a constant larger than the beam due to its inherent degeneracy with the amplitudes. \vspace{-0.3cm} 
	}
\end{deluxetable*}

\begin{figure}[ht]
	\centering
	\includegraphics[width=1\linewidth]{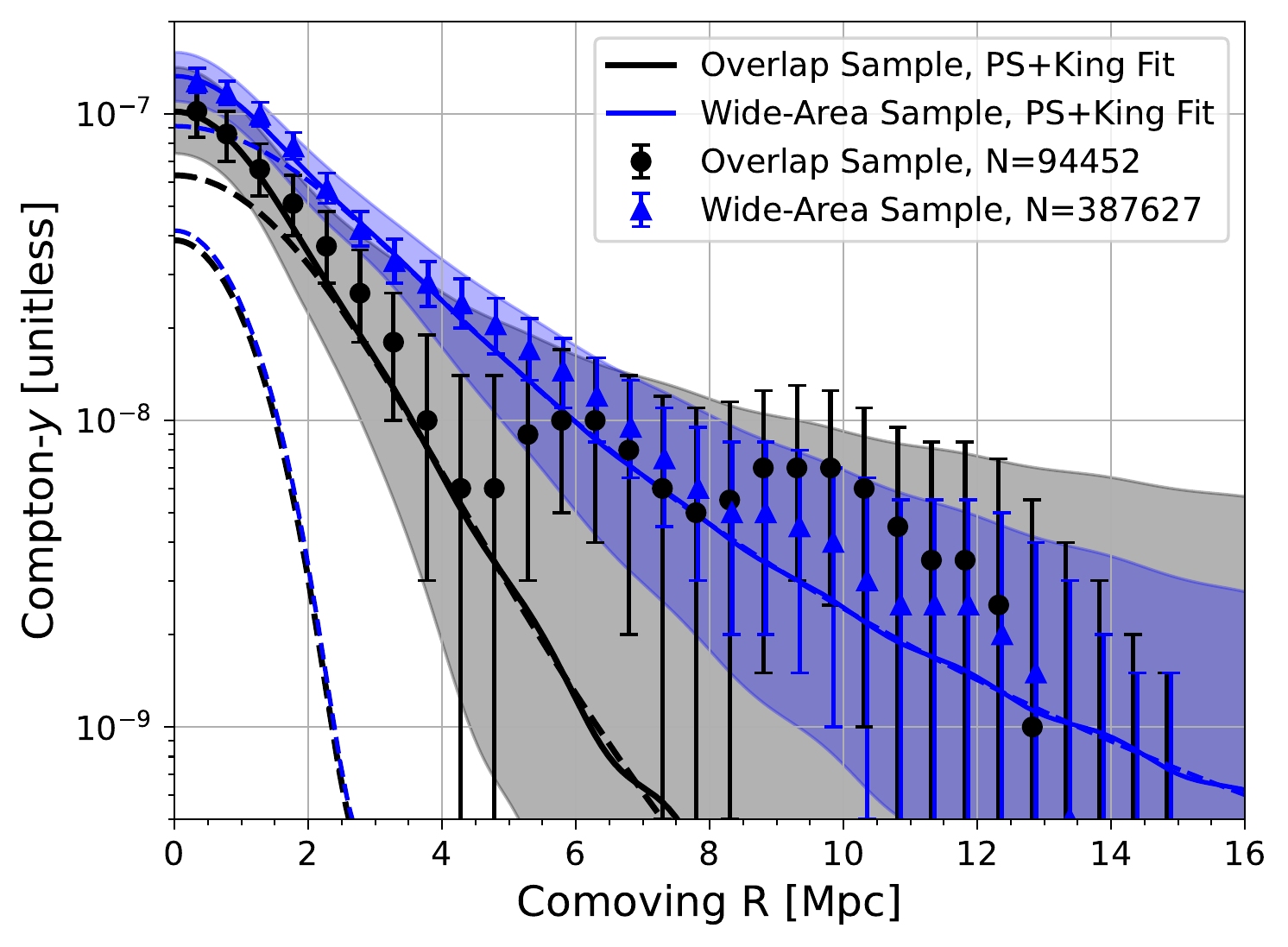}
	\caption{Compton-$y$ radial profile for our quiescent galaxy catalogs; Overlap (\textit{black, triangles}) and Wide-Area (\textit{blue, circles}).  Alongside their best fit (\textit{solid}), shaded $2\sigma$ bounds, and individual point source and King components (\textit{dashed}).}
	\label{fig:tSZ_profile}
\end{figure}

The Compton-y component from our two-component fit of eq.~(\ref{eq:2compFit}) per radial bin is shown in Fig.~\ref{fig:tSZ_profile} for each complete catalog.  Here the difference in sample size is apparent, as the centermost radial bins for Overlap Sample ($N=94,452$) detect the tSZ at up to $5.4\sigma$, while the Wide-Area Sample ($N=387,627$) is up to $11\sigma$.  Of equal importance is the distance at which the $S/N$ drops below $2\sigma$.  This occurs at a radius of $4.0\arcmin$ ($4.0$ comoving Mpc) for the Overlap Sample, versus $8.0\arcmin$ ($8.1$ comoving Mpc) for the Wide-Area Sample.  So while we do observe extended tSZ larger than the beam, noise begins to dominate much quicker than observed with dust, especially for the Overlap Sample.  Therefore we elect to only fit our profiles up to $10\arcmin$ ($10.1$ comoving Mpc).

We assume tSZ profile fit priors given in Table~\ref{tab:tSZ_profile_fit}.  Due to degeneracy between the central point source and King model, we assume a core radius again of $r_{0}=3.0$ comoving Mpc.  It should be noted that just as with the dust, this profile fit does not fully isolate the one- and two- component contributions due to our inherent central degeneracy between the King and point source models as a result of the beam.  Our main goal in applying this fit is to demonstrate the presence of extended tSZ, and compare the resultant King slope to that found for dust.  We again checked the effect of using different core radii and see similar trends as with the dust; increasing core radius to $r_{0}=5.0$ comoving Mpc yields a $\approx25\%$ increase in tSZ point source amplitude ($A_{\rm{ps}}$), $\leq 5\%$ decrease in King amplitude ($A_{\rm{k}}$), and $\approx40\%$ increase in slope ($\gamma$).  The change in slope with core radius here is larger than observed with dust, due to the faster rate at which our tSZ profile $S/N$ drops.

Our fit results are shown in Table~\ref{tab:tSZ_profile_fit} and plotted alongside our measurements in Fig.~\ref{fig:tSZ_profile}.  The marginalized posterior distributions are shown in Fig.~\ref{fig:tSZ_profile_fit_corner} within Appendix~\ref{appendix:profile_fit_posteriors}. Indicative of the quick $S/N$ drop-off, the King model for the Overlap Sample is poorly constrained. Point source amplitudes ($A_{\rm{ps}}$) are detected with $1.7\sigma$ and $2.8\sigma$ significance for Overlap and Wide-Area samples, respectively. They are also within $1\sigma$ of each other, showing overall consistency.  The King slopes of $\gamma=6.6_{-2.1}^{+2.1}$ and $4.1_{-0.5}^{+0.7}$ indicate a sharper decline in tSZ two-point correlation than that of dust, as possible from a nonlinear relationship between ionized gas and lower mass neighbors.  An uneven presence of radio contamination in the profile's outer vs inner radius, could also increase our reported slope via tSZ fit underestimation.

\subsection{Stellar Mass Binning}\label{subsec:stellar_mass_binning}

We also wish to measure the dust mass and thermal energy from our galaxies as a function of stellar mass, similar to previous studies \citep{Planck2014-XIV,Greco2015,Meinke2021}.  Hence we no longer are concerned with a profile fit, but rather the total integrated signal over a solid angle expected to be dominated by the primary central source. 

A circular top-hat aperture with radius of $R=2.0\arcmin$ is selected to integrate within, on all frequency maps per stellar mass bin.  The two-component fit of eq.~(\ref{eq:2compFit}) is then applied to each sample and bin.  Errors are calculated via bootstrap resampling from the same resample catalogs as Section~\ref{subsec:Uncertainties}.  

We separate our catalogs into stellar mass bins with widths of $0.1$ in $\log_{10}(M_{\star}/\text{M}_{\odot})$, over a range from $10.9-12.0$ and $10.8-12.1$ dex for our Overlap and Wide-Area samples, respectively.  Additional bins were possible in the latter due to its larger number of total galaxies.  The impact of bin size was checked and found to be negligible, as wider $0.2$ dex-wide bins produced similar results, but created fewer points of measurement for the subsequent stellar mass uncertainty correction to be applied in Section~\ref{subsubsec:stellarMassUncertainty}.

\begin{figure*}[ht]
	\centering
	\includegraphics[width=0.65\linewidth, keepaspectratio]{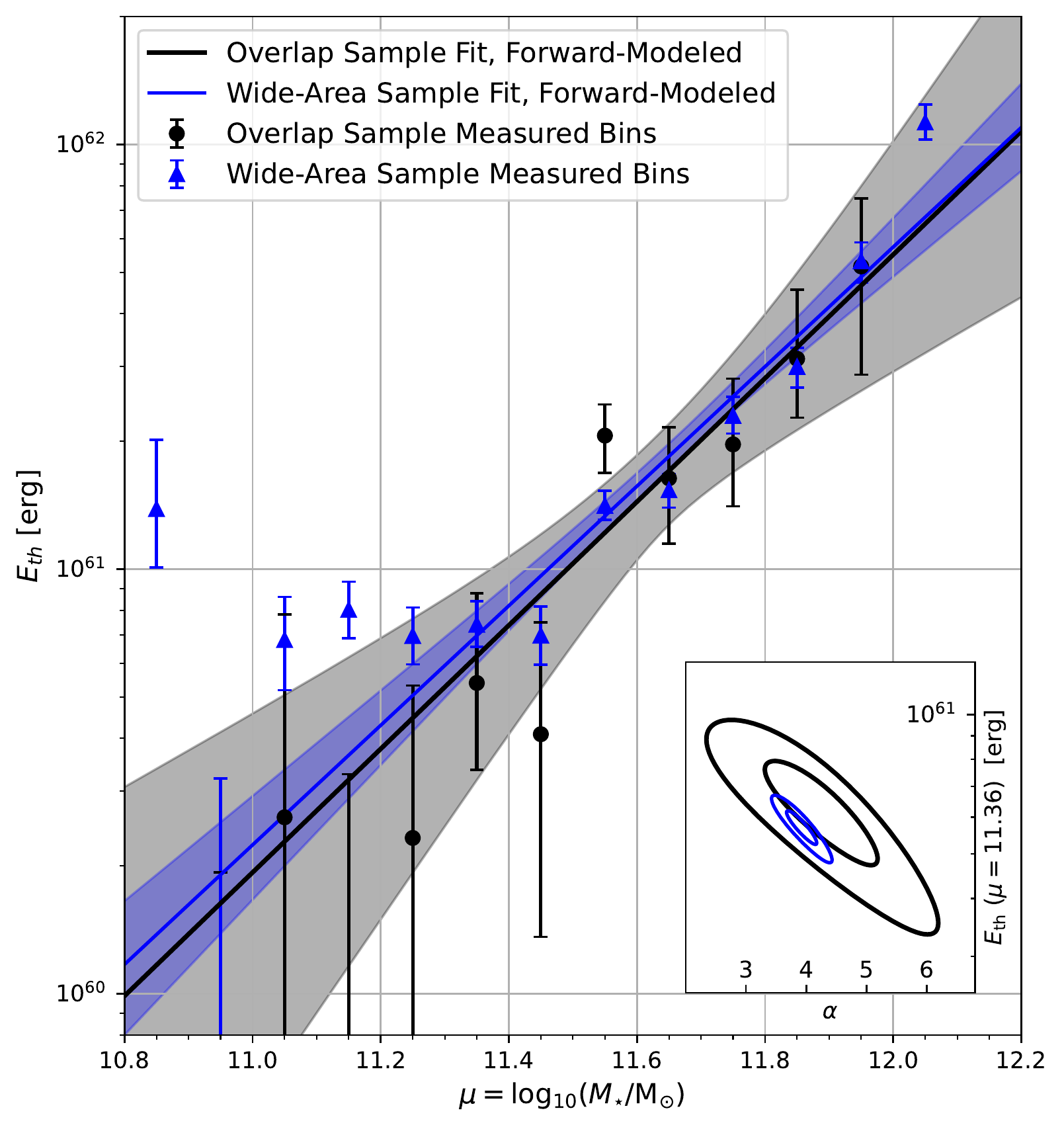}
	\caption{Overlap (\textit{black, circles}) and Wide-Area (\textit{blue, triangles}) galaxies' energy in 0.1 dex stellar mass bins with associated energy-mass fit as described in Table~\ref{tab:E_M_fits} after forward-modeling our stellar mass uncertainty  (Section~\ref{subsubsec:stellarMassUncertainty}).  The shaded fit regions correspond to forward-modeled $2\sigma$ levels.  Inset: 1- and 2-$\sigma$ bounds of the (non-forward-modeled) fit parameters $E_{\rm{th}}(\mu=11.36 \text{~dex})$, and slope $\alpha$.  We show $E_{\rm{th}}(\mu=11.36$ dex$)$ instead of $E_{\rm{pk}}$ here as the samples contain different peak masses.}
	\label{fig:Eth_vs_M_wFit}
\end{figure*}

Integrated Compton-$y$ values are converted to thermal energies via eq.~(\ref{eq:EthrmT}) and are shown versus stellar mass in Fig.~\ref{fig:Eth_vs_M_wFit}.  These align closely to the previous investigation in \citet{Meinke2021}, showing a clear trend of increasing thermal energy versus stellar mass.  For our mass range we expect the relation between thermal energy ($E_{\rm{th}}$) and stellar mass to be sufficiently described by a simple power-law model.  As our analysis is conducted in terms of $\mu=\log_{10}(M_{\star}/\text{M}_{\odot})$, we write this energy-mass relation as a log-log model,
\begin{equation}
	\mathcal{E}(\mu)=\log_{10}(E_{\rm{th}})(\mu)=\log_{10}(E_{\rm{pk}})+\alpha \left(\mu-\mu_{pk}\right)
	\label{eq:Eth_vs_M_log},
\end{equation}
where $\alpha$ is the slope, $\mu_{pk}=\log_{10}(M_{\star,\rm{pk}}/\text{M}_{\odot})$ is the $\log_{10}$ peak stellar mass, and $E_{\rm{pk}}$ is the thermal energy at the peak stellar mass.

We conduct a similar analysis using the two-component fit's dust result to determine our dust mass (via Section~\ref{subsec:dust_mass}) as a function of stellar mass. These are shown in Fig.~\ref{fig:Md_vs_M_wFit}.  Here we again assume a log-log power-law relation,  
\begin{equation}
	\mathcal{M}_{d}(\mu)=\log_{10}(M_{\rm{d}})(\mu)=\log_{10}(M_{\rm{d,pk}})+\alpha_{\rm{d}}\left(\mu-\mu_{\rm{pk}}\right)
	\label{eq:Md_vs_M_log},
\end{equation}
where $\alpha_{\rm{d}}$ is the slope, and $M_{\rm{d,pk}}$ is the dust mass at the peak stellar mass.  Both power-law equations of eqs.~(\ref{eq:Eth_vs_M_log})~\&~(\ref{eq:Md_vs_M_log}) describe the expected relation versus stellar mass prior to any contributions that may arise from stellar mass uncertainty, discussed below.

\subsubsection{Stellar Mass Uncertainty}\label{subsubsec:stellarMassUncertainty}

\begin{deluxetable*}{cccccc}[ht]
	\tablecaption{Forward-modeled energy and dust mass versus stellar mass fits of eqs.~(\ref{eq:Eth_vs_M_log})~\&~(\ref{eq:Md_vs_M_log}) following the inclusion stellar mass uncertainty via eqs.~(\ref{eq:Ebin}), (\ref{eq:Mdbin}), \& (\ref{eq:bin_w}). For our two quiescent galaxy samples and \textsuperscript{\textdagger}previous energy-stellar mass fit from \citet{Meinke2021}. \label{tab:E_M_fits}}
	\tablehead{
		\colhead{Catalog} & \colhead{$\mu_{\rm{pk}}$} & \colhead{$E_{\rm{pk}}$} & \colhead{$\alpha$} & \colhead{$M_{\rm{d,pk}}$} & \colhead{$\alpha_{\rm{d}}$} \\
		\colhead{} & \colhead{$\left[\log_{10}(M_{\star}/\text{M}_{\odot})\, \right]$} & \colhead{$[10^{60}\text{erg}]$} & \colhead{[unitless]} & \colhead{$[10^{8}\text{M}_{\odot}]$} & \colhead{[unitless]}
	}
	\startdata
	Overlap Sample & $11.36$ & $6.45_{-1.52}^{+1.67}$ & $4.04_{-0.92}^{+0.94}$ & $6.23_{-0.67}^{+0.67}$ & $2.59_{-0.44}^{+0.46}$\\
	Wide-Area Sample & $11.40$ & $8.20_{-0.52}^{+0.52}$ & $3.91_{-0.25}^{+0.25}$ & $6.76_{-0.56}^{+0.56}$ & $2.22_{-0.34}^{+0.35}$ \\
	Meinke et al. 2021\textsuperscript{\textdagger} & $11.36$ & $5.98_{-1.00}^{+1.02}$ & $3.77_{-0.74}^{+0.60}$ & -- & -- \\
	\enddata
	\tablecomments{\textsuperscript{\textdagger}Our methods differ slightly from those in \citet{Meinke2021} due to changes in beam, map processing, and $S/N<1\sigma$ cut.  Dust mass was calculated from eq.~(\ref{eq:dust_mass}) for a $\kappa(\nu_{0})=0.08$~m$^{2}$kg$^{-1}$, which we recognize might fluctuate by a further factor of two or $0.30$ dex. \vspace{-0.6cm}}
\end{deluxetable*}

The main caveat in the stellar mass bin approach is our catalogs' inherent stellar mass uncertainty.  We find our SED fitting in Section~\ref{sec:sample_selection} has a stellar mass uncertainty of $\sigma_{\rm SED}=0.16$ dex, due in part from our high redshift and use of only photometric data \citep{Meinke2021}.  Thus, to accurately fit measured stellar mass bins with the energy and dust mass vs stellar mass functions of eqs.~(\ref{eq:Eth_vs_M_log})~\&~(\ref{eq:Md_vs_M_log}), we must correctly incorporate our stellar mass uncertainty.  Luckily our quiescent galaxy mass distributions are well fit by Gaussians of the form ${G}(\mu_{\rm{pk}},\sigma_{q}^{2})$, with $\sigma_{q}=0.20$ dex for both and $\mu_{\rm{pk}}$ listed in Table~\ref{tab:E_M_fits}. Applying uncertainty, the average $\log_{10}$ thermal energy within a stellar mass bin centered on $\log_{10}$ mass $\mu_i$ becomes,
\begin{equation}
	\overline{\mathcal{E}}(E_{\rm{pk}},\alpha,\mu_{i})=\frac{\int_{8}^{15}\mathcal{E}(\mu)~w(\mu,\mu_{i})~{d}\mu}{\int_{8}^{15}w(\mu,\mu_{i})~d\mu}
	\label{eq:Ebin}
\end{equation}
and similarly for average $\log_{10}$ dust mass,
\begin{equation}
	\overline{\mathcal{M}}_{d}(M_{\rm{d,pk}},\alpha_{\rm{d}},\mu_i)=\frac{\int_{8}^{15}\mathcal{M}_{d}(\mu)~w(\mu,\mu_{i})~d\mu}{\int_{8}^{15}w(\mu,\mu_{i})~d\mu}
	\label{eq:Mdbin},
\end{equation}
where $w(\mu,\mu_{i})$ is the effective weight of a galaxy with $\log_{10}$ stellar mass $\mu$ to appear within the mass bin defined from $\mu_{i-1/2}$ to $\mu_{i+1/2}$,
\begin{equation}
	w(\mu,\mu_{i})={G}(\mu-\mu_{\rm{pk}},\sigma_{\mu}^2)\int_{\mu_{i-1/2}}^{\mu_{i+1/2}}{G}(\mu^{\prime}-\mu,\sigma_{\rm{SED}}^2)~d\mu^\prime
	\label{eq:bin_w},
\end{equation}
with $\sigma_{\mu}^2=\sigma_q^2-\sigma_{\rm{SED}}^2$, corresponding to the standard deviation of our expected true mass distribution if no stellar mass uncertainty was present.  The first Gaussian term is the weight of a galaxy selected with the true mass $\mu$, while the integral and second Gaussian term is the chance that said galaxy actually appears in the mass bin between $\mu_{i-1/2}$ and $\mu_{i+1/2}$ due to our stellar mass uncertainty.

Equations~(\ref{eq:Ebin})~\&~(\ref{eq:Mdbin}) take the ideal generalized power-law functions of eqs.~(\ref{eq:Eth_vs_M_log})~\&~(\ref{eq:Md_vs_M_log}) and forward-model them into expected observations within a stellar mass bin.  To clarify, this method is synonymous with the past energy-mass approach in \citet{Meinke2021}, which was not described in as much detail.

Equation~(\ref{eq:Ebin}) was fit to the energy-mass bins found for each catalog.  We refrain from fitting any bins with $S/N<1\sigma$ to avoid introducing spurious bias.  The forward-modeled best fits and $2\sigma$ uncertainties are shown in Fig.~\ref{fig:Eth_vs_M_wFit}.  The inset plot shows the posterior distributions of $\alpha$ and $E_{\rm{th}}(\mu=11.36$ dex).  We display $E_{\rm{th}}(\mu=11.36$ dex) instead of $E_{\rm{pk}}$ in order to compare catalogs, as they have different peak masses ($\mu_{\rm{pk}}$).  Best fit values for $E_{\rm{pk}}$ and $\alpha$ are shown in Table~\ref{tab:E_M_fits}, compared to previous SPT results \citep{Meinke2021}.  All three catalogs show agreeing slopes ($\alpha$) within $1\sigma$, $4.04_{-0.92}^{+0.94}$ for the Overlap and $3.91_{-0.25}^{+0.25}$ for the Wide-Area Sample.  These slopes uphold a strong trend of observations that indicate only CGM in the most massive galaxies and clusters produce significant levels of thermal energy \citep{Greco2015}.  Some plateauing at lower stellar mass may be present as well, evident by the low mass bin outliers in our Wide-Area measurements.  Energies at peak mass ($E_{\rm{pk}}$) are also significant, at a level of $4\sigma$ for our Overlap Sample and $16\sigma$ for the Wide-Area.

\begin{figure*}[t]
	\centering
	\includegraphics[width=0.65\linewidth, keepaspectratio]{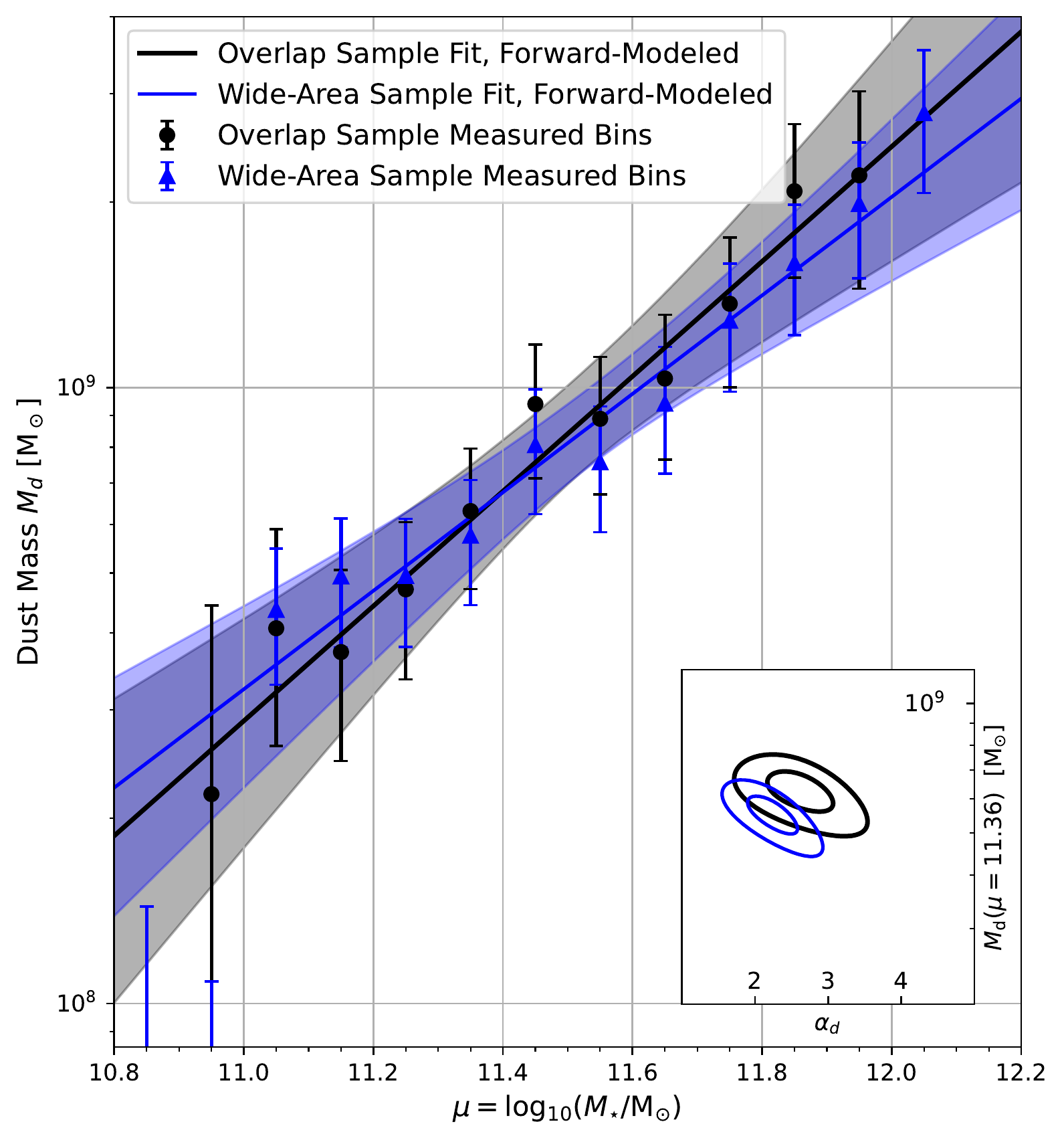}
	\caption{Overlap (\textit{black, circles}) and Wide-Area (\textit{blue, triangles}) galaxies' dust mass in 0.1 dex stellar mass bins with associated dust-stellar mass fit as in Table~\ref{tab:E_M_fits} after forward-modeling our stellar mass uncertainty  (Section~\ref{subsubsec:stellarMassUncertainty}).  The shaded fit regions correspond to forward-modeled $2\sigma$ levels.  Inset: 1- and 2-$\sigma$ bounds of the non-forward-modeled fit parameters $M_{\rm{d}}(\mu=11.36~\text{dex})$, $\alpha_{\rm{d}}$.  We show $M_{\rm d}(\mu=11.36~\text{dex})$ instead of $M_{\rm{d,pk}}$ here as the samples contain different peak stellar masses.}
	\label{fig:Md_vs_M_wFit}
\end{figure*}

In similar fashion, eq.~(\ref{eq:Mdbin}) was fit to the measured dust versus stellar mass bins found for each catalog.  We again refrain from fitting any bins with $S/N<1\sigma$ to avoid introducing spurious bias.  The forward-modeled best fits and $2\sigma$ uncertainties are shown in Fig.~\ref{fig:Md_vs_M_wFit}.  The inset plot shows the posterior distributions of $\alpha_{\rm{d}}$ and $M_{\rm{d}}(\mu=11.36$ dex).  Fits for $M_{\rm{d,pk}}$ and $\alpha_{\rm{d}}$ are shown in Table~\ref{tab:E_M_fits}.

Dust mass at the peak stellar masses were detected at a $9.3\sigma$ level for the Overlap and $12\sigma$ for the Wide-Area Sample.  Both catalogs show agreeing slopes ($\alpha_{\rm{d}}$) within $1\sigma$, $2.59_{-0.44}^{+0.46}$ for Overlap and $2.22_{-0.34}^{+0.35}$ for Wide-Area. As these slopes are greater than $1$, they highlight a non-linear relationship between dust and stellar mass that indicates increasingly massive quiescent galaxies have a higher dust-to-stellar mass ratio.  However, this trend may only be present in the high stellar mass regime due to our narrow galaxy mass distribution.

We also consider the potential of neighboring two-halo contributions that may falsely inflate these measurements.  Discussed in more detail in \citet{Meinke2021} with the same $R=2.0\arcmin$ aperture corresponding to a radius of $\approx2.0$ comoving Mpc, a central halo will be expected to dominate the tSZ signal for thermal energies exceeding $\approx3\times10^{60}$~erg or halo masses larger than $\approx10^{13}\text{M}_\odot$. This is determined under the assumption that the gas in all neighboring halos is heated to virial temperature $T_{\rm{vir}}$.  As all our reported thermal energies with $S/N>1\sigma$ in Fig.~\ref{fig:Eth_vs_M_wFit} reside above $3\times10^{60}$~erg, we conclude the two-halo contribution within them are negligible compared to their respective measured uncertainties.  

We also consider our dust mass measurements in Fig.~\ref{fig:Md_vs_M_wFit} to be the expected upper limit.  As discussed in Section~\ref{subsec:dust_mass}, the $R=2.0\arcmin$ aperture produces a result roughly $0.30$~dex greater than a separate conservative estimate via evaluation of the dust profile in Section~\ref{subsec:dust}.  As the assumed dust mass opacity coefficient $\kappa(\nu_{0})=0.08$~m$^{2}$kg$^{-1}$ contains a further factor of two or $0.30$~dex uncertainty, any contributions from neighbors are likely within this uncertainty.  Hence, we refrain from drawing any large conclusions aside from the relation indicated by our dust-to-stellar mass slope fit, as we expect $\kappa(\nu_{0})$ to not vary significantly between our stellar mass bins.

\subsection{Implications for AGN Feedback}\label{subsec:Feedback}

Our constraints on $E_{\rm{th}}$ allow us to glean information about AGN feedback, though detailed comparisons with AGN models are best carried out alongside full numerical simulations.  First, comparisons with previous work in \citet{Meinke2021} show strong similarities in their energy-mass fit.  This is as expected, due to an overlap of target galaxy samples and use of SPT data.  However, as before we also see significant similarities with the lower redshift ($z\approx0.1$) results from \citet{Greco2015}, even though they used locally bright galaxies as opposed to our age~$>1$ Gyr, $SSFR <0.01$ Gyr$^{-1}$, quiescent galaxies.  There are a variety of theoretical models that suggest a good match between the most massive quiescent galaxies at moderate redshifts and the central galaxies of massive halos in the nearby universe \citep[e.g,][]{Moster2013,Schaye2015,Pillepich2018}.

Such lack of thermal energy evolution in the CGM around massive galaxies since $z\approx 1$ mirrors what occurs for the luminosity function of these galaxies \cite[e.g.][]{vanDokkum2010,Muzzin2013}.  This trend could be more indicative of radio-mode AGN feedback, where gas accretion contributes to CGM heating and radiative losses to CGM cooling.  Whenever cooling surpasses heating, jets will arise that quickly push the gas up to a constant temperature and entropy at which cooling is inefficient.  On the other hand, quasar models instead produce an energy input from feedback which occurs once at high redshift, heating the gas such that cooling is extremely inefficient up until today.  As a result, gravitational heating will increase $E_{\rm{th}}$ without any significant mechanism to oppose it. However, specifics of this evolution are highly dependent on the history of galaxy and halo mergers between $0<z\lesssim{1}$.  Hence, it is possible that some types of quasar dominated models may be compatible with our measurements. 

A second major inference is the overall level of feedback. To estimate the magnitude of gravitational heating, we can assume that the gas collapses and virializes along with an encompassing spherical dark matter halo, and is heated to the virial temperature $T_{\rm{vir}}$.  This gives 
\begin{equation}
E_{\rm{th,halo}}(M_{13},z)=1.5\times10^{60}~{\rm erg}~M_{13}^{5/3}~(1+z),
\label{eq:Egravhalo}
\end{equation}
where $M_{13}$ is the mass of the halo in units of $10^{13}\text{M}_{\odot}$ \citep{Spacek2016}. We can convert from halo mass to galaxy stellar mass using the observed relation between black hole mass and halo circular velocity for massive quiescent galaxies \citep{Ferrarese2002}, and the relation between black hole mass and bulge dynamical \citep{Marconi2003}. As shown in \citet{Spacek2016}, this gives
\begin{equation}
E_{\rm{th,gravity}}(M_{\star},z)\approx{5}\times10^{60}~{\rm{erg}}~\frac{M_{\star}}{10^{11}\text{M}_{\odot}}~(1+z)^{-3/2},
\label{eq:Egrav}
\end{equation}
representing the expected total thermal energy around a galaxy of stellar mass $M_{\star}$ ignoring both radiative cooling and feedback.  For a mean redshift of $z\approx1.1$ this yields $\approx3.8\times10^{60}~{\rm{erg}}$ and $\approx4.1\times10^{60}~{\rm{erg}}$ for our $M_{\rm{\star,pk}}=2.29\times10^{11}\text{M}_{\odot}$ (Overlap) and $2.51\times10^{11}\text{M}_{\odot}$ (Wide-Area), respectively.  
Unfortunately this estimate has an uncertainty of about a factor of two, which is significantly larger than the uncertainty in our measurements. 
Regardless, these are lower than the $E_{\rm{pk}}=6.45_{-1.52}^{+1.67}\times10^{60}~{\rm{erg}}$ and $8.20_{-0.52}^{+0.52}\times10^{60}~{\rm{erg}}$ respectively, that we found in Section~\ref{subsec:stellar_mass_binning}.  These suggest the presence of additional non-gravitational heating, particularly as cooling losses are not included in eq.~(\ref{eq:Egrav}). 

To estimate quasar-mode feedback heating we use a simple model described in \citet{Scannapieco2004}, given as
\begin{equation}
E_{\rm{th,feedback}}(M_{\star},z)\approx4\times10^{60}~{\rm{erg}}~\epsilon_{k,0.05}~\frac{M_{\star}}{10^{11}\text{M}_{\odot}}~(1+z)^{-3/2},
\label{eq:EAGN}
\end{equation}
where $\epsilon_{k,0.05}$ is the fraction of bolometric luminosity from the quasar associated with an outburst, normalized by a fiducial value of $5\%$, which is typical of quasar models  \cite[e.g.][]{Scannapieco2004,Thacker2006,Costa2014}.  Taking $\epsilon_{k,0.05}=1$ for our samples' mean redshifts and peak masses, this gives $\approx3.0\times10^{60}~{\rm{erg}}$ (Overlap) and $\approx3.3\times10^{60}~{\rm erg}$ (Wide-Area).  Adding these to the contributions from $E_{\rm{th,gravity}}$ above gives a total energy of $\approx6.8\times10^{60}~{\rm{erg}}$ and $\approx7.4\times10^{60}~{\rm{erg}}$, respectively.  Including this additional energy from quasar-mode AGN feedback better matches our results of $E_{\rm{pk}}=6.45_{-1.52}^{+1.67}\times10^{60}~{\rm{erg}}$ and $8.20_{-0.52}^{+0.52}\times10^{60}~{\rm{erg}}$ than heating from gravity alone. It also does not account for any energy losses. 

Meanwhile, radio mode models are expected to fall somewhere between these two limits, with jets supplying power to roughly balance cooling processes, but never adding a large burst of additional energy near that of eq.~(\ref{eq:EAGN}).  This would suggest values slightly below our measurements, but again with too much theoretical uncertainty to draw any definite conclusions.

A third major inference from our measurements comes from the slope of eq.~(\ref{eq:Eth_vs_M_log}), which is significantly steeper than in our simple models.  This is most likely due to uncertainties in the halo-mass stellar mass relation, which are particularly large for massive $z\approx1$ galaxies \citep{Wang2013,Lu2015,Moster2018,Kravtsov2018,Behroozi2010,Behroozi2019}.  Recent studies alongside our own \citep{Schaan2021,Amodeo2021,Meinke2021,Vavagiakis2021} make it clear that observations are now fast outpacing theoretical estimates, a major change from several years ago when only galaxy cluster sized halos were capable of being moderately detected. Future comparisons between measurements and full simulations will yield key new insights into the processes behind AGN feedback.

\section{Discussion}\label{sec:Discussion}

Many galaxies from $z\approx1$ to present day, starting with the most massive, undergo a process that quenches new star formation.  The proposed likely culprit is feedback from accretion onto supermassive black holes, which would have a noticeable impact on the surrounding CGM.  By probing the CGM for signs of heating via the redshift-independent tSZ effect, we can begin to differentiate between various AGN accretion models and provide much needed constraints for theoretical simulations.

Here we have selected $N=387,627$ old quiescent galaxies with low SFR at $0.5\leq{z}\leq1.5$ from DES and WISE within the ACT millimeter telescope field (Wide-Area Sample).  A subset of $N=94,452$ galaxies are further used to incorporate data from SPT for an analysis across multiple instruments (Overlap Sample).  These quiescent galaxies are ideal candidates to show strong heating via feedback.  A detailed set of map processing (Section~\ref{subsec:MapProcessing}) is conducted to mitigate any systematic differences between SPT and ACT, applying a uniform $2.1\arcmin$ FWHM Gaussian beam across all maps that reside near $95/150/220$ GHz.  We then subtract a $5.0\arcmin$ resolution \textit{Planck} SMICA SZ-Free CMB map to remove large-scale CMB fluctuations uncorrelated with our target galaxies.  

When stacked, we observe separable dust and tSZ profiles from both galaxy catalogs.  Further split into stellar mass bins, we show a clear thermal energy versus stellar mass relation influenced by our photometric uncertainty in stellar mass.  Often simply discarded in tSZ analysis, we also use the dust to estimate the associated dust mass for our samples.

This work builds off of previous $z\approx1$ quiescent galaxy stacking conducted by \citet{Spacek2017,Meinke2021}.  Our analysis here is enhanced from the prior via use of the recent ACT data release \citep{Naess2020,MallabyKay2021}, improved map processing, and a heightened focus on the radial profile and dust mass of our target galaxies.  Others have also begun a more concerted effort to analyze the galactic structure of the tSZ and kSZ  \citep{Schaan2021,Amodeo2021,Calafut2021,Vavagiakis2021,Lokken2022}.

Firstly, the dust profile of our Overlap and Wide-Area galaxies produce up to $16\sigma$ and $20\sigma$ detection respectively, for radial bins with widths of $0.5\arcmin$.  Profile detection with $S/N\geq{2}\sigma$ is found out to $15\arcmin$ ($15.2$ comoving Mpc).  We observe a dust profile shape for each sample indicative of a central point source associated with our galaxies and an extended profile that traces the two-point correlation function of neighboring galaxies and structure.  To obtain a slope for the extended dust, we fit a point source plus King model as described in Section~\ref{subsec:profileFits}, finding slopes of $\gamma=2.60_{-0.15}^{+0.16}$ and $2.95_{-0.14}^{+0.16}$.  These are $20-90\%$ greater than power-law fits conducted in galaxy cluster studies \citep[$\gamma\approx1.5-2.0$,][]{Eftekharzadeh2015,Coil2017}.  We attribute most of this discrepancy to a divergence between the King and power-law models when near or below our core radius of $r_{0}=3.0$ comoving Mpc.  

Such dust profile analysis might also provide a novel method to constrain a catalog's intergalactic medium (IGM) and central halo mass, wherein a similar catalog of known halo mass or bias factor is used to compare two-point correlation terms traced by the observed extended dust.  However a correct comparison requires careful consideration of all systematic differences in catalog selection and accurate removal of dust associated with the central source(s).

Secondly, the high $S/N$ detection of dust allows us to convert our dust intensity fit in the $\nu_{0}=353$~GHz rest frame to a dust mass as shown in eq.~(\ref{eq:dust_mass}).  The primary difficulty in this approach is an existing uncertainty in the dust mass opacity or absorption cross-section coefficient, where we take an intermediate value of $\kappa(\nu_{0})=0.08$~m$^{2}$kg$^{-1}$ while acknowledging this may vary by a factor of two \citep{Draine2003,Dunne2003,Casey2012}.  We then consider reasonable lower and upper limits to isolate the dust solely associated with our central galaxies: the lower limit from the point source fit of our aforementioned profile fit, which has noted degeneracy with the King model at small radii; and an upper limit through integration within a circular aperture of $R=2.0\arcmin$ radius.  

These result in a $\log_{10}$ dust mass range from $8.43_{-0.12}^{+0.10}$ to $8.82_{-0.11}^{+0.09}$ $\log_{10}(\text{M}_{\odot})$ for the Overlap Sample and $8.46_{-0.12}^{+0.09}$ to $8.83_{-0.11}^{+0.09}$ $\log_{10}(\text{M}_{\odot})$ for the Wide-Area Sample.  As a dust-to-stellar mass ratio, these become $-2.98_{-0.12}^{+0.10}$ to $-2.59_{-0.11}^{+0.09}$ $\log_{10}(M_{\rm{d}}/\text{M}_{\star})$ and $-2.98_{-0.12}^{+0.09}$ to $-2.61_{-0.11}^{+0.09}$ $\log_{10}(M_{\rm{d}}/\text{M}_{\star})$, respectively.  Other studies involving massive or quiescent galaxies at $z\approx1$ have found $\log_{10}(M_{\rm{d}}/\text{M}_{\star})\approx-3.5$ to $-2.7$ \citep{Gobat2018,Magdis2021}.  As our dust mass contains an additional $0.30$ dex uncertainty from $\kappa(\nu_{0})$, we conclude our values are in agreement, but do not draw any larger inferences.  This consistency is notable however, as it echoes reports of higher dust-to-stellar mass ratios for massive galaxies at $z\approx1$ than those at nearby lower redshifts \citep{Santini2015,Magdis2021}.

Thirdly, we inspect our tSZ radial profile and obtain a clear central detection, up to $5.4\sigma$ in our Overlap Sample and $11\sigma$ in the Wide-Area Sample.  However our detection falls off much more rapidly than for dust, dropping below $2\sigma$ at $4.0\arcmin$ ($4.0$ comoving Mpc) and $8.0\arcmin$ ($8.1$ comoving Mpc), respectively.  As a result compared to dust, we find steeper King slopes of $\gamma=6.6_{-2.1}^{+2.1}$ and $4.1_{-0.5}^{+0.7}$, which indicate a sharper decline in the tSZ two-point correlation or two-halo term.  This is within expectations, since the neighboring lower mass galaxies should contain reduced or cooler levels of ionized gas at a nonlinear relationship to stellar mass \citep{Hill2018}.  We also note that radio contamination would produce an underestimated fit of the tSZ, while an uneven relation of radio contaminants versus radii could affect fit slopes as well.  This effect is likely marginal for our redshift and frequency bands.

We also fit the tSZ point source amplitudes at $1.7\sigma$ significance for the Overlap Sample and $2.8\sigma$ for the Wide-Area Sample.  These profiles and fits as shown in Fig.~\ref{fig:tSZ_profile} indicate an extended tSZ signal.  However, also evident is the inherent degeneracy between our combined point source plus King model brought about by the map resolution.  This results in an inability to accurately separate the central one-halo tSZ from its two-halo counterpart and limit further detailed analysis.

Fourthly, we focused on measurements split into $0.1$ dex stellar mass bins.  In a more generalized approach than our profiles above, we separated the tSZ and dust integrated within a $R=2.0\arcmin$ radius circular aperture.  These signals were then converted into thermal energy (eq.~\ref{eq:EthrmT}) and dust mass (eq.~\ref{eq:dust_mass}), respectively.  Power-law relations were defined for both thermal energy and dust mass versus stellar mass (eqs.~\ref{eq:Eth_vs_M_log}~\&~\ref{eq:Md_vs_M_log}), scaled with respect to peak mass ($M_{\rm{\star,pk}}$) of $2.29\times10^{11}~\text{M}_{\odot}$ for Overlap and $2.51\times10^{11}~\text{M}_{\odot}$ for Wide-Area Sample.  However, to accurately fit our measurements we also incorporated and forward-modeled a stellar mass uncertainty of $0.16$ dex that arises from our SED fitting of photometric data.  

Our thermal energy to stellar mass power-law fit produces energies of $E_{\rm{pk}}=6.45_{-1.52}^{+1.67}\times10^{60}~{\rm{erg}}$ for Overlap and $8.20_{-0.52}^{+0.52}\times10^{60}~{\rm{erg}}$ for Wide-Area, at their peak mass.  These values only appear inconsistent due to their different peak masses.
The power-law slopes are found to be within $1\sigma$ of each other, with $\alpha=4.04_{-0.92}^{+0.94}$ and $3.91_{-0.25}^{+0.25}$, respectively.  These slopes are significantly steeper than our simple feedback models in Section~\ref{subsec:Feedback}.  This can likely be attributed to model uncertainties in the halo-to-stellar mass relation for massive $z\approx1$ galaxies \citep{Wang2013,Moster2018,Behroozi2019}.  Our fits, shown in Fig.~\ref{fig:Eth_vs_M_wFit}, are also consistent with the previous investigation of \citet{Meinke2021} and lower redshift measurements by \citet{Greco2015}.

Meanwhile, our dust to stellar mass power-law fit produces dust masses of $M_{\rm{d,pk}}=6.23_{-0.67}^{+0.67}\times10^{8}\rm{~M_{\odot}}$ for the Overlap Sample and $6.76_{-0.56}^{+0.56}\times10^{8}\rm{~M_{\odot}}$ for the Wide-Area Sample, at peak stellar mass.  With power-law slopes of $\alpha_{\rm{d}}=2.59_{-0.44}^{+0.46}$ and  $2.22_{-0.34}^{+0.35}$ for the Overlap and Wide-Area samples, respectively.  Our slope fits are more trustworthy than the aforementioned dust masses due to the uncertainties in dust mass opacity $\kappa(\nu_{0})$ that would only scale our measurements and not affect the fit slope $\alpha_{\rm{d}}$.  As our slopes indicate a greater than linear relation ($\alpha_{\rm{d}}>1$), we conclude that massive $z\approx1$ quiescent galaxies have an increasing dust-to-stellar mass ratio for our sample.  Notably this may only be valid for our high and narrow stellar mass range.

Finally, we compare the stellar mass binned energy fit to those predicted by simple theoretical feedback models in Section~\ref{subsec:Feedback}.  Our values more closely align with heating due to quasar-mode feedback rather than from gravity alone.  However, both theoretical models have uncertainties of roughly a factor of two that result in the models overlapping in the same regime that our energy fit is found.  Additionally, a third option of radio-mode feedback would also be situated in-between.  Hence, we conclude our values are strong indicators that some form of AGN feedback is present, but the exact process and amount is unable to be determined when compared to theory.  This highlights the need for improved theoretical and simulation models to keep pace with observations.

With the development of better instruments in both noise, resolution, and sky coverage, observations will continue to improve the characterization of galactic structures.  We have demonstrated here that such detailed analysis at $z\approx1$ is currently possible and will greatly benefit from improved resolution for future analysis. The latest generation of telescopes includes SPT-3G \citep{SPT3G-2014,SPT3G-2022} and TolTEC \citep{TolTEC2018,TolTEC2020} which are more than capable of improving upon this work.  TolTEC in particular, currently being deployed on the $50$ m Large Millimeter Telescope, will grant a $\geq{5}\times$ better resolution. This will enable the ability to resolve the tSZ mainly associated with the CGM and separate it from the potentially still unresolved dust which comes primarily from the underlying galaxy.

\acknowledgments

We would like to thank Peter Behroozi and Elena Pierpaoli for useful discussions that improved the manuscript.
This research used the High Performance Computing (HPC) \textit{Agave} Research Computing at Arizona State University.  The galaxy data used here was from DES and WISE, while the maps are publicly obtained from SPT, ACT, and \textit{Planck}.  This publication makes use of data products from the Wide-field Infrared Survey Explorer, which is a joint project of the University of California, Los Angeles, and the Jet Propulsion Laboratory/California Institute of Technology, and NEOWISE, which is a project of the Jet Propulsion Laboratory/California Institute of Technology. WISE and NEOWISE are funded by the National Aeronautics and Space Administration.

The footprint map of Fig.~\ref{fig:footprints} was based off of code from the \textit{cmb-footprints} public Python code. Manipulation of the maps was aided by the \textit{Healpy} and \textit{Pixell} Python modules for map operations.  
A summary with examples for many of the Python methods used here, including our custom Bayesian estimation code, is publicly available online\footnote{\url{https://github.com/JeremyMeinke/mm_astronomy_stacking}}.
We also acknowledge the indigenous peoples of Arizona, including the Akimel O’odham (Pima) and Pee Posh (Maricopa) Indian Communities, whose care and keeping of the land has enabled us to be at ASU’s Tempe campus in the Salt River Valley, where most of our work was conducted.

\appendix
\section{Correlation Matrices}\label{appendix:tSZ_dust_corr}

\begin{figure*}[ht]
	\gridline{
		\fig{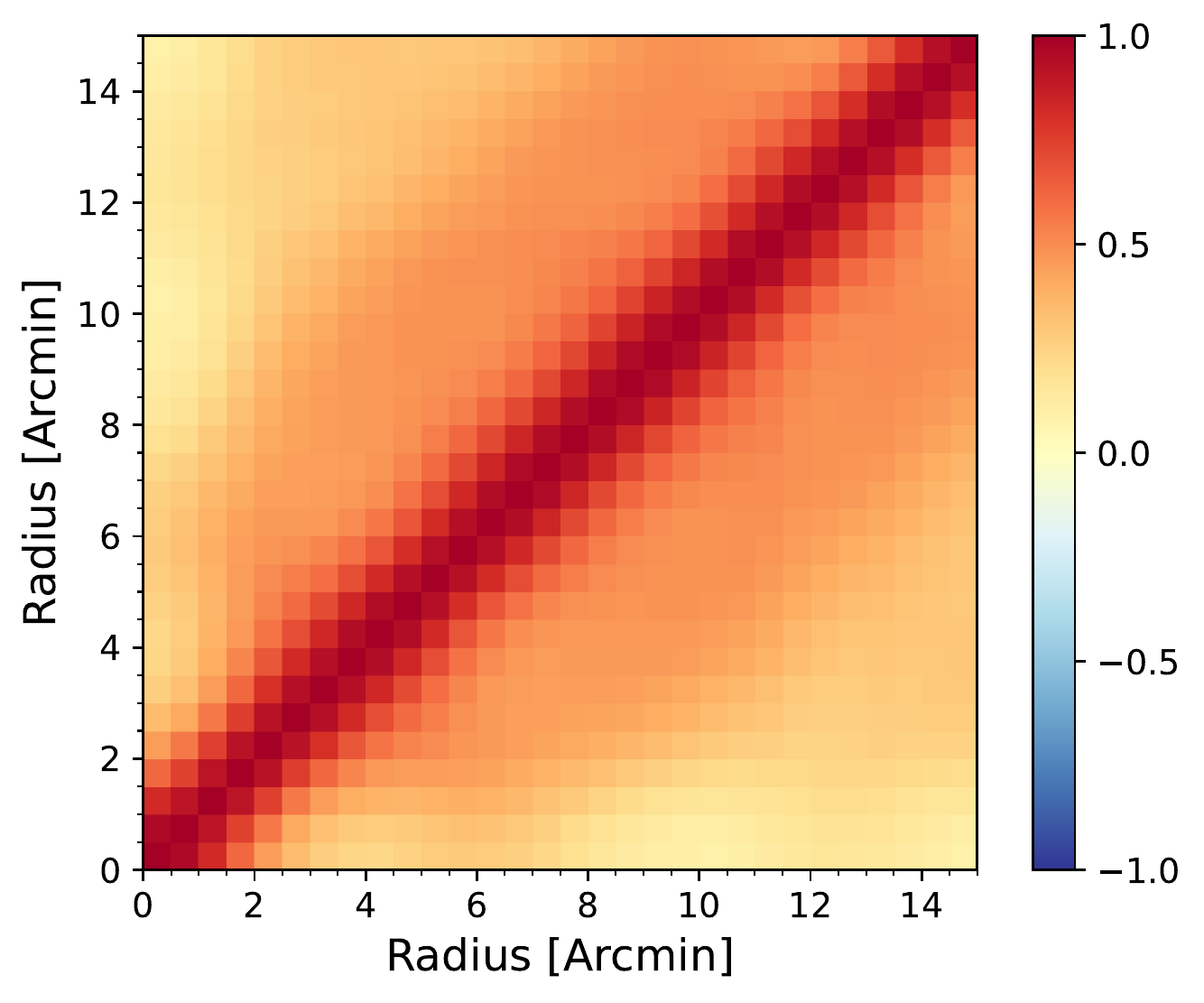}{0.45\textwidth}{ (a)}
		\fig{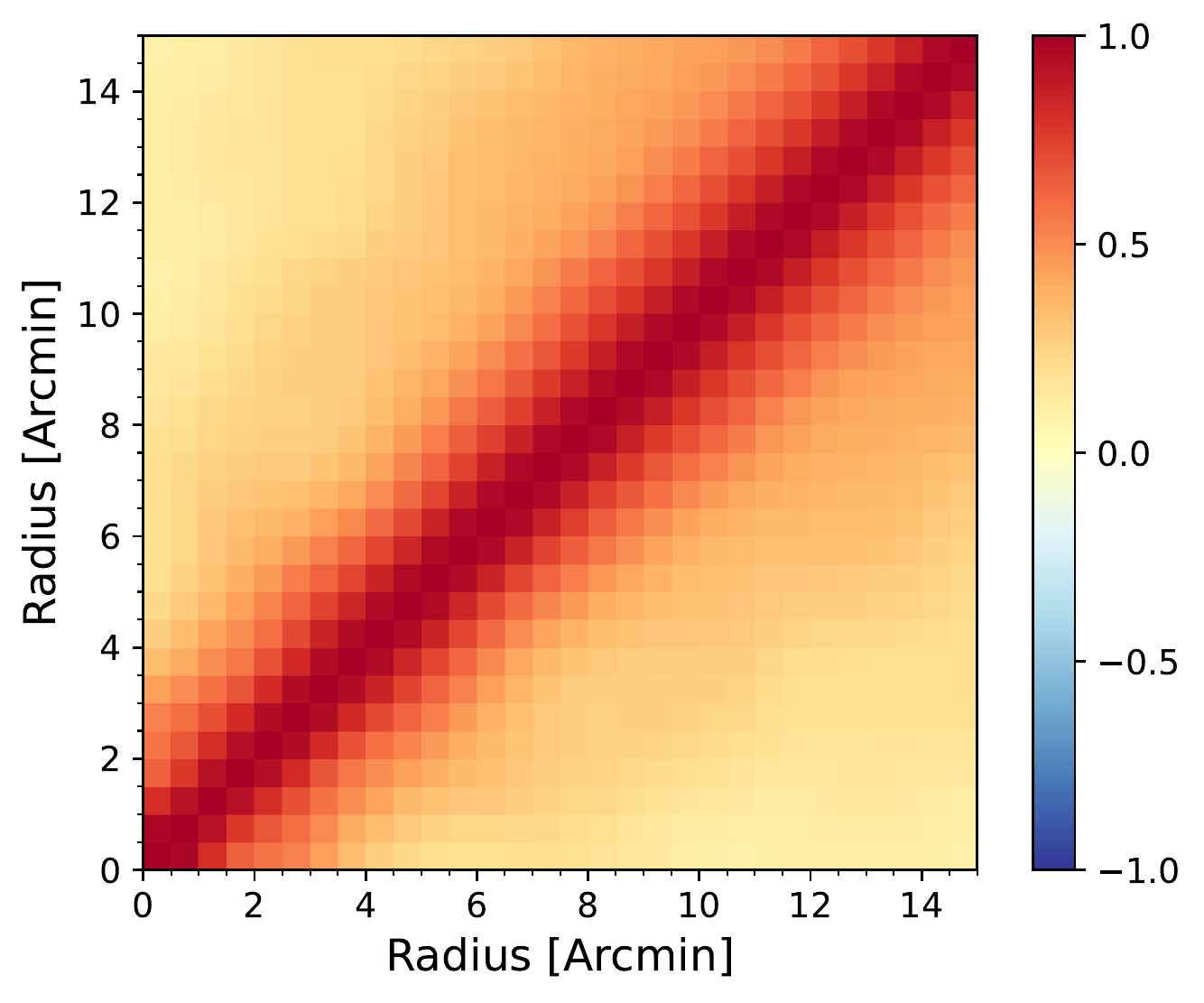}{0.45\textwidth}{ (b)}
	}
	\gridline{ \fig{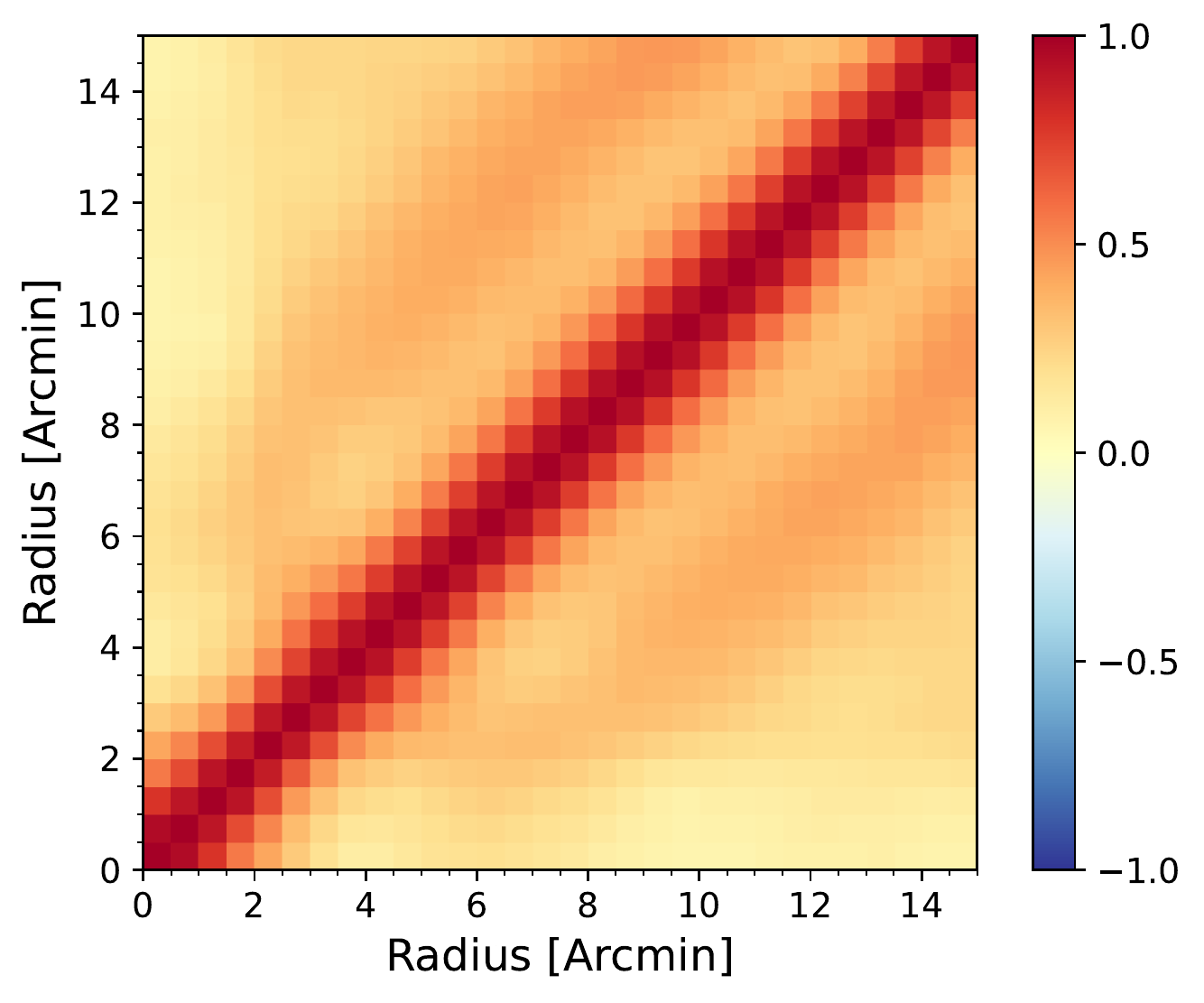}{0.45\textwidth}{ (c)}
	\fig{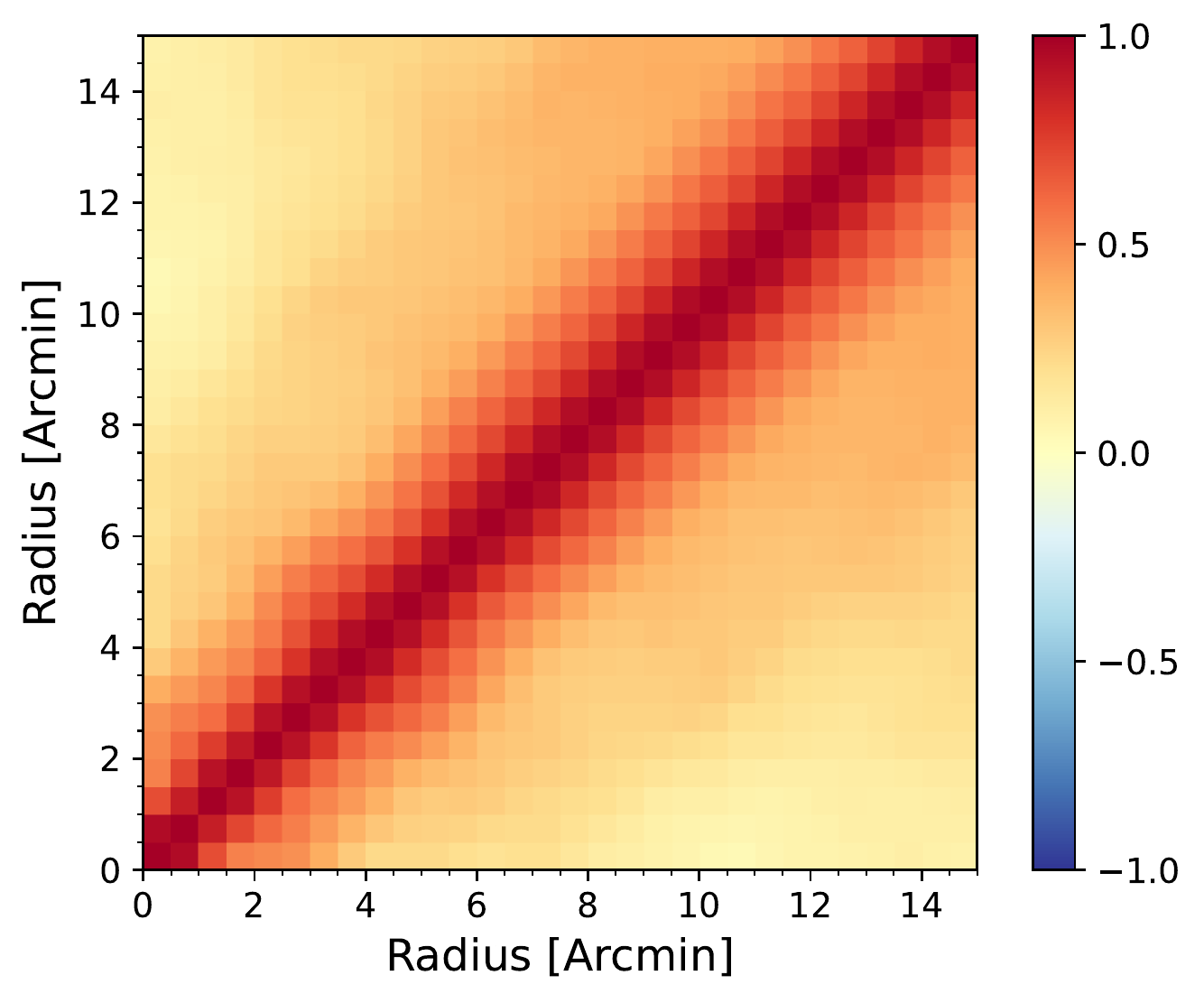}{0.45\textwidth}{(d)}
	}
	\caption{Radial bin ($0-15\arcmin$) correlation matrices for the dust (\textit{left}) and tSZ (\textit{right}) two-component fit (Section~\ref{subsec:TwoCompFitting}) of $4000$ bootstrap resamples outlined in Section~\ref{subsec:Uncertainties} for Overlap (\textit{top}) and Wide-Area (\textit{bottom}) galaxy catalogs.}
	\label{fig:radial_correlations_tSZ_dust}
\end{figure*}

As discussed in Section~\ref{subsec:Uncertainties}, a large number ($4000$) of bootstrap resampled catalogs were constructed from each galaxy catalog.  Covariance matrices for each respective frequency map were then made from these bootstrap resamples and used through our two-component fitting procedure of Section~\ref{subsec:TwoCompFitting}.  Correlation between radial bins is also important for our profile analysis described in Section~\ref{subsec:RadialProfile}.  Thus, we fit our two-component tSZ and dust model to all aforementioned bootstrap resamples to correctly estimate the correlation between neighboring radial bins.  Fig.~\ref{fig:radial_correlations_tSZ_dust} shows the radial correlation matrices for each catalog (Overlap, Wide-Area) and fit component (dust, tSZ).  The most significant effect would be from our beam resolution, with minor contributions likely as a result of residual foreground components not accounted for and structure of the surrounding signals.  As shown in Fig.~\ref{fig:radial_correlations_tSZ_dust}, neighboring radial bins are not independent from one another.

\section{Radial Profile Fit Posteriors}\label{appendix:profile_fit_posteriors}

\begin{figure*}[ht]
	\centering
	\includegraphics[width=0.7\linewidth, keepaspectratio]{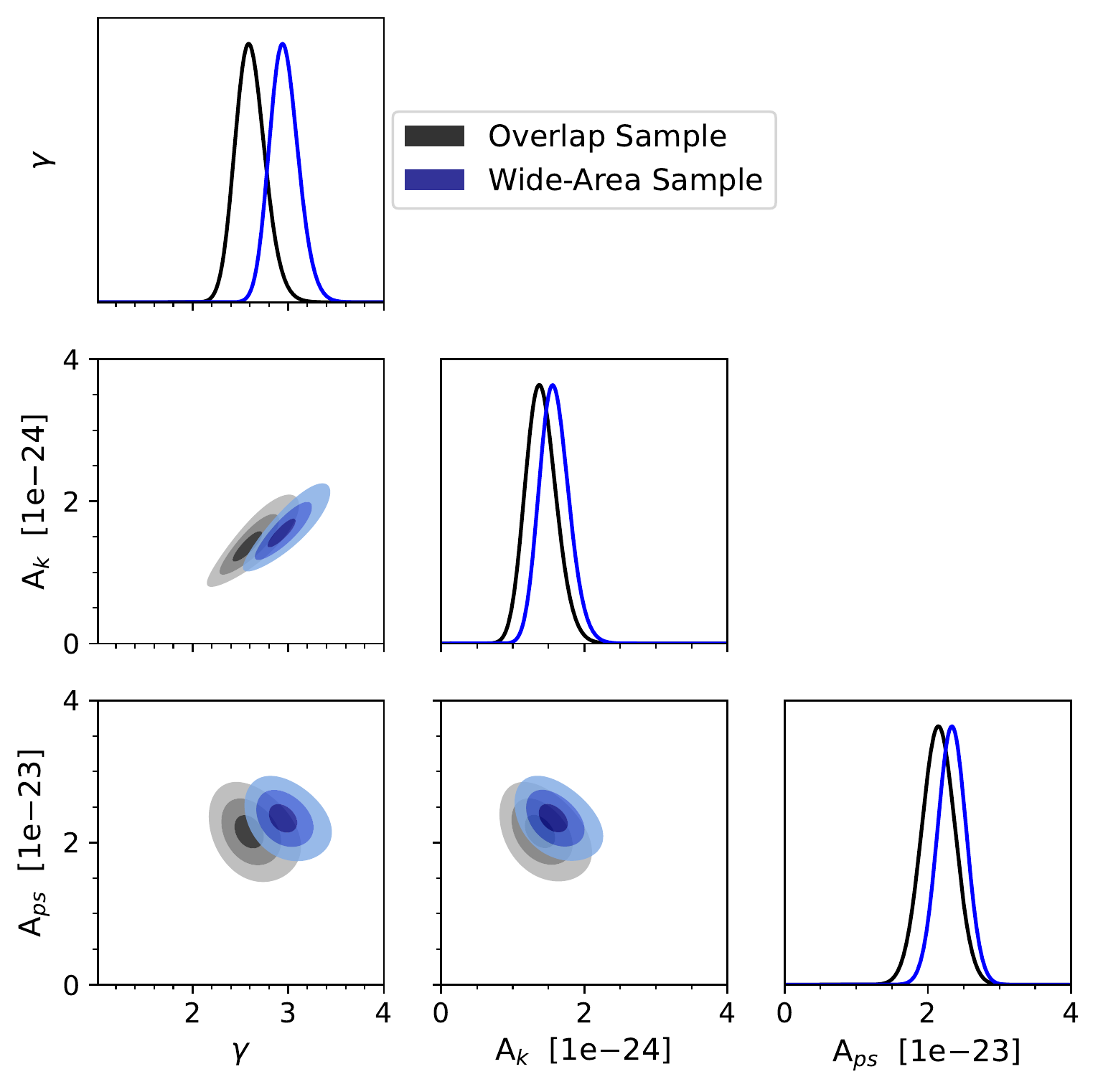}

	\caption{Posterior corner plot of the dust profile fit for our Overlap (\textit{blue}) and Wide-Area (\textit{black}) galaxy catalogs.  From dark to light, the three shaded contour regions correspond to the $1\sigma$, $2\sigma$, and $3\sigma$ levels.}
	\label{fig:dust_profile_fit_corner}
\end{figure*}

The radial profile fits conducted in Sections~\ref{subsec:dust}~\&~\ref{subsec:compton_y} used Bayesian estimation to fit a point source plus King profile as described in Section~\ref{subsec:profileFits}.  Covariance matrices were constructed from the correlation matrices in Appendix~\ref{appendix:tSZ_dust_corr} and passed through the fit procedure.  The marginalized posterior probability distributions of the dust profile fits are shown in Fig.~\ref{fig:dust_profile_fit_corner} for our Overlap and Wide-Area galaxy catalogs.  These correspond to the profile fits shown in Fig.~\ref{fig:dust_profile} and Table~\ref{tab:dust_profile_fit}.

\begin{figure*}[ht]
	\centering
	\includegraphics[width=0.7\linewidth, keepaspectratio]{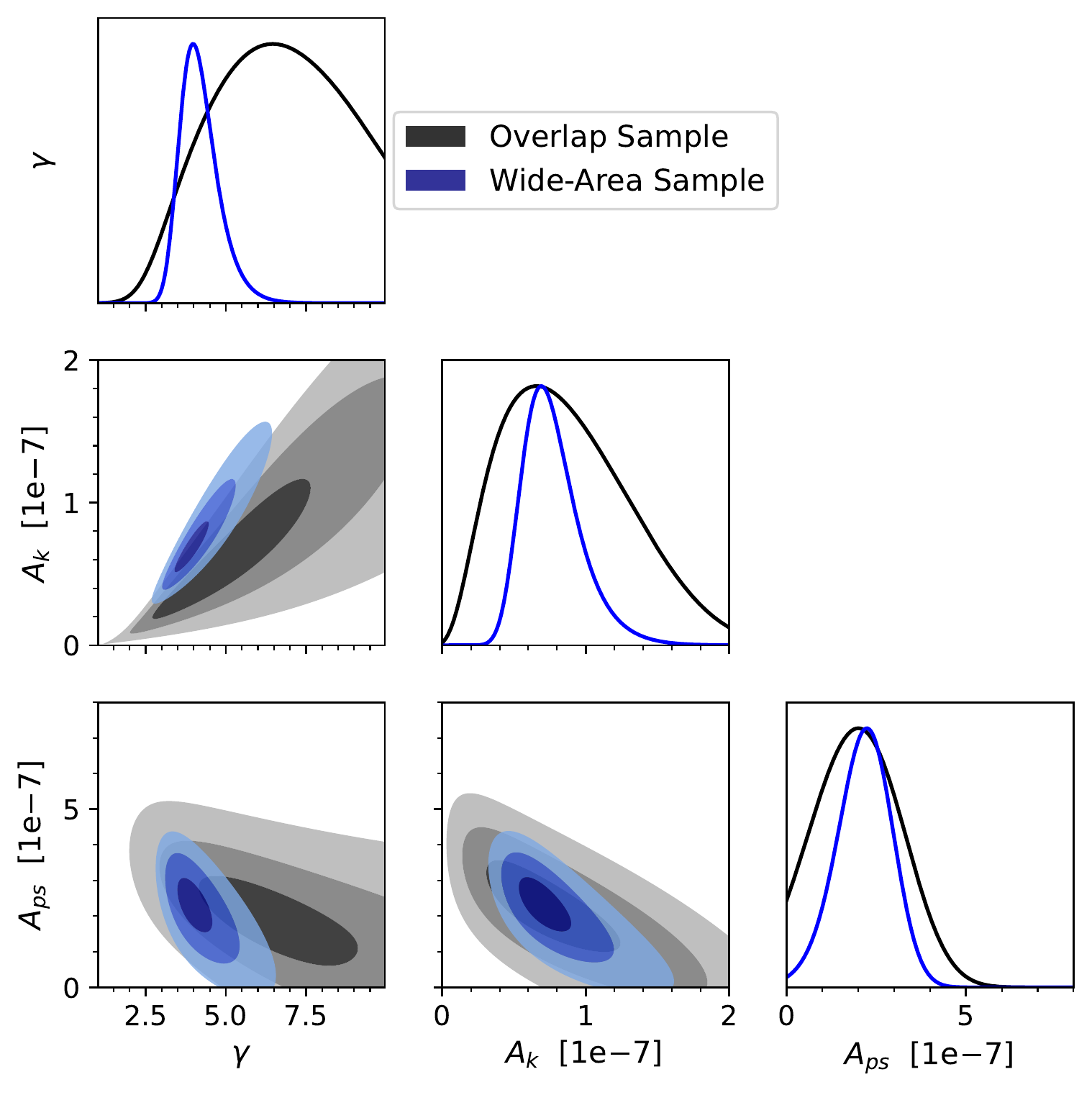}
	
	\caption{Posterior corner plot of the tSZ profile fit for our Overlap (\textit{blue}) and Wide-Area (\textit{black}) galaxy catalogs.  From dark to light, the three shaded contour regions correspond to the $1\sigma$, $2\sigma$, and $3\sigma$ levels.}
	\label{fig:tSZ_profile_fit_corner}
\end{figure*}

Similarly, Fig.~\ref{fig:tSZ_profile_fit_corner} shows the marginalized posterior probability distributions for our tSZ profile fits.  These correspond to the results shown in Fig.~\ref{fig:tSZ_profile} and Table~\ref{tab:tSZ_profile_fit}.

\bibliographystyle{apj}

\bibliography{spt_act_sz_stack}

\begin{thebibliography}{}
\expandafter\ifx\csname natexlab\endcsname\relax\def\natexlab#1{#1}\fi

\bibitem[{Abbott {et~al.}(2018)Abbott, Abdalla, Allam, Amara, Annis, Asorey,
  Avila, Ballester, Banerji, Barkhouse, \& et~al.}]{Abbott2018}
Abbott, T. M.~C., Abdalla, F.~B., Allam, S., {et~al.} 2018, \apjs, 239, 18

\bibitem[{Addison {et~al.}(2013)Addison, Dunkley, \& Bond}]{Addison2013}
Addison, G.~E., Dunkley, J., \& Bond, J.~R. 2013, \mnras, 436, 1896

\bibitem[{Amodeo {et~al.}(2021)Amodeo, Battaglia, Schaan, Ferraro, Moser,
  Aiola, Austermann, Beall, Bean, Becker, Bond, Calabrese, Calafut, Choi,
  Denison, Devlin, Duff, Duivenvoorden, Dunkley, D\"unner, Gallardo, Hall, Han,
  Hill, Hilton, Hilton, Hlo\ifmmode~\check{z}\else \v{z}\fi{}ek, Hubmayr,
  Huffenberger, Hughes, Koopman, MacInnis, McMahon, Madhavacheril, Moodley,
  Mroczkowski, Naess, Nati, Newburgh, Niemack, Page, Partridge, Schillaci,
  Sehgal, Sif\'on, Spergel, Staggs, Storer, Ullom, Vale, van Engelen,
  Van~Lanen, Vavagiakis, Wollack, \& Xu}]{Amodeo2021}
Amodeo, S., Battaglia, N., Schaan, E., {et~al.} 2021, \prd, 103, 063514

\bibitem[{Amvrosiadis {et~al.}(2018)Amvrosiadis, Valiante, Gonzalez-Nuevo,
  Maddox, Negrello, Eales, Dunne, Wang, van Kampen, De Zotti, Smith,
  Andreani, Greenslade, Tai-An, \& Michałowski}]{Amvrosiadis2018}
Amvrosiadis, A., Valiante, E., Gonzalez-Nuevo, J., {et~al.} 2018, \mnras, 483,
  4649

\bibitem[{Battaglia {et~al.}(2010)Battaglia, Bond, Pfrommer, Sievers, \&
  Sijacki}]{Battaglia2010}
Battaglia, N., Bond, J.~R., Pfrommer, C., Sievers, J.~L., \& Sijacki, D. 2010,
  \apj, 725, 91

\bibitem[{{Behroozi} {et~al.}(2019){Behroozi}, {Wechsler}, {Hearin}, \&
  {Conroy}}]{Behroozi2019}
{Behroozi}, P., {Wechsler}, R.~H., {Hearin}, A.~P., \& {Conroy}, C. 2019,
  \mnras, 488, 3143

\bibitem[{{Behroozi} {et~al.}(2010){Behroozi}, {Conroy}, \&
  {Wechsler}}]{Behroozi2010}
{Behroozi}, P.~S., {Conroy}, C., \& {Wechsler}, R.~H. 2010, \apj, 717, 379

\bibitem[{{Benson} {et~al.}(2014){Benson}, {Ade}, {Ahmed}, {Allen}, {Arnold},
  {Austermann}, {Bender}, {Bleem}, {Carlstrom}, {Chang}, {Cho}, {Cliche},
  {Crawford}, {Cukierman}, {de Haan}, {Dobbs}, {Dutcher}, {Everett}, {Gilbert},
  {Halverson}, {Hanson}, {Harrington}, {Hattori}, {Henning}, {Hilton},
  {Holder}, {Holzapfel}, {Irwin}, {Keisler}, {Knox}, {Kubik}, {Kuo}, {Lee},
  {Leitch}, {Li}, {McDonald}, {Meyer}, {Montgomery}, {Myers}, {Natoli},
  {Nguyen}, {Novosad}, {Padin}, {Pan}, {Pearson}, {Reichardt}, {Ruhl},
  {Saliwanchik}, {Simard}, {Smecher}, {Sayre}, {Shirokoff}, {Stark}, {Story},
  {Suzuki}, {Thompson}, {Tucker}, {Vanderlinde}, {Vieira}, {Vikhlinin}, {Wang},
  {Yefremenko}, \& {Yoon}}]{SPT3G-2014}
{Benson}, B.~A., {Ade}, P.~A.~R., {Ahmed}, Z., {et~al.} 2014, in \procspie,
  Vol. 9153, Millimeter, Submillimeter, and Far-Infrared Detectors and
  Instrumentation for Astronomy VII, 91531P

\bibitem[{{Berta} {et~al.}(2016){Berta}, {Lutz}, {Genzel},
  {F{\"o}rster-Schreiber}, \& {Tacconi}}]{Berta2016}
{Berta}, S., {Lutz}, D., {Genzel}, R., {F{\"o}rster-Schreiber}, N.~M., \&
  {Tacconi}, L.~J. 2016, \aap, 587, A73

\bibitem[{{Bocquet} {et~al.}(2019){Bocquet}, {Dietrich}, {Schrabback}, {Bleem},
  {Klein}, {Allen}, {Applegate}, {Ashby}, {Bautz}, {Bayliss}, {Benson},
  {Brodwin}, {Bulbul}, {Canning}, {Capasso}, {Carlstrom}, {Chang}, {Chiu},
  {Cho}, {Clocchiatti}, {Crawford}, {Crites}, {de Haan}, {Desai}, {Dobbs},
  {Foley}, {Forman}, {Garmire}, {George}, {Gladders}, {Gonzalez}, {Grandis},
  {Gupta}, {Halverson}, {Hlavacek-Larrondo}, {Hoekstra}, {Holder}, {Holzapfel},
  {Hou}, {Hrubes}, {Huang}, {Jones}, {Khullar}, {Knox}, {Kraft}, {Lee}, {von
  der Linden}, {Luong-Van}, {Mantz}, {Marrone}, {McDonald}, {McMahon}, {Meyer},
  {Mocanu}, {Mohr}, {Morris}, {Padin}, {Patil}, {Pryke}, {Rapetti},
  {Reichardt}, {Rest}, {Ruhl}, {Saliwanchik}, {Saro}, {Sayre}, {Schaffer},
  {Shirokoff}, {Stalder}, {Stanford}, {Staniszewski}, {Stark}, {Story},
  {Strazzullo}, {Stubbs}, {Vanderlinde}, {Vieira}, {Vikhlinin}, {Williamson},
  \& {Zenteno}}]{Bocquet2019}
{Bocquet}, S., {Dietrich}, J.~P., {Schrabback}, T., {et~al.} 2019, \apj, 878,
  55

\bibitem[{{Bower} {et~al.}(2006){Bower}, {Benson}, {Malbon}, {Helly}, {Frenk},
  {Baugh}, {Cole}, \& {Lacey}}]{Bower2006}
{Bower}, R.~G., {Benson}, A.~J., {Malbon}, R., {et~al.} 2006, \mnras, 370, 645

\bibitem[{{Brammer} {et~al.}(2008){Brammer}, {van Dokkum}, \& {Coppi}}]{eazy08}
{Brammer}, G.~B., {van Dokkum}, P.~G., \& {Coppi}, P. 2008, \apj, 686, 1503

\bibitem[{{Bregman} {et~al.}(2022){Bregman}, {Hodges-Kluck}, {Qu}, {Pratt},
  {Li}, \& {Yun}}]{Bregman2022}
{Bregman}, J.~N., {Hodges-Kluck}, E., {Qu}, Z., {et~al.} 2022, \apj, 928, 14

\bibitem[{{Brownson} {et~al.}(2019){Brownson}, {Maiolino}, {Tazzari},
  {Carniani}, \& {Henden}}]{Brownson2019}
{Brownson}, S., {Maiolino}, R., {Tazzari}, M., {Carniani}, S., \& {Henden}, N.
  2019, \mnras, 490, 5134

\bibitem[{{Bruzual} \& {Charlot}(2003)}]{bc03}
{Bruzual}, G., \& {Charlot}, S. 2003, \mnras, 344, 1000

\bibitem[{{Bryan} {et~al.}(2018){Bryan}, {Austermann}, {Ferrusca}, {Mauskopf},
  {McMahon}, {Monta{\~n}a}, {Simon}, {Novak}, {S{\'a}nchez-Arg{\"u}elles}, \&
  {Wilson}}]{TolTEC2018}
{Bryan}, S., {Austermann}, J., {Ferrusca}, D., {et~al.} 2018, in \procspie,
  Vol. 10708, Millimeter, Submillimeter, and Far-Infrared Detectors and
  Instrumentation for Astronomy IX, 107080J

\bibitem[{{Calafut} {et~al.}(2021){Calafut}, {Gallardo}, {Vavagiakis},
  {Amodeo}, {Aiola}, {Austermann}, {Battaglia}, {Battistelli}, {Beall}, {Bean},
  {Bond}, {Calabrese}, {Choi}, {Cothard}, {Devlin}, {Duell}, {Duff},
  {Duivenvoorden}, {Dunkley}, {Dunner}, {Ferraro}, {Guan}, {Hill}, {Hilton},
  {Hilton}, {Hlo{\v{z}}ek}, {Huber}, {Hubmayr}, {Huffenberger}, {Hughes},
  {Koopman}, {Kosowsky}, {Li}, {Lokken}, {Madhavacheril}, {McMahon}, {Moodley},
  {Naess}, {Nati}, {Newburgh}, {Niemack}, {Page}, {Partridge}, {Schaan},
  {Schillaci}, {Sif{\'o}n}, {Spergel}, {Staggs}, {Ullom}, {Vale}, {Van
  Engelen}, {Van Lanen}, {Wollack}, \& {Xu}}]{Calafut2021}
{Calafut}, V., {Gallardo}, P.~A., {Vavagiakis}, E.~M., {et~al.} 2021, \prd,
  104, 043502

\bibitem[{Calura {et~al.}(2016)Calura, Pozzi, Cresci, Santini, Gruppioni,
  Pozzetti, Gilli, Matteucci, \& Maiolino}]{Calura2016}
Calura, F., Pozzi, F., Cresci, G., {et~al.} 2016, \mnras, 465, 54

\bibitem[{{Casey}(2012)}]{Casey2012}
{Casey}, C.~M. 2012, \mnras, 425, 3094

\bibitem[{{Chamberlain} {et~al.}(2015){Chamberlain}, {Arav}, \&
  {Benn}}]{Chamberlain2015}
{Chamberlain}, C., {Arav}, N., \& {Benn}, C. 2015, \mnras, 450, 1085

\bibitem[{{Chartas} {et~al.}(2007){Chartas}, {Brandt}, {Gallagher}, \&
  {Proga}}]{Chartas2007}
{Chartas}, G., {Brandt}, W.~N., {Gallagher}, S.~C., \& {Proga}, D. 2007, \aj,
  133, 1849

\bibitem[{{Chatterjee} {et~al.}(2010){Chatterjee}, {Ho}, {Newman}, \&
  {Kosowsky}}]{Chatterjee2010}
{Chatterjee}, S., {Ho}, S., {Newman}, J.~A., \& {Kosowsky}, A. 2010, \apj, 720,
  299

\bibitem[{Chown {et~al.}(2018)Chown, Omori, Aylor, Benson, Bleem, Carlstrom,
  Chang, Cho, Crawford, Crites, de~Haan, Dobbs, Everett, George, Henning,
  Halverson, Harrington, Holder, Holzapfel, Hou, Hrubes, Knox, Lee, Luong-Van,
  Marrone, McMahon, Meyer, Millea, Mocanu, Mohr, Natoli, Padin, Pryke,
  Reichardt, Ruhl, Sayre, Schaffer, Shirokoff, Simard, Staniszewski, Stark,
  Story, Vanderlinde, Vieira, Williamson, \& and}]{Chown2018}
Chown, R., Omori, Y., Aylor, K., {et~al.} 2018, \apjs, 239, 10

\bibitem[{{Churazov} {et~al.}(2001){Churazov}, {Br{\"u}ggen}, {Kaiser},
  {B{\"o}hringer}, \& {Forman}}]{Churazov2001}
{Churazov}, E., {Br{\"u}ggen}, M., {Kaiser}, C.~R., {B{\"o}hringer}, H., \&
  {Forman}, W. 2001, \apj, 554, 261

\bibitem[{Coil {et~al.}(2017)Coil, Mendez, Eisenstein, \& Moustakas}]{Coil2017}
Coil, A.~L., Mendez, A.~J., Eisenstein, D.~J., \& Moustakas, J. 2017, \apj,
  838, 87

\bibitem[{{Coleman} {et~al.}(1980){Coleman}, {Wu}, \& {Weedman}}]{cww80}
{Coleman}, G.~D., {Wu}, C.~C., \& {Weedman}, D.~W. 1980, \apjs, 43, 393

\bibitem[{{Costa} {et~al.}(2014){Costa}, {Sijacki}, \& {Haehnelt}}]{Costa2014}
{Costa}, T., {Sijacki}, D., \& {Haehnelt}, M.~G. 2014, \mnras, 444, 2355

\bibitem[{{Cowie} {et~al.}(1996){Cowie}, {Songaila}, {Hu}, \&
  {Cohen}}]{Cowie1996}
{Cowie}, L.~L., {Songaila}, A., {Hu}, E.~M., \& {Cohen}, J.~G. 1996, \aj, 112,
  839

\bibitem[{{Crichton} {et~al.}(2016){Crichton}, {Gralla}, {Hall}, {Marriage},
  {Zakamska}, {Battaglia}, {Bond}, {Devlin}, {Hill}, {Hilton}, {Hincks},
  {Huffenberger}, {Hughes}, {Kosowsky}, {Moodley}, {Niemack}, {Page},
  {Partridge}, {Sievers}, {Sif{\'o}n}, {Staggs}, {Viero}, \&
  {Wollack}}]{Crichton2016}
{Crichton}, D., {Gralla}, M.~B., {Hall}, K., {et~al.} 2016, \mnras, 458, 1478

\bibitem[{{Croton} {et~al.}(2006){Croton}, {Springel}, {White}, {De Lucia},
  {Frenk}, {Gao}, {Jenkins}, {Kauffmann}, {Navarro}, \& {Yoshida}}]{Croton2006}
{Croton}, D.~J., {Springel}, V., {White}, S.~D.~M., {et~al.} 2006, \mnras, 365,
  11

\bibitem[{{de Kool} {et~al.}(2001){de Kool}, {Arav}, {Becker}, {Gregg},
  {White}, {Laurent-Muehleisen}, {Price}, \& {Korista}}]{deKool2001}
{de Kool}, M., {Arav}, N., {Becker}, R.~H., {et~al.} 2001, \apj, 548, 609

\bibitem[{{Donevski} {et~al.}(2020){Donevski}, {Lapi}, {Ma{\l}ek}, {Liu},
  {G{\'o}mez-Guijarro}, {Dav{\'e}}, {Kraljic}, {Pantoni}, {Man}, {Fujimoto},
  {Feltre}, {Pearson}, {Li}, \& {Narayanan}}]{Donevski2020}
{Donevski}, D., {Lapi}, A., {Ma{\l}ek}, K., {et~al.} 2020, \aap, 644, A144

\bibitem[{{Draine}(2003)}]{Draine2003}
{Draine}, B.~T. 2003, \araa, 41, 241

\bibitem[{{Draine}(2011)}]{Draine2011}
---. 2011, {Physics of the Interstellar and Intergalactic Medium} (Princeton,
  NJ: Princeton Univ. Press)

\bibitem[{{Drory} \& {Alvarez}(2008)}]{Drory2008}
{Drory}, N., \& {Alvarez}, M. 2008, \apj, 680, 41

\bibitem[{{Dunn} {et~al.}(2010){Dunn}, {Bautista}, {Arav}, {Moe}, {Korista},
  {Costantini}, {Benn}, {Ellison}, \& {Edmonds}}]{Dunn2010}
{Dunn}, J.~P., {Bautista}, M., {Arav}, N., {et~al.} 2010, \apj, 709, 611

\bibitem[{{Dunne} {et~al.}(2003){Dunne}, {Eales}, {Ivison}, {Morgan}, \&
  {Edmunds}}]{Dunne2003}
{Dunne}, L., {Eales}, S., {Ivison}, R., {Morgan}, H., \& {Edmunds}, M. 2003,
  \nat, 424, 285

\bibitem[{{Dunne} \& {Eales}(2001)}]{Dunne2001-SCUBA-II}
{Dunne}, L., \& {Eales}, S.~A. 2001, \mnras, 327, 697

\bibitem[{{Eftekharzadeh} {et~al.}(2015){Eftekharzadeh}, {Myers}, {White},
  {Weinberg}, {Schneider}, {Shen}, {Font-Ribera}, {Ross}, {Paris}, \&
  {Streblyanska}}]{Eftekharzadeh2015}
{Eftekharzadeh}, S., {Myers}, A.~D., {White}, M., {et~al.} 2015, \mnras, 453,
  2779

\bibitem[{{Fabian}(2012)}]{Fabian2012}
{Fabian}, A.~C. 2012, \araa, 50, 455

\bibitem[{{Ferrarese}(2002)}]{Ferrarese2002}
{Ferrarese}, L. 2002, \apj, 578, 90

\bibitem[{{Feruglio} {et~al.}(2010){Feruglio}, {Maiolino}, {Piconcelli},
  {Menci}, {Aussel}, {Lamastra}, \& {Fiore}}]{Feruglio2010}
{Feruglio}, C., {Maiolino}, R., {Piconcelli}, E., {et~al.} 2010, \aap, 518,
  L155

\bibitem[{Gobat {et~al.}(2018)Gobat, Daddi, Magdis, Bournaud, Sargent, Martig,
  Jin, Finoguenov, B{\'e}thermin, Hwang, Renzini, Wilson, Aretxaga, Yun,
  Strazzullo, \& Valentino}]{Gobat2018}
Gobat, R., Daddi, E., Magdis, G., {et~al.} 2018, \natas, 2, 239

\bibitem[{{Gralla} {et~al.}(2014){Gralla}, {Crichton}, {Marriage}, {Mo},
  {Aguirre}, {Addison}, {Asboth}, {Battaglia}, {Bock}, {Bond}, {Devlin},
  {D{\"u}nner}, {Hajian}, {Halpern}, {Hilton}, {Hincks}, {Hlozek},
  {Huffenberger}, {Hughes}, {Ivison}, {Kosowsky}, {Lin}, {Marsden},
  {Menanteau}, {Moodley}, {Morales}, {Niemack}, {Oliver}, {Page}, {Partridge},
  {Reese}, {Rojas}, {Sehgal}, {Sievers}, {Sif{\'o}n}, {Spergel}, {Staggs},
  {Switzer}, {Viero}, {Wollack}, \& {Zemcov}}]{Gralla2014}
{Gralla}, M.~B., {Crichton}, D., {Marriage}, T.~A., {et~al.} 2014, \mnras, 445,
  460

\bibitem[{{Granato} {et~al.}(2004){Granato}, {De Zotti}, {Silva}, {Bressan}, \&
  {Danese}}]{Granato2004}
{Granato}, G.~L., {De Zotti}, G., {Silva}, L., {Bressan}, A., \& {Danese}, L.
  2004, \apj, 600, 580

\bibitem[{{Greco} {et~al.}(2015){Greco}, {Hill}, {Spergel}, \&
  {Battaglia}}]{Greco2015}
{Greco}, J.~P., {Hill}, J.~C., {Spergel}, D.~N., \& {Battaglia}, N. 2015, \apj,
  808, 151

\bibitem[{{Greene} {et~al.}(2014){Greene}, {Pooley}, {Zakamska}, {Comerford},
  \& {Sun}}]{Greene2014}
{Greene}, J.~E., {Pooley}, D., {Zakamska}, N.~L., {Comerford}, J.~M., \& {Sun},
  A.-L. 2014, \apj, 788, 54

\bibitem[{{Hall} {et~al.}(2019){Hall}, {Zakamska}, {Addison}, {Battaglia},
  {Crichton}, {Devlin}, {Dunkley}, {Gralla}, {Hill}, {Hilton}, {Hubmayr},
  {Hughes}, {Huffenberger}, {Kosowsky}, {Marriage}, {Maurin}, {Moodley},
  {Niemack}, {Page}, {Partridge}, {D{\"u}nner Planella}, {Schillaci},
  {Sif{\'o}n}, {Staggs}, {Wollack}, \& {Xu}}]{Hall2019}
{Hall}, K.~R., {Zakamska}, N.~L., {Addison}, G.~E., {et~al.} 2019, \mnras, 490,
  2315

\bibitem[{{Hand} {et~al.}(2011){Hand}, {Appel}, {Battaglia}, {Bond}, {Das},
  {Devlin}, {Dunkley}, {D{\"u}nner}, {Essinger-Hileman}, {Fowler}, {Hajian},
  {Halpern}, {Hasselfield}, {Hilton}, {Hincks}, {Hlozek}, {Hughes}, {Irwin},
  {Klein}, {Kosowsky}, {Lin}, {Marriage}, {Marsden}, {McLaren}, {Menanteau},
  {Moodley}, {Niemack}, {Nolta}, {Page}, {Parker}, {Partridge}, {Plimpton},
  {Reese}, {Rojas}, {Sehgal}, {Sherwin}, {Sievers}, {Spergel}, {Staggs},
  {Swetz}, {Switzer}, {Thornton}, {Trac}, {Visnjic}, \& {Wollack}}]{Hand2011}
{Hand}, N., {Appel}, J.~W., {Battaglia}, N., {et~al.} 2011, \apj, 736, 39

\bibitem[{{Harrison} {et~al.}(2014){Harrison}, {Alexander}, {Mullaney}, \&
  {Swinbank}}]{Harrison2014}
{Harrison}, C.~M., {Alexander}, D.~M., {Mullaney}, J.~R., \& {Swinbank}, A.~M.
  2014, \mnras, 441, 3306

\bibitem[{Hill {et~al.}(2018)Hill, Baxter, Lidz, Greco, \& Jain}]{Hill2018}
Hill, J.~C., Baxter, E.~J., Lidz, A., Greco, J.~P., \& Jain, B. 2018, \prd, 97,
  083501

\bibitem[{{Hilton} {et~al.}(2018){Hilton}, {Hasselfield}, {Sif{\'o}n},
  {Battaglia}, {Aiola}, {Bharadwaj}, {Bond}, {Choi}, {Crichton}, {Datta},
  {Devlin}, {Dunkley}, {D{\"u}nner}, {Gallardo}, {Gralla}, {Hincks}, {Ho},
  {Hubmayr}, {Huffenberger}, {Hughes}, {Koopman}, {Kosowsky}, {Louis},
  {Madhavacheril}, {Marriage}, {Maurin}, {McMahon}, {Miyatake}, {Moodley},
  {N{\ae}ss}, {Nati}, {Newburgh}, {Niemack}, {Oguri}, {Page}, {Partridge},
  {Schmitt}, {Sievers}, {Spergel}, {Staggs}, {Trac}, {van Engelen},
  {Vavagiakis}, \& {Wollack}}]{Hilton2018}
{Hilton}, M., {Hasselfield}, M., {Sif{\'o}n}, C., {et~al.} 2018, \apjs, 235, 20

\bibitem[{Hojjati {et~al.}(2017)Hojjati, Tröster, Harnois-Déraps, McCarthy,
  van Waerbeke, Choi, Erben, Heymans, Hildebrandt, Hinshaw, Ma, Miller, Viola,
  \& Tanimura}]{Hojjati2016}
Hojjati, A., Tröster, T., Harnois-Déraps, J., {et~al.} 2017, \mnras, 471,
  1565

\bibitem[{{King}(1962)}]{King1962}
{King}, I. 1962, \aj, 67, 471

\bibitem[{{Kinney} {et~al.}(1996){Kinney}, {Calzetti}, {Bohlin}, {McQuade},
  {Storchi-Bergmann}, \& {Schmitt}}]{kinney96}
{Kinney}, A.~L., {Calzetti}, D., {Bohlin}, R.~C., {et~al.} 1996, \apj, 467, 38

\bibitem[{{Kravtsov} {et~al.}(2018){Kravtsov}, {Vikhlinin}, \&
  {Meshcheryakov}}]{Kravtsov2018}
{Kravtsov}, A.~V., {Vikhlinin}, A.~A., \& {Meshcheryakov}, A.~V. 2018, \astl,
  44, 8

\bibitem[{{Lacy} {et~al.}(2019){Lacy}, {Mason}, {Sarazin}, {Chatterjee},
  {Nyland}, {Kimball}, {Rocha}, {Rowe}, \& {Surace}}]{Lacy2019}
{Lacy}, M., {Mason}, B., {Sarazin}, C., {et~al.} 2019, \mnras, 483, L22

\bibitem[{{Lansbury} {et~al.}(2018){Lansbury}, {Jarvis}, {Harrison},
  {Alexander}, {Del Moro}, {Edge}, {Mullaney}, \& {Thomson}}]{Lansbury2018}
{Lansbury}, G.~B., {Jarvis}, M.~E., {Harrison}, C.~M., {et~al.} 2018, \apjl,
  856, L1

\bibitem[{{Li} \& {Draine}(2001)}]{Draine2001}
{Li}, A., \& {Draine}, B.~T. 2001, \apj, 554, 778

\bibitem[{{Lokken} {et~al.}(2022){Lokken}, {Hlo{\v{z}}ek}, {Engelen},
  {Madhavacheril}, {Baxter}, {DeRose}, {Doux}, {Pandey}, {Rykoff}, {Stein},
  {To}, {Abbott}, {Adhikari}, {Aguena}, {Allam}, {Andrade-Oliveira}, {Annis},
  {Battaglia}, {Bernstein}, {Bertin}, {Bond}, {Brooks}, {Calabrese}, {Rosell},
  {Kind}, {Carretero}, {Cawthon}, {Choi}, {Costanzi}, {Crocce}, {Costa},
  {Pereira}, {Vicente}, {Desai}, {Dietrich}, {Doel}, {Dunkley}, {Everett},
  {Evrard}, {Ferraro}, {Flaugher}, {Fosalba}, {Frieman}, {Gallardo},
  {Garc{\'\i}a-Bellido}, {Gaztanaga}, {Gerdes}, {Giannantonio}, {Gruen},
  {Gruendl}, {Gschwend}, {Gutierrez}, {Hill}, {Hilton}, {Hincks}, {Hinton},
  {Hollowood}, {Honscheid}, {Hoyle}, {Huang}, {Hughes}, {Huterer}, {Jain},
  {James}, {Jeltema}, {Kuehn}, {Lima}, {Maia}, {Marshall}, {McMahon},
  {Melchior}, {Menanteau}, {Miquel}, {Mohr}, {Moodley}, {Morgan}, {Nati},
  {Page}, {Ogando}, {Palmese}, {Paz-Chinch{\'o}n}, {Malag{\'o}n}, {Pieres},
  {Romer}, {Rozo}, {Sanchez}, {Scarpine}, {Schillaci}, {Schubnell}, {Serrano},
  {Sevilla-Noarbe}, {Sheldon}, {Shin}, {Sif{\'o}n}, {Smith}, {Soares-Santos},
  {Suchyta}, {Swanson}, {Tarle}, {Thomas}, {Tucker}, {Varga}, {Weller},
  {Wechsler}, {Wilkinson}, {Wollack}, \& {Xu}}]{Lokken2022}
{Lokken}, M., {Hlo{\v{z}}ek}, R., {Engelen}, A.~v., {et~al.} 2022, \apj, 933,
  134

\bibitem[{{Lu} {et~al.}(2015){Lu}, {Mo}, {Lu}, {Katz}, {Weinberg}, {van den
  Bosch}, \& {Yang}}]{Lu2015}
{Lu}, Z., {Mo}, H.~J., {Lu}, Y., {et~al.} 2015, \mnras, 450, 1604

\bibitem[{{Magdis} {et~al.}(2021){Magdis}, {Gobat}, {Valentino}, {Daddi},
  {Zanella}, {Kokorev}, {Toft}, {Jin}, \& {Whitaker}}]{Magdis2021}
{Magdis}, G.~E., {Gobat}, R., {Valentino}, F., {et~al.} 2021, \aap, 647, A33

\bibitem[{{Mainzer} {et~al.}(2011){Mainzer}, {Bauer}, {Grav}, {Masiero},
  {Cutri}, {Dailey}, {Eisenhardt}, {McMillan}, {Wright}, {Walker}, {Jedicke},
  {Spahr}, {Tholen}, {Alles}, {Beck}, {Brand enburg}, {Conrow}, {Evans},
  {Fowler}, {Jarrett}, {Marsh}, {Masci}, {McCallon}, {Wheelock}, {Wittman},
  {Wyatt}, {DeBaun}, {Elliott}, {Elsbury}, {Gautier}, {Gomillion}, {Leisawitz},
  {Maleszewski}, {Micheli}, \& {Wilkins}}]{neowise11}
{Mainzer}, A., {Bauer}, J., {Grav}, T., {et~al.} 2011, \apj, 731, 53

\bibitem[{{Mallaby-Kay} {et~al.}(2021){Mallaby-Kay}, {Atkins}, {Aiola},
  {Amodeo}, {Austermann}, {Beall}, {Becker}, {Bond}, {Calabrese}, {Chesmore},
  {Choi}, {Crowley}, {Darwish}, {Denison}, {Devlin}, {Duff}, {Duivenvoorden},
  {Dunkley}, {Ferraro}, {Fichman}, {Gallardo}, {Golec}, {Guan}, {Han},
  {Hasselfield}, {Hill}, {Hilton}, {Hilton}, {Hlo{\v{z}}ek}, {Hubmayr},
  {Huffenberger}, {Hughes}, {Koopman}, {Louis}, {MacInnis}, {Madhavacheril},
  {McMahon}, {Moodley}, {Naess}, {Namikawa}, {Nati}, {Newburgh}, {Nibarger},
  {Niemack}, {Page}, {Salatino}, {Schaan}, {Schillaci}, {Sehgal}, {Sherwin},
  {Sif{\'o}n}, {Simon}, {Staggs}, {Storer}, {Ullom}, {Van Engelen}, {Van
  Lanen}, {Vale}, {Wollack}, \& {Xu}}]{MallabyKay2021}
{Mallaby-Kay}, M., {Atkins}, Z., {Aiola}, S., {et~al.} 2021, \apjs, 255, 11

\bibitem[{{Marconi} \& {Hunt}(2003)}]{Marconi2003}
{Marconi}, A., \& {Hunt}, L.~K. 2003, \apjl, 589, L21

\bibitem[{{McNamara} {et~al.}(2016){McNamara}, {Russell}, {Nulsen}, {Hogan},
  {Fabian}, {Pulido}, \& {Edge}}]{McNamara2016}
{McNamara}, B.~R., {Russell}, H.~R., {Nulsen}, P.~E.~J., {et~al.} 2016, \apj,
  830, 79

\bibitem[{{McNamara} {et~al.}(2000){McNamara}, {Wise}, {Nulsen}, {David},
  {Sarazin}, {Bautz}, {Markevitch}, {Vikhlinin}, {Forman}, {Jones}, \&
  {Harris}}]{McNamara2000}
{McNamara}, B.~R., {Wise}, M., {Nulsen}, P.~E.~J., {et~al.} 2000, \apjl, 534,
  L135

\bibitem[{{Meinke} {et~al.}(2021){Meinke}, {B{\"o}ckmann}, {Cohen}, {Mauskopf},
  {Scannapieco}, {Sarmento}, {Lunde}, \& {Cottle}}]{Meinke2021}
{Meinke}, J., {B{\"o}ckmann}, K., {Cohen}, S., {et~al.} 2021, \apj, 913, 88

\bibitem[{Miller {et~al.}(2020)Miller, Arav, Xu, \& Kriss}]{Miller2020}
Miller, T.~R., Arav, N., Xu, X., \& Kriss, G.~A. 2020, \mnras, 499, 1522

\bibitem[{{Moster} {et~al.}(2013){Moster}, {Naab}, \& {White}}]{Moster2013}
{Moster}, B.~P., {Naab}, T., \& {White}, S. D.~M. 2013, \mnras, 428, 3121

\bibitem[{{Moster} {et~al.}(2018){Moster}, {Naab}, \& {White}}]{Moster2018}
---. 2018, \mnras, 477, 1822

\bibitem[{{Mroczkowski} {et~al.}(2019){Mroczkowski}, {Nagai}, {Basu}, {Chluba},
  {Sayers}, {Adam}, {Churazov}, {Crites}, {Di Mascolo}, {Eckert},
  {Macias-Perez}, {Mayet}, {Perotto}, {Pointecouteau}, {Romero}, {Ruppin},
  {Scannapieco}, \& {ZuHone}}]{Mroczkowski2019}
{Mroczkowski}, T., {Nagai}, D., {Basu}, K., {et~al.} 2019, \ssr, 215, 17

\bibitem[{{Muzzin} {et~al.}(2013){Muzzin}, {Marchesini}, {Stefanon}, {Franx},
  {McCracken}, {Milvang-Jensen}, {Dunlop}, {Fynbo}, {Brammer}, {Labb{\'e}}, \&
  {van Dokkum}}]{Muzzin2013}
{Muzzin}, A., {Marchesini}, D., {Stefanon}, M., {et~al.} 2013, \apj, 777, 18

\bibitem[{Naess {et~al.}(2020)Naess, Aiola, Austermann, Battaglia, Beall,
  Becker, Bond, Calabrese, Choi, Cothard, Crowley, Darwish, Datta, Denison,
  Devlin, Duell, Duff, Duivenvoorden, Dunkley, Dünner, Fox, Gallardo, Halpern,
  Han, Hasselfield, Hill, Hilton, Hilton, Hincks, Hlo{\v{z}}ek, Ho, Hubmayr,
  Huffenberger, Hughes, Kosowsky, Louis, Madhavacheril, McMahon, Moodley, Nati,
  Nibarger, Niemack, Page, Partridge, Salatino, Schaan, Schillaci, Schmitt,
  Sherwin, Sehgal, Sif{\'{o}}n, Spergel, Staggs, Stevens, Storer, Ullom, Vale,
  Engelen, Lanen, Vavagiakis, Wollack, \& Xu}]{Naess2020}
Naess, S., Aiola, S., Austermann, J.~E., {et~al.} 2020, \jcap, 2020, 046

\bibitem[{{Oke} \& {Gunn}(1983)}]{Oke1983}
{Oke}, J.~B., \& {Gunn}, J.~E. 1983, \apj, 266, 713

\bibitem[{{Pillepich} {et~al.}(2018){Pillepich}, {Nelson}, {Hernquist},
  {Springel}, {Pakmor}, {Torrey}, {Weinberger}, {Genel}, {Naiman}, {Marinacci},
  \& {Vogelsberger}}]{Pillepich2018}
{Pillepich}, A., {Nelson}, D., {Hernquist}, L., {et~al.} 2018, \mnras, 475, 648

\bibitem[{{Planck Collaboration} {et~al.}(2011){Planck Collaboration}, {Ade},
  {Aghanim}, {Arnaud}, {Ashdown}, {Aumont}, {Baccigalupi}, {Balbi}, {Banday},
  {Barreiro}, {Bartelmann}, {Bartlett}, {Battaner}, {Battye}, {Benabed},
  {Beno{\^\i}t}, {Bernard}, {Bersanelli}, {Bhatia}, {Bock}, {Bonaldi}, {Bond},
  {Borrill}, {Bouchet}, {Brown}, {Bucher}, {Burigana}, {Cabella}, {Cantalupo},
  {Cardoso}, {Carvalho}, {Catalano}, {Cay{\'o}n}, {Challinor}, {Chamballu},
  {Chary}, {Chiang}, {Chiang}, {Chon}, {Christensen}, {Churazov}, {Clements},
  {Colafrancesco}, {Colombi}, {Couchot}, {Coulais}, {Crill}, {Cuttaia}, {da
  Silva}, {Dahle}, {Danese}, {Davis}, {de Bernardis}, {de Gasperis}, {de Rosa},
  {de Zotti}, {Delabrouille}, {Delouis}, {D{\'e}sert}, {Dickinson}, {Diego},
  {Dolag}, {Dole}, {Donzelli}, {Dor{\'e}}, {D{\"o}rl}, {Douspis}, {Dupac},
  {Efstathiou}, {Eisenhardt}, {En{\ss}lin}, {Feroz}, {Finelli}, {Flores-Cacho},
  {Forni}, {Fosalba}, {Frailis}, {Franceschi}, {Fromenteau}, {Galeotta},
  {Ganga}, {G{\'e}nova-Santos}, {Giard}, {Giardino}, {Giraud-H{\'e}raud},
  {Gonz{\'a}lez-Nuevo}, {Gonz{\'a}lez-Riestra}, {G{\'o}rski}, {Grainge},
  {Gratton}, {Gregorio}, {Gruppuso}, {Harrison}, {Hein{\"a}m{\"a}ki},
  {Henrot-Versill{\'e}}, {Hern{\'a}ndez-Monteagudo}, {Herranz}, {Hildebrand t},
  {Hivon}, {Hobson}, {Holmes}, {Hovest}, {Hoyland}, {Huffenberger}, {Hurier},
  {Hurley-Walker}, {Jaffe}, {Jones}, {Juvela}, {Keih{\"a}nen}, {Keskitalo},
  {Kisner}, {Kneissl}, {Knox}, {Kurki-Suonio}, {Lagache}, {Lamarre}, {Lasenby},
  {Laureijs}, {Lawrence}, {Le Jeune}, {Leach}, {Leonardi}, {Li}, {Liddle},
  {Lilje}, {Linden-V{\o}rnle}, {L{\'o}pez-Caniego}, {Lubin},
  {Mac{\'\i}as-P{\'e}rez}, {MacTavish}, {Maffei}, {Maino}, {Mandolesi}, {Mann},
  {Maris}, {Marleau}, {Mart{\'\i}nez-Gonz{\'a}lez}, {Masi}, {Matarrese},
  {Matthai}, {Mazzotta}, {Mei}, {Meinhold}, {Melchiorri}, {Melin}, {Mendes},
  {Mennella}, {Mitra}, {Miville-Desch{\^e}nes}, {Moneti}, {Montier},
  {Morgante}, {Mortlock}, {Munshi}, {Murphy}, {Naselsky}, {Nati}, {Natoli},
  {Netterfield}, {N{\o}rgaard-Nielsen}, {Noviello}, {Novikov}, {Novikov},
  {Olamaie}, {Osborne}, {Pajot}, {Pasian}, {Patanchon}, {Pearson}, {Perdereau},
  {Perotto}, {Perrotta}, {Piacentini}, {Piat}, {Pierpaoli}, {Piffaretti},
  {Plaszczynski}, {Pointecouteau}, {Polenta}, {Ponthieu}, {Poutanen}, {Pratt},
  {Pr{\'e}zeau}, {Prunet}, {Puget}, {Rachen}, {Reach}, {Rebolo}, {Reinecke},
  {Renault}, {Ricciardi}, {Riller}, {Ristorcelli}, {Rocha}, {Rosset},
  {Rubi{\~n}o-Mart{\'\i}n}, {Rusholme}, {Saar}, {Sandri}, {Santos}, {Saunders},
  {Savini}, {Schaefer}, {Scott}, {Seiffert}, {Shellard}, {Smoot}, {Stanford},
  {Starck}, {Stivoli}, {Stolyarov}, {Stompor}, {Sudiwala}, {Sunyaev}, {Sutton},
  {Sygnet}, {Taburet}, {Tauber}, {Terenzi}, {Toffolatti}, {Tomasi}, {Torre},
  {Tristram}, {Tuovinen}, {Valenziano}, {Vibert}, {Vielva}, {Villa},
  {Vittorio}, {Wade}, {Wandelt}, {Weller}, {White}, {White}, {Yvon}, {Zacchei},
  \& {Zonca}}]{Planck2011-VIII}
{Planck Collaboration}, {Ade}, P.~A.~R., {Aghanim}, N., {et~al.} 2011, \aap,
  536, A8

\bibitem[{{Planck Collaboration:} {et~al.}(2014){Planck Collaboration:}, {Ade,
  P. A. R.}, {Aghanim, N.}, {Alves, M. I. R.}, {Arnaud, M.}, {Ashdown, M.},
  {Atrio-Barandela, F.}, {Aumont, J.}, {Baccigalupi, C.}, {Banday, A. J.},
  {Barreiro, R. B.}, {Bartlett, J. G.}, {Battaner, E.}, {Benabed, K.},
  {Benoit-L\'evy, A.}, {Bernard, J.-P.}, {Bersanelli, M.}, {Bielewicz, P.},
  {Bobin, J.}, {Bonaldi, A.}, {Bond, J. R.}, {Borrill, J.}, {Bouchet, F. R.},
  {Boulanger, F.}, {Bucher, M.}, {Burigana, C.}, {Butler, R. C.}, {Cardoso,
  J.-F.}, {Catalano, A.}, {Chamballu, A.}, {Chiang, H. C.}, {Chiang, L.-Y.},
  {Christensen, P. R.}, {Clements, D. L.}, {Colombi, S.}, {Colombo, L. P. L.},
  {Couchot, F.}, {Crill, B. P.}, {Curto, A.}, {Cuttaia, F.}, {Danese, L.},
  {Davies, R. D.}, {Davis, R. J.}, {de Bernardis, P.}, {de Rosa, A.}, {de
  Zotti, G.}, {Delabrouille, J.}, {Dickinson, C.}, {Diego, J. M.}, {Dole, H.},
  {Donzelli, S.}, {Dor\'e, O.}, {Douspis, M.}, {Dupac, X.}, {En\ss{}lin, T.
  A.}, {Eriksen, H. K.}, {Falgarone, E.}, {Finelli, F.}, {Forni, O.}, {Frailis,
  M.}, {Franceschi, E.}, {Galeotta, S.}, {Ganga, K.}, {Ghosh, T.}, {Giard, M.},
  {Giardino, G.}, {Gonz\'alez-Nuevo, J.}, {G\'orski, K. M.}, {Gregorio, A.},
  {Gruppuso, A.}, {Hansen, F. K.}, {Harrison, D. L.}, {Hern\'andez-Monteagudo,
  C.}, {Herranz, D.}, {Hildebrandt, S. R.}, {Hivon, E.}, {Holmes, W. A.},
  {Hornstrup, A.}, {Hovest, W.}, {Jaffe, A. H.}, {Jones, W. C.}, {Juvela, M.},
  {Keih\"anen, E.}, {Keskitalo, R.}, {Kisner, T. S.}, {Kneissl, R.}, {Knoche,
  J.}, {Kunz, M.}, {Kurki-Suonio, H.}, {Lagache, G.}, {L\"ahteenm\"aki, A.},
  {Lamarre, J.-M.}, {Lasenby, A.}, {Laureijs, R. J.}, {Lawrence, C. R.},
  {Leonardi, R.}, {Levrier, F.}, {Liguori, M.}, {Lilje, P. B.},
  {Linden-V\o{}rnle, M.}, {L\'opez-Caniego, M.}, {Mac\'{\i}as-P\'erez, J. F.},
  {Maffei, B.}, {Maino, D.}, {Mandolesi, N.}, {Maris, M.}, {Marshall, D. J.},
  {Martin, P. G.}, {Mart\'{\i}nez-Gonz\'alez, E.}, {Masi, S.}, {Matarrese, S.},
  {Mazzotta, P.}, {Melchiorri, A.}, {Mendes, L.}, {Mennella, A.}, {Migliaccio,
  M.}, {Mitra, S.}, {Miville-Desch\^enes, M.-A.}, {Moneti, A.}, {Montier, L.},
  {Morgante, G.}, {Mortlock, D.}, {Munshi, D.}, {Murphy, J. A.}, {Naselsky,
  P.}, {Nati, F.}, {Natoli, P.}, {N\o{}rgaard-Nielsen, H. U.}, {Noviello, F.},
  {Novikov, D.}, {Novikov, I.}, {Oxborrow, C. A.}, {Pagano, L.}, {Pajot, F.},
  {Paladini, R.}, {Paoletti, D.}, {Pasian, F.}, {Patanchon, G.}, {Peel, M.},
  {Perdereau, O.}, {Perrotta, F.}, {Piacentini, F.}, {Piat, M.}, {Pierpaoli,
  E.}, {Pietrobon, D.}, {Plaszczynski, S.}, {Pointecouteau, E.}, {Polenta, G.},
  {Ponthieu, N.}, {Popa, L.}, {Pratt, G. W.}, {Prunet, S.}, {Puget, J.-L.},
  {Rachen, J. P.}, {Reach, W. T.}, {Rebolo, R.}, {Reinecke, M.}, {Remazeilles,
  M.}, {Renault, C.}, {Ricciardi, S.}, {Riller, T.}, {Ristorcelli, I.}, {Rocha,
  G.}, {Rosset, C.}, {Rubi\~no-Mart\'{\i}n, J. A.}, {Rusholme, B.}, {Sandri,
  M.}, {Savini, G.}, {Scott, D.}, {Spencer, L. D.}, {Starck, J.-L.},
  {Stolyarov, V.}, {Sureau, F.}, {Sutton, D.}, {Suur-Uski, A.-S.}, {Sygnet,
  J.-F.}, {Tauber, J. A.}, {Tavagnacco, D.}, {Terenzi, L.}, {Toffolatti, L.},
  {Tomasi, M.}, {Tristram, M.}, {Tucci, M.}, {Valenziano, L.}, {Valiviita, J.},
  {Van Tent, B.}, {Verstraete, L.}, {Vielva, P.}, {Villa, F.}, {Vittorio, N.},
  {Wade, L. A.}, {Wandelt, B. D.}, {Yvon, D.}, {Zacchei, A.}, \& {Zonca,
  A.}}]{Planck2014-XIV}
{Planck Collaboration:}, {Ade, P. A. R.}, {Aghanim, N.}, {et~al.} 2014, A\&A,
  564, A45

\bibitem[{{Planck Collaboration} {et~al.}(2016){Planck Collaboration}, {Ade},
  {Aghanim}, {Arnaud}, {Ashdown}, {Aumont}, {Baccigalupi}, {Banday},
  {Barreiro}, {Bartlett}, {Bartolo}, {Battaner}, {Battye}, {Benabed},
  {Beno{\^\i}t}, {Benoit-L{\'e}vy}, {Bernard}, {Bersanelli}, {Bielewicz},
  {Bock}, {Bonaldi}, {Bonavera}, {Bond}, {Borrill}, {Bouchet}, {Bucher},
  {Burigana}, {Butler}, {Calabrese}, {Cardoso}, {Catalano}, {Challinor},
  {Chamballu}, {Chary}, {Chiang}, {Christensen}, {Church}, {Clements},
  {Colombi}, {Colombo}, {Combet}, {Comis}, {Couchot}, {Coulais}, {Crill},
  {Curto}, {Cuttaia}, {Danese}, {Davies}, {Davis}, {de Bernardis}, {de Rosa},
  {de Zotti}, {Delabrouille}, {D{\'e}sert}, {Diego}, {Dolag}, {Dole},
  {Donzelli}, {Dor{\'e}}, {Douspis}, {Ducout}, {Dupac}, {Efstathiou}, {Elsner},
  {En{\ss}lin}, {Eriksen}, {Falgarone}, {Fergusson}, {Finelli}, {Forni},
  {Frailis}, {Fraisse}, {Franceschi}, {Frejsel}, {Galeotta}, {Galli}, {Ganga},
  {Giard}, {Giraud-H{\'e}raud}, {Gjerl{\o}w}, {Gonz{\'a}lez-Nuevo},
  {G{\'o}rski}, {Gratton}, {Gregorio}, {Gruppuso}, {Gudmundsson}, {Hansen},
  {Hanson}, {Harrison}, {Henrot-Versill{\'e}}, {Hern{\'a}ndez-Monteagudo},
  {Herranz}, {Hildebrandt}, {Hivon}, {Hobson}, {Holmes}, {Hornstrup}, {Hovest},
  {Huffenberger}, {Hurier}, {Jaffe}, {Jaffe}, {Jones}, {Juvela},
  {Keih{\"a}nen}, {Keskitalo}, {Kisner}, {Kneissl}, {Knoche}, {Kunz},
  {Kurki-Suonio}, {Lagache}, {L{\"a}hteenm{\"a}ki}, {Lamarre}, {Lasenby},
  {Lattanzi}, {Lawrence}, {Leonardi}, {Lesgourgues}, {Levrier}, {Liguori},
  {Lilje}, {Linden-V{\o}rnle}, {L{\'o}pez-Caniego}, {Lubin},
  {Mac{\'\i}as-P{\'e}rez}, {Maggio}, {Maino}, {Mand olesi}, {Mangilli},
  {Maris}, {Martin}, {Mart{\'\i}nez-Gonz{\'a}lez}, {Masi}, {Matarrese},
  {McGehee}, {Meinhold}, {Melchiorri}, {Melin}, {Mendes}, {Mennella},
  {Migliaccio}, {Mitra}, {Miville-Desch{\^e}nes}, {Moneti}, {Montier},
  {Morgante}, {Mortlock}, {Moss}, {Munshi}, {Murphy}, {Naselsky}, {Nati},
  {Natoli}, {Netterfield}, {N{\o}rgaard-Nielsen}, {Noviello}, {Novikov},
  {Novikov}, {Oxborrow}, {Paci}, {Pagano}, {Pajot}, {Paoletti}, {Partridge},
  {Pasian}, {Patanchon}, {Pearson}, {Perdereau}, {Perotto}, {Perrotta},
  {Pettorino}, {Piacentini}, {Piat}, {Pierpaoli}, {Pietrobon}, {Plaszczynski},
  {Pointecouteau}, {Polenta}, {Popa}, {Pratt}, {Pr{\'e}zeau}, {Prunet},
  {Puget}, {Rachen}, {Rebolo}, {Reinecke}, {Remazeilles}, {Renault}, {Renzi},
  {Ristorcelli}, {Rocha}, {Roman}, {Rosset}, {Rossetti}, {Roudier},
  {Rubi{\~n}o-Mart{\'\i}n}, {Rusholme}, {Sandri}, {Santos}, {Savelainen},
  {Savini}, {Scott}, {Seiffert}, {Shellard}, {Spencer}, {Stolyarov}, {Stompor},
  {Sudiwala}, {Sunyaev}, {Sutton}, {Suur-Uski}, {Sygnet}, {Tauber}, {Terenzi},
  {Toffolatti}, {Tomasi}, {Tristram}, {Tucci}, {Tuovinen}, {T{\"u}rler},
  {Umana}, {Valenziano}, {Valiviita}, {Van Tent}, {Vielva}, {Villa}, {Wade},
  {Wandelt}, {Wehus}, {Weller}, {White}, {Yvon}, {Zacchei}, \&
  {Zonca}}]{Planck2016-XXIV}
{Planck Collaboration}, {Ade}, P.~A.~R., {Aghanim}, N., {et~al.} 2016, \aap,
  594, A24

\bibitem[{{Planck Collaboration} {et~al.}(2020{\natexlab{a}}){Planck
  Collaboration}, {Aghanim, N.}, {Akrami, Y.}, {Arroja, F.}, {Ashdown, M.},
  {Aumont, J.}, {Baccigalupi, C.}, {Ballardini, M.}, {Banday, A. J.},
  {Barreiro, R. B.}, {Bartolo, N.}, {Basak, S.}, {Battye, R.}, {Benabed, K.},
  {Bernard, J.-P.}, {Bersanelli, M.}, {Bielewicz, P.}, {Bock, J. J.}, {Bond, J.
  R.}, {Borrill, J.}, {Bouchet, F. R.}, {Boulanger, F.}, {Bucher, M.},
  {Burigana, C.}, {Butler, R. C.}, {Calabrese, E.}, {Cardoso, J.-F.}, {Carron,
  J.}, {Casaponsa, B.}, {Challinor, A.}, {Chiang, H. C.}, {Colombo, L. P. L.},
  {Combet, C.}, {Contreras, D.}, {Crill, B. P.}, {Cuttaia, F.}, {de Bernardis,
  P.}, {de Zotti, G.}, {Delabrouille, J.}, {Delouis, J.-M.}, {D\'esert, F.-X.},
  {Di Valentino, E.}, {Dickinson, C.}, {Diego, J. M.}, {Donzelli, S.}, {Dor\'e,
  O.}, {Douspis, M.}, {Ducout, A.}, {Dupac, X.}, {Efstathiou, G.}, {Elsner,
  F.}, {En\ss{}lin, T. A.}, {Eriksen, H. K.}, {Falgarone, E.}, {Fantaye, Y.},
  {Fergusson, J.}, {Fernandez-Cobos, R.}, {Finelli, F.}, {Forastieri, F.},
  {Frailis, M.}, {Franceschi, E.}, {Frolov, A.}, {Galeotta, S.}, {Galli, S.},
  {Ganga, K.}, {G\'enova-Santos, R. T.}, {Gerbino, M.}, {Ghosh, T.},
  {Gonz\'alez-Nuevo, J.}, {G\'orski, K. M.}, {Gratton, S.}, {Gruppuso, A.},
  {Gudmundsson, J. E.}, {Hamann, J.}, {Handley, W.}, {Hansen, F. K.}, {Helou,
  G.}, {Herranz, D.}, {Hildebrandt, S. R.}, {Hivon, E.}, {Huang, Z.}, {Jaffe,
  A. H.}, {Jones, W. C.}, {Karakci, A.}, {Keih\"anen, E.}, {Keskitalo, R.},
  {Kiiveri, K.}, {Kim, J.}, {Kisner, T. S.}, {Knox, L.}, {Krachmalnicoff, N.},
  {Kunz, M.}, {Kurki-Suonio, H.}, {Lagache, G.}, {Lamarre, J.-M.}, {Langer,
  M.}, {Lasenby, A.}, {Lattanzi, M.}, {Lawrence, C. R.}, {Le Jeune, M.},
  {Leahy, J. P.}, {Lesgourgues, J.}, {Levrier, F.}, {Lewis, A.}, {Liguori, M.},
  {Lilje, P. B.}, {Lilley, M.}, {Lindholm, V.}, {L\'opez-Caniego, M.}, {Lubin,
  P. M.}, {Ma, Y.-Z.}, {Mac\'{\i}as-P\'erez, J. F.}, {Maggio, G.}, {Maino, D.},
  {Mandolesi, N.}, {Mangilli, A.}, {Marcos-Caballero, A.}, {Maris, M.},
  {Martin, P. G.}, {Martinelli, M.}, {Mart\'{\i}nez-Gonz\'alez, E.},
  {Matarrese, S.}, {Mauri, N.}, {McEwen, J. D.}, {Meerburg, P. D.}, {Meinhold,
  P. R.}, {Melchiorri, A.}, {Mennella, A.}, {Migliaccio, M.}, {Millea, M.},
  {Mitra, S.}, {Miville-Desch\^enes, M.-A.}, {Molinari, D.}, {Moneti, A.},
  {Montier, L.}, {Morgante, G.}, {Moss, A.}, {Mottet, S.}, {M\"unchmeyer, M.},
  {Natoli, P.}, {N\o{}rgaard-Nielsen, H. U.}, {Oxborrow, C. A.}, {Pagano, L.},
  {Paoletti, D.}, {Partridge, B.}, {Patanchon, G.}, {Pearson, T. J.}, {Peel,
  M.}, {Peiris, H. V.}, {Perrotta, F.}, {Pettorino, V.}, {Piacentini, F.},
  {Polastri, L.}, {Polenta, G.}, {Puget, J.-L.}, {Rachen, J. P.}, {Reinecke,
  M.}, {Remazeilles, M.}, {Renault, C.}, {Renzi, A.}, {Rocha, G.}, {Rosset,
  C.}, {Roudier, G.}, {Rubi\~no-Mart\'{\i}n, J. A.}, {Ruiz-Granados, B.},
  {Salvati, L.}, {Sandri, M.}, {Savelainen, M.}, {Scott, D.}, {Shellard, E. P.
  S.}, {Shiraishi, M.}, {Sirignano, C.}, {Sirri, G.}, {Spencer, L. D.},
  {Sunyaev, R.}, {Suur-Uski, A.-S.}, {Tauber, J. A.}, {Tavagnacco, D.}, {Tenti,
  M.}, {Terenzi, L.}, {Toffolatti, L.}, {Tomasi, M.}, {Trombetti, T.},
  {Valiviita, J.}, {Van Tent, B.}, {Vibert, L.}, {Vielva, P.}, {Villa, F.},
  {Vittorio, N.}, {Wandelt, B. D.}, {Wehus, I. K.}, {White, M.}, {White, S. D.
  M.}, {Zacchei, A.}, \& {Zonca, A.}}]{Planck2020-I}
{Planck Collaboration}, {Aghanim, N.}, {Akrami, Y.}, {et~al.}
  2020{\natexlab{a}}, A\&A, 641, A1

\bibitem[{{Planck Collaboration} {et~al.}(2020{\natexlab{b}}){Planck
  Collaboration}, {Akrami, Y.}, {Ashdown, M.}, {Aumont, J.}, {Baccigalupi, C.},
  {Ballardini, M.}, {Banday, A. J.}, {Barreiro, R. B.}, {Bartolo, N.}, {Basak,
  S.}, {Benabed, K.}, {Bersanelli, M.}, {Bielewicz, P.}, {Bond, J. R.},
  {Borrill, J.}, {Bouchet, F. R.}, {Boulanger, F.}, {Bucher, M.}, {Burigana,
  C.}, {Calabrese, E.}, {Cardoso, J.-F.}, {Carron, J.}, {Casaponsa, B.},
  {Challinor, A.}, {Colombo, L. P. L.}, {Combet, C.}, {Crill, B. P.}, {Cuttaia,
  F.}, {de Bernardis, P.}, {de Rosa, A.}, {de Zotti, G.}, {Delabrouille, J.},
  {Delouis, J.-M.}, {Di Valentino, E.}, {Dickinson, C.}, {Diego, J. M.},
  {Donzelli, S.}, {Dor\'e, O.}, {Ducout, A.}, {Dupac, X.}, {Efstathiou, G.},
  {Elsner, F.}, {En\ss{}lin, T. A.}, {Eriksen, H. K.}, {Falgarone, E.},
  {Fernandez-Cobos, R.}, {Finelli, F.}, {Forastieri, F.}, {Frailis, M.},
  {Fraisse, A. A.}, {Franceschi, E.}, {Frolov, A.}, {Galeotta, S.}, {Galli,
  S.}, {Ganga, K.}, {G\'enova-Santos, R. T.}, {Gerbino, M.}, {Ghosh, T.},
  {Gonz\'alez-Nuevo, J.}, {G\'orski, K. M.}, {Gratton, S.}, {Gruppuso, A.},
  {Gudmundsson, J. E.}, {Handley, W.}, {Hansen, F. K.}, {Helou, G.}, {Herranz,
  D.}, {Hildebrandt, S. R.}, {Huang, Z.}, {Jaffe, A. H.}, {Karakci, A.},
  {Keih\"anen, E.}, {Keskitalo, R.}, {Kiiveri, K.}, {Kim, J.}, {Kisner, T. S.},
  {Krachmalnicoff, N.}, {Kunz, M.}, {Kurki-Suonio, H.}, {Lagache, G.},
  {Lamarre, J.-M.}, {Lasenby, A.}, {Lattanzi, M.}, {Lawrence, C. R.}, {Le
  Jeune, M.}, {Levrier, F.}, {Liguori, M.}, {Lilje, P. B.}, {Lindholm, V.},
  {L\'opez-Caniego, M.}, {Lubin, P. M.}, {Ma, Y.-Z.}, {Mac\'{\i}as-P\'erez, J.
  F.}, {Maggio, G.}, {Maino, D.}, {Mandolesi, N.}, {Mangilli, A.},
  {Marcos-Caballero, A.}, {Maris, M.}, {Martin, P. G.},
  {Mart\'{\i}nez-Gonz\'alez, E.}, {Matarrese, S.}, {Mauri, N.}, {McEwen, J.
  D.}, {Meinhold, P. R.}, {Melchiorri, A.}, {Mennella, A.}, {Migliaccio, M.},
  {Miville-Desch\^enes, M.-A.}, {Molinari, D.}, {Moneti, A.}, {Montier, L.},
  {Morgante, G.}, {Natoli, P.}, {Oppizzi, F.}, {Pagano, L.}, {Paoletti, D.},
  {Partridge, B.}, {Peel, M.}, {Pettorino, V.}, {Piacentini, F.}, {Polenta,
  G.}, {Puget, J.-L.}, {Rachen, J. P.}, {Reinecke, M.}, {Remazeilles, M.},
  {Renzi, A.}, {Rocha, G.}, {Roudier, G.}, {Rubi\~no-Mart\'{\i}n, J. A.},
  {Ruiz-Granados, B.}, {Salvati, L.}, {Sandri, M.}, {Savelainen, M.}, {Scott,
  D.}, {Seljebotn, D. S.}, {Sirignano, C.}, {Spencer, L. D.}, {Suur-Uski,
  A.-S.}, {Tauber, J. A.}, {Tavagnacco, D.}, {Tenti, M.}, {Thommesen, H.},
  {Toffolatti, L.}, {Tomasi, M.}, {Trombetti, T.}, {Valiviita, J.}, {Van Tent,
  B.}, {Vielva, P.}, {Villa, F.}, {Vittorio, N.}, {Wandelt, B. D.}, {Wehus, I.
  K.}, {Zacchei, A.}, \& {Zonca, A.}}]{Planck2020-IV}
{Planck Collaboration}, {Akrami, Y.}, {Ashdown, M.}, {et~al.}
  2020{\natexlab{b}}, A\&A, 641, A4

\bibitem[{{Planck Collaboration} {et~al.}(2020{\natexlab{c}}){Planck
  Collaboration}, {Aghanim}, {Akrami}, {Ashdown}, {Aumont}, {Baccigalupi},
  {Ballardini}, {Banday}, {Barreiro}, {Bartolo}, {Basak}, {Battye}, {Benabed},
  {Bernard}, {Bersanelli}, {Bielewicz}, {Bock}, {Bond}, {Borrill}, {Bouchet},
  {Boulanger}, {Bucher}, {Burigana}, {Butler}, {Calabrese}, {Cardoso},
  {Carron}, {Challinor}, {Chiang}, {Chluba}, {Colombo}, {Combet}, {Contreras},
  {Crill}, {Cuttaia}, {de Bernardis}, {de Zotti}, {Delabrouille}, {Delouis},
  {Di Valentino}, {Diego}, {Dor{\'e}}, {Douspis}, {Ducout}, {Dupac}, {Dusini},
  {Efstathiou}, {Elsner}, {En{\ss}lin}, {Eriksen}, {Fantaye}, {Farhang},
  {Fergusson}, {Fernandez-Cobos}, {Finelli}, {Forastieri}, {Frailis},
  {Fraisse}, {Franceschi}, {Frolov}, {Galeotta}, {Galli}, {Ganga},
  {G{\'e}nova-Santos}, {Gerbino}, {Ghosh}, {Gonz{\'a}lez-Nuevo}, {G{\'o}rski},
  {Gratton}, {Gruppuso}, {Gudmundsson}, {Hamann}, {Handley}, {Hansen},
  {Herranz}, {Hildebrandt}, {Hivon}, {Huang}, {Jaffe}, {Jones}, {Karakci},
  {Keih{\"a}nen}, {Keskitalo}, {Kiiveri}, {Kim}, {Kisner}, {Knox},
  {Krachmalnicoff}, {Kunz}, {Kurki-Suonio}, {Lagache}, {Lamarre}, {Lasenby},
  {Lattanzi}, {Lawrence}, {Le Jeune}, {Lemos}, {Lesgourgues}, {Levrier},
  {Lewis}, {Liguori}, {Lilje}, {Lilley}, {Lindholm}, {L{\'o}pez-Caniego},
  {Lubin}, {Ma}, {Mac{\'\i}as-P{\'e}rez}, {Maggio}, {Maino}, {Mandolesi},
  {Mangilli}, {Marcos-Caballero}, {Maris}, {Martin}, {Martinelli},
  {Mart{\'\i}nez-Gonz{\'a}lez}, {Matarrese}, {Mauri}, {McEwen}, {Meinhold},
  {Melchiorri}, {Mennella}, {Migliaccio}, {Millea}, {Mitra},
  {Miville-Desch{\^e}nes}, {Molinari}, {Montier}, {Morgante}, {Moss}, {Natoli},
  {N{\o}rgaard-Nielsen}, {Pagano}, {Paoletti}, {Partridge}, {Patanchon},
  {Peiris}, {Perrotta}, {Pettorino}, {Piacentini}, {Polastri}, {Polenta},
  {Puget}, {Rachen}, {Reinecke}, {Remazeilles}, {Renzi}, {Rocha}, {Rosset},
  {Roudier}, {Rubi{\~n}o-Mart{\'\i}n}, {Ruiz-Granados}, {Salvati}, {Sandri},
  {Savelainen}, {Scott}, {Shellard}, {Sirignano}, {Sirri}, {Spencer},
  {Sunyaev}, {Suur-Uski}, {Tauber}, {Tavagnacco}, {Tenti}, {Toffolatti},
  {Tomasi}, {Trombetti}, {Valenziano}, {Valiviita}, {Van Tent}, {Vibert},
  {Vielva}, {Villa}, {Vittorio}, {Wand elt}, {Wehus}, {White}, {White},
  {Zacchei}, \& {Zonca}}]{Planck2020-VI}
{Planck Collaboration}, {Aghanim}, N., {Akrami}, Y., {et~al.}
  2020{\natexlab{c}}, \aap, 641, A6

\bibitem[{Pratt {et~al.}(2021)Pratt, Qu, \& Bregman}]{Pratt2021}
Pratt, C.~T., Qu, Z., \& Bregman, J.~N. 2021, \apj, 920, 104

\bibitem[{{Rees} \& {Ostriker}(1977)}]{Rees1977}
{Rees}, M.~J., \& {Ostriker}, J.~P. 1977, \mnras, 179, 541

\bibitem[{{Reichardt} {et~al.}(2013){Reichardt}, {Stalder}, {Bleem}, {Montroy},
  {Aird}, {Andersson}, {Armstrong}, {Ashby}, {Bautz}, {Bayliss}, {Bazin},
  {Benson}, {Brodwin}, {Carlstrom}, {Chang}, {Cho}, {Clocchiatti}, {Crawford},
  {Crites}, {de Haan}, {Desai}, {Dobbs}, {Dudley}, {Foley}, {Forman}, {George},
  {Gladders}, {Gonzalez}, {Halverson}, {Harrington}, {High}, {Holder},
  {Holzapfel}, {Hoover}, {Hrubes}, {Jones}, {Joy}, {Keisler}, {Knox}, {Lee},
  {Leitch}, {Liu}, {Lueker}, {Luong-Van}, {Mantz}, {Marrone}, {McDonald},
  {McMahon}, {Mehl}, {Meyer}, {Mocanu}, {Mohr}, {Murray}, {Natoli}, {Padin},
  {Plagge}, {Pryke}, {Rest}, {Ruel}, {Ruhl}, {Saliwanchik}, {Saro}, {Sayre},
  {Schaffer}, {Shaw}, {Shirokoff}, {Song}, {Spieler}, {Staniszewski}, {Stark},
  {Story}, {Stubbs}, {{\v S}uhada}, {van Engelen}, {Vanderlinde}, {Vieira},
  {Vikhlinin}, {Williamson}, {Zahn}, \& {Zenteno}}]{Reichardt2013}
{Reichardt}, C.~L., {Stalder}, B., {Bleem}, L.~E., {et~al.} 2013, \apj, 763,
  127

\bibitem[{{Ruan} {et~al.}(2015){Ruan}, {McQuinn}, \& {Anderson}}]{Ruan2015}
{Ruan}, J.~J., {McQuinn}, M., \& {Anderson}, S.~F. 2015, \apj, 802, 135

\bibitem[{{Santini} {et~al.}(2014){Santini}, {Maiolino}, {Magnelli}, {Lutz},
  {Lamastra}, {Li Causi}, {Eales}, {Andreani}, {Berta}, {Buat}, {Cooray},
  {Cresci}, {Daddi}, {Farrah}, {Fontana}, {Franceschini}, {Genzel}, {Granato},
  {Grazian}, {Le Floc'h}, {Magdis}, {Magliocchetti}, {Mannucci}, {Menci},
  {Nordon}, {Oliver}, {Popesso}, {Pozzi}, {Riguccini}, {Rodighiero}, {Rosario},
  {Salvato}, {Scott}, {Silva}, {Tacconi}, {Viero}, {Wang}, {Wuyts}, \&
  {Xu}}]{Satini2014}
{Santini}, P., {Maiolino}, R., {Magnelli}, B., {et~al.} 2014, \aap, 562, A30

\bibitem[{{Santini} {et~al.}(2015){Santini}, {Ferguson}, {Fontana}, {Mobasher},
  {Barro}, {Castellano}, {Finkelstein}, {Grazian}, {Hsu}, {Lee}, {Lee},
  {Pforr}, {Salvato}, {Wiklind}, {Wuyts}, {Almaini}, {Cooper}, {Galametz},
  {Weiner}, {Amorin}, {Boutsia}, {Conselice}, {Dahlen}, {Dickinson},
  {Giavalisco}, {Grogin}, {Guo}, {Hathi}, {Kocevski}, {Koekemoer},
  {Kurczynski}, {Merlin}, {Mortlock}, {Newman}, {Paris}, {Pentericci},
  {Simons}, \& {Willner}}]{Santini2015}
{Santini}, P., {Ferguson}, H.~C., {Fontana}, A., {et~al.} 2015, \apj, 801, 97

\bibitem[{{Scannapieco} \& {Oh}(2004)}]{Scannapieco2004}
{Scannapieco}, E., \& {Oh}, S.~P. 2004, \apj, 608, 62

\bibitem[{{Scannapieco} {et~al.}(2008){Scannapieco}, {Thacker}, \&
  {Couchman}}]{Scannapieco2008}
{Scannapieco}, E., {Thacker}, R.~J., \& {Couchman}, H.~M.~P. 2008, \apj, 678,
  674

\bibitem[{Schaan {et~al.}(2021)Schaan, Ferraro, Amodeo, Battaglia, Aiola,
  Austermann, Beall, Bean, Becker, Bond, Calabrese, Calafut, Choi, Denison,
  Devlin, Duff, Duivenvoorden, Dunkley, D\"unner, Gallardo, Guan, Han, Hill,
  Hilton, Hilton, Hlo\ifmmode~\check{z}\else \v{z}\fi{}ek, Hubmayr,
  Huffenberger, Hughes, Koopman, MacInnis, McMahon, Madhavacheril, Moodley,
  Mroczkowski, Naess, Nati, Newburgh, Niemack, Page, Partridge, Salatino,
  Sehgal, Schillaci, Sif\'on, Smith, Spergel, Staggs, Storer, Trac, Ullom,
  Van~Lanen, Vale, van Engelen, Maga\~na, Vavagiakis, Wollack, \&
  Xu}]{Schaan2021}
Schaan, E., Ferraro, S., Amodeo, S., {et~al.} 2021, \prd, 103, 063513

\bibitem[{{Schaye} {et~al.}(2015){Schaye}, {Crain}, {Bower}, {Furlong},
  {Schaller}, {Theuns}, {Dalla Vecchia}, {Frenk}, {McCarthy}, {Helly},
  {Jenkins}, {Rosas-Guevara}, {White}, {Baes}, {Booth}, {Camps}, {Navarro},
  {Qu}, {Rahmati}, {Sawala}, {Thomas}, \& {Trayford}}]{Schaye2015}
{Schaye}, J., {Crain}, R.~A., {Bower}, R.~G., {et~al.} 2015, \mnras, 446, 521

\bibitem[{Schlafly {et~al.}(2019)Schlafly, Meisner, \& Green}]{Schlafly2019}
Schlafly, E.~F., Meisner, A.~M., \& Green, G.~M. 2019, \apjs, 240, 30

\bibitem[{{Silk} \& {Rees}(1998)}]{Silk98}
{Silk}, J., \& {Rees}, M.~J. 1998, \aap, 331, L1

\bibitem[{{Sobrin} {et~al.}(2022){Sobrin}, {Anderson}, {Bender}, {Benson},
  {Dutcher}, {Foster}, {Goeckner-Wald}, {Montgomery}, {Nadolski}, {Rahlin},
  {Ade}, {Ahmed}, {Anderes}, {Archipley}, {Austermann}, {Avva}, {Aylor},
  {Balkenhol}, {Barry}, {Thakur}, {Benabed}, {Bianchini}, {Bleem}, {Bouchet},
  {Bryant}, {Byrum}, {Carlstrom}, {Carter}, {Cecil}, {Chang}, {Chaubal},
  {Chen}, {Cho}, {Chou}, {Cliche}, {Crawford}, {Cukierman}, {Daley}, {Haan},
  {Denison}, {Dibert}, {Ding}, {Dobbs}, {Everett}, {Feng}, {Ferguson}, {Fu},
  {Galli}, {Gambrel}, {Gardner}, {Gualtieri}, {Guns}, {Gupta}, {Guyser},
  {Halverson}, {Harke-Hosemann}, {Harrington}, {Henning}, {Hilton}, {Hivon},
  {Holder}, {Holzapfel}, {Hood}, {Howe}, {Huang}, {Irwin}, {Jeong}, {Jonas},
  {Jones}, {Khaire}, {Knox}, {Kofman}, {Korman}, {Kubik}, {Kuhlmann}, {Kuo},
  {Lee}, {Leitch}, {Lowitz}, {Lu}, {Meyer}, {Michalik}, {Millea}, {Natoli},
  {Nguyen}, {Noble}, {Novosad}, {Omori}, {Padin}, {Pan}, {Paschos}, {Pearson},
  {Posada}, {Prabhu}, {Quan}, {Reichardt}, {Riebel}, {Riedel}, {Rouble},
  {Ruhl}, {Saliwanchik}, {Sayre}, {Schiappucci}, {Shirokoff}, {Smecher},
  {Stark}, {Stephen}, {Story}, {Suzuki}, {Tandoi}, {Thompson}, {Thorne},
  {Tucker}, {Umilta}, {Vale}, {Vanderlinde}, {Vieira}, {Wang}, {Whitehorn},
  {Wu}, {Yefremenko}, {Yoon}, \& {Young}}]{SPT3G-2022}
{Sobrin}, J.~A., {Anderson}, A.~J., {Bender}, A.~N., {et~al.} 2022, \apjs, 258,
  42

\bibitem[{{Somerville} \& {Dav{\'e}}(2015)}]{Somerville2015}
{Somerville}, R.~S., \& {Dav{\'e}}, R. 2015, \araa, 53, 51

\bibitem[{{Spacek} {et~al.}(2016){Spacek}, {Scannapieco}, {Cohen}, {Joshi}, \&
  {Mauskopf}}]{Spacek2016}
{Spacek}, A., {Scannapieco}, E., {Cohen}, S., {Joshi}, B., \& {Mauskopf}, P.
  2016, \apj, 819, 128

\bibitem[{{Spacek} {et~al.}(2017){Spacek}, {Scannapieco}, {Cohen}, {Joshi}, \&
  {Mauskopf}}]{Spacek2017}
---. 2017, \apj, 834, 102

\bibitem[{{Springel} {et~al.}(2018){Springel}, {Pakmor}, {Pillepich},
  {Weinberger}, {Nelson}, {Hernquist}, {Vogelsberger}, {Genel}, {Torrey},
  {Marinacci}, \& {Naiman}}]{Springel2018}
{Springel}, V., {Pakmor}, R., {Pillepich}, A., {et~al.} 2018, \mnras, 475, 676

\bibitem[{{Sunyaev} \& {Zeldovich}(1972)}]{Sunyaev1972}
{Sunyaev}, R.~A., \& {Zeldovich}, Y.~B. 1972, \coasp, 4, 173

\bibitem[{{Sunyaev} \& {Zeldovich}(1980)}]{Sunyaev1980}
---. 1980, \mnras, 190, 413

\bibitem[{{Thacker} {et~al.}(2006){Thacker}, {Scannapieco}, \&
  {Couchman}}]{Thacker2006}
{Thacker}, R.~J., {Scannapieco}, E., \& {Couchman}, H.~M.~P. 2006, \apj, 653,
  86

\bibitem[{{Treu} {et~al.}(2005){Treu}, {Ellis}, {Liao}, \& {van
  Dokkum}}]{Treu2005}
{Treu}, T., {Ellis}, R.~S., {Liao}, T.~X., \& {van Dokkum}, P.~G. 2005, \apjl,
  622, L5

\bibitem[{{van Dokkum} {et~al.}(2010){van Dokkum}, {Whitaker}, {Brammer},
  {Franx}, {Kriek}, {Labb{\'e}}, {Marchesini}, {Quadri}, {Bezanson},
  {Illingworth}, {Muzzin}, {Rudnick}, {Tal}, \& {Wake}}]{vanDokkum2010}
{van Dokkum}, P.~G., {Whitaker}, K.~E., {Brammer}, G., {et~al.} 2010, \apj,
  709, 1018

\bibitem[{{Vavagiakis} {et~al.}(2021){Vavagiakis}, {Gallardo}, {Calafut},
  {Amodeo}, {Aiola}, {Austermann}, {Battaglia}, {Battistelli}, {Beall}, {Bean},
  {Bond}, {Calabrese}, {Choi}, {Cothard}, {Devlin}, {Duell}, {Duff},
  {Duivenvoorden}, {Dunkley}, {Dunner}, {Ferraro}, {Guan}, {Hill}, {Hilton},
  {Hilton}, {Hlo{\v{z}}ek}, {Huber}, {Hubmayr}, {Huffenberger}, {Hughes},
  {Koopman}, {Kosowsky}, {Li}, {Lokken}, {Madhavacheril}, {McMahon}, {Moodley},
  {Naess}, {Nati}, {Newburgh}, {Niemack}, {Page}, {Partridge}, {Schaan},
  {Schillaci}, {Sif{\'o}n}, {Spergel}, {Staggs}, {Ullom}, {Vale}, {Van
  Engelen}, {Van Lanen}, {Wollack}, \& {Xu}}]{Vavagiakis2021}
{Vavagiakis}, E.~M., {Gallardo}, P.~A., {Calafut}, V., {et~al.} 2021, \prd,
  104, 043503

\bibitem[{{Veilleux} {et~al.}(2013){Veilleux}, {Mel{\'e}ndez}, {Sturm},
  {Gracia-Carpio}, {Fischer}, {Gonz{\'a}lez-Alfonso}, {Contursi}, {Lutz},
  {Poglitsch}, {Davies}, {Genzel}, {Tacconi}, {de Jong}, {Sternberg}, {Netzer},
  {Hailey-Dunsheath}, {Verma}, {Rupke}, {Maiolino}, {Teng}, \&
  {Polisensky}}]{Veilleux2013}
{Veilleux}, S., {Mel{\'e}ndez}, M., {Sturm}, E., {et~al.} 2013, \apj, 776, 27

\bibitem[{{Wampler} {et~al.}(1995){Wampler}, {Chugai}, \&
  {Petitjean}}]{Wampler1995}
{Wampler}, E.~J., {Chugai}, N.~N., \& {Petitjean}, P. 1995, \apj, 443, 586

\bibitem[{{Wang} {et~al.}(2013){Wang}, {Farrah}, {Oliver}, {Amblard},
  {B{\'e}thermin}, {Bock}, {Conley}, {Cooray}, {Halpern}, {Heinis}, {Ibar},
  {Ilbert}, {Ivison}, {Marsden}, {Roseboom}, {Rowan-Robinson}, {Schulz},
  {Smith}, {Viero}, \& {Zemcov}}]{Wang2013}
{Wang}, L., {Farrah}, D., {Oliver}, S.~J., {et~al.} 2013, \mnras, 431, 648

\bibitem[{{Werner} {et~al.}(2019){Werner}, {McNamara}, {Churazov}, \&
  {Scannapieco}}]{Werner2019}
{Werner}, N., {McNamara}, B.~R., {Churazov}, E., \& {Scannapieco}, E. 2019,
  \ssr, 215, 5

\bibitem[{{White} \& {Frenk}(1991)}]{White1991}
{White}, S.~D.~M., \& {Frenk}, C.~S. 1991, \apj, 379, 52

\bibitem[{{Wilson} {et~al.}(2020){Wilson}, {Abi-Saad}, {Ade}, {Aretxaga},
  {Austermann}, {Ban}, {Bardin}, {Beall}, {Berthoud}, {Bryan}, {Bussan},
  {Castillo}, {Chavez}, {Contente}, {DeNigris}, {Dober}, {Eiben}, {Ferrusca},
  {Fissel}, {Gao}, {Golec}, {Golina}, {Gomez}, {Gordon}, {Gutermuth}, {Hilton},
  {Hosseini}, {Hubmayr}, {Hughes}, {Kuczarski}, {Lee}, {Lunde}, {Ma}, {Mani},
  {Mauskopf}, {McCrackan}, {McKenney}, {McMahon}, {Novak}, {Pisano}, {Pope},
  {Ralston}, {Rodriguez}, {S{\'a}nchez-Arg{\"u}elles}, {Schloerb}, {Simon},
  {Sinclair}, {Souccar}, {Torres Campos}, {Tucker}, {Ullom}, {Van Camp}, {Van
  Lanen}, {Velazquez}, {Vissers}, {Weeks}, \& {Yun}}]{TolTEC2020}
{Wilson}, G.~W., {Abi-Saad}, S., {Ade}, P., {et~al.} 2020, in \procspie, Vol.
  11453, Society of Photo-Optical Instrumentation Engineers (SPIE) Conference
  Series, 1145302

\bibitem[{{Wright} {et~al.}(2010){Wright}, {Eisenhardt}, {Mainzer}, {Ressler},
  {Cutri}, {Jarrett}, {Kirkpatrick}, {Padgett}, {McMillan}, {Skrutskie},
  {Stanford}, {Cohen}, {Walker}, {Mather}, {Leisawitz}, {Gautier}, {McLean},
  {Benford}, {Lonsdale}, {Blain}, {Mendez}, {Irace}, {Duval}, {Liu}, {Royer},
  {Heinrichsen}, {Howard}, {Shannon}, {Kendall}, {Walsh}, {Larsen}, {Cardon},
  {Schick}, {Schwalm}, {Abid}, {Fabinsky}, {Naes}, \& {Tsai}}]{wise10}
{Wright}, E.~L., {Eisenhardt}, P. R.~M., {Mainzer}, A.~K., {et~al.} 2010, \aj,
  140, 1868

\end{thebibliography}

\end{document}